\newif\ifpreprint
\preprinttrue 

\newif\ifSupplementary
\Supplementaryfalse

\ifpreprint
\documentclass[12pt]{natureprintstyle}
\else
\documentclass[12pt]{nature}
\fi

\usepackage[font=small,labelfont=bf]{caption}
\usepackage{booktabs,fixltx2e}
\usepackage[flushleft]{threeparttable}
\usepackage[breaklinks,colorlinks,linkcolor=black, citecolor=black, urlcolor=black]{hyperref}  

\usepackage{epsfig}
\usepackage{xcolor}
\usepackage{bm}
\usepackage{graphicx}
\usepackage{longtable}
\usepackage{amssymb}
\usepackage{tikz}
\usepackage{paralist}
\usepackage{lscape}
\usepackage[para,symbol]{footmisc}
\usepackage{aas_macros} 

\usepackage{amsmath}

\usepackage{soul}
\setstcolor{red}

\usepackage{placeins}
\usepackage{afterpage}

\DeclareMathOperator{\sech}{sech}

\newcommand{\be}{\begin{equation}}
\newcommand{\ee}{\end{equation}}

\newcommand\T{\rule{0pt}{2.6ex}}       

\def \basi{Bull.~Astron.~Soc.~India}

\def \jcap{J.~Cosm.~\&~Astropart.~Phys.}

\def \prl{Phys.~Rev.~Lett.}
\def \pr{Phys.~Rev.}

\def \aj{Astron.~J.}

\def\msun{{\,M_\odot}}

\def\lesssim{\lower.5ex\hbox{$\; \buildrel < \over \sim \;$}}
\def\gtrsim{\lower.5ex\hbox{$\; \buildrel > \over \sim \;$}}
\newcommand{\lsim}{\,\rlap{\raise 0.35ex\hbox{$<$}}{\lower 0.7ex\hbox{$\sim$}}\,}
\newcommand{\gsim}{\,\rlap{\raise 0.35ex\hbox{$>$}}{\lower 0.7ex\hbox{$\sim$}}\,}

\newcommand{\bea}{\begin{eqnarray}}
\newcommand{\eea}{\end{eqnarray}}

\renewcommand{\apj} {Astrophys. J.}
\renewcommand{\apjl} {Astrophys. J. Lett.}
\renewcommand{\apjs} {Astrophys. J. Suppl.}
\renewcommand{\mnras} {Mon. Not. R. Astron. Soc.}

\renewcommand{\araa} {Ann. Rev. Astron. Astrophys.}
\renewcommand{\aap} {Astron. Astrophys.}

\title{Galactic Bulge Preferred Over Dark Matter for
the Galactic Center Gamma-Ray Excess}

\author{
Oscar~Macias$^{1}$,
Chris~Gordon$^{2}$,
Roland~M.~Crocker$^{3}$,
Brendan~Coleman$^{2}$,
Dylan~Paterson$^{2}$,
Shunsaku~Horiuchi$^{1}$\&
Martin Pohl$^{4,5}$}

\begin{document}
\onecolumn
\maketitle

\begin{affiliations}
 \item Center for Neutrino Physics, Department of Physics, Virginia Tech, Blacksburg, VA 24061, USA
 \item School of Physical and Chemical Sciences,  University of Canterbury,
 Christchurch 8140, New Zealand
 \item Research School of Astronomy and Astrophysics, Australian National University, Canberra, Australia
 \item Institute of Physics and Astronomy, University of Potsdam, 14476 Potsdam-Golm, Germany
\item DESY, Platanenallee 6, 15738 Zeuthen, Germany
\end{affiliations}

\vspace{0.5 cm}

\begin{abstract}
\textbf{An anomalous gamma-ray excess emission has been found in Fermi Large Area Telescope 
data\cite{FermiInstrument2009} covering the centre of the Galaxy\cite{Goodenough2009gk,
FermiGCE2017}. 
Several theories have been proposed for this `Galactic Centre Excess'.
They include self-annihilation of dark matter particles\cite{
CaloreCholisWeniger2015
}, an unresolved population of millisecond pulsars\cite{AbazajianKaplinghat2012
}, an unresolved population of young pulsars\cite{OLeary2015}, or a series of burst events\cite{Cholis2015}. Here we report on a new analysis that exploits hydrodynamical modelling to register the position of interstellar gas associated with diffuse Galactic gamma-ray emission. We find evidence that the Galactic Centre Excess gamma rays are statistically better described by the  stellar over-density in the Galactic bulge and the nuclear stellar bulge, rather than a spherical excess. Given its non-spherical nature, we argue that the Galactic Centre Excess is not a dark matter phenomenon but rather associated with the stellar population of the Galactic bulge and nuclear bulge. }

\end{abstract}

The main challenge in pinning down the properties of the Galactic Centre Excess (GCE) is the  modelling of diffuse Galactic emission from the interaction of cosmic rays with interstellar gas and radiation fields, by far the dominant source of gamma-rays in this region. The Fermi-Large Area Telescope (LAT) Collaboration designed a diffuse Galactic emission model based on a 
 template\cite{Casandjian:andFermiLat2016} approach that is optimized to single-out gamma-ray point sources.
 This  approach presupposes that the diffuse Galactic emission can be modelled as a linear combination of interstellar gas, inverse Compton 
 maps, and several other diffuse components. Due to the limited kinematic resolution of gas tracers towards the Galactic Centre (GC), interstellar gas correlated gamma rays from the GC direction are  difficult to disentangle. Previous studies\cite{MaciasGordon2014,CaloreCholisWeniger2015,FermiGCE2017} utilized interstellar gas maps that were constructed with an interpolation approach that assumed circular motion of interstellar gas. This kinematic assumption provides for an estimate of the distance to a 
 part of the interstellar gas. However, it is  well established that the Galaxy contains a central bar which causes non-circular motion of interstellar gas in its inner regions, so assuming circularity introduces a significant and avoidable bias to gamma-ray analyses of the GC region\cite{Pohl2008}. 

We use Fermi-LAT data accumulated between August 4, 2008 and September 4, 2015 in the $15^{\circ}\times15^{\circ}$ region around the GC. Hydrodynamical simulations\cite{Pohl2008}
that account for the  effects of the Galactic bar were used to better determine the diffuse Galactic gamma-ray emission.  
To  evaluate the impact that the choice of interstellar gas models have on our results, we also constructed atomic and molecular hydrogen gas maps using an interpolation approach that reproduced those used in most previous gamma-ray analyses of the GC. We split each into 4 concentric rings, each with its own normalization parameter.
Details of the model components and approach are provided in the Methods section.

\begin{figure*}
\centerline{
\includegraphics[width=1.0\linewidth,trim=4pt 10pt 5pt 5pt, clip=false]{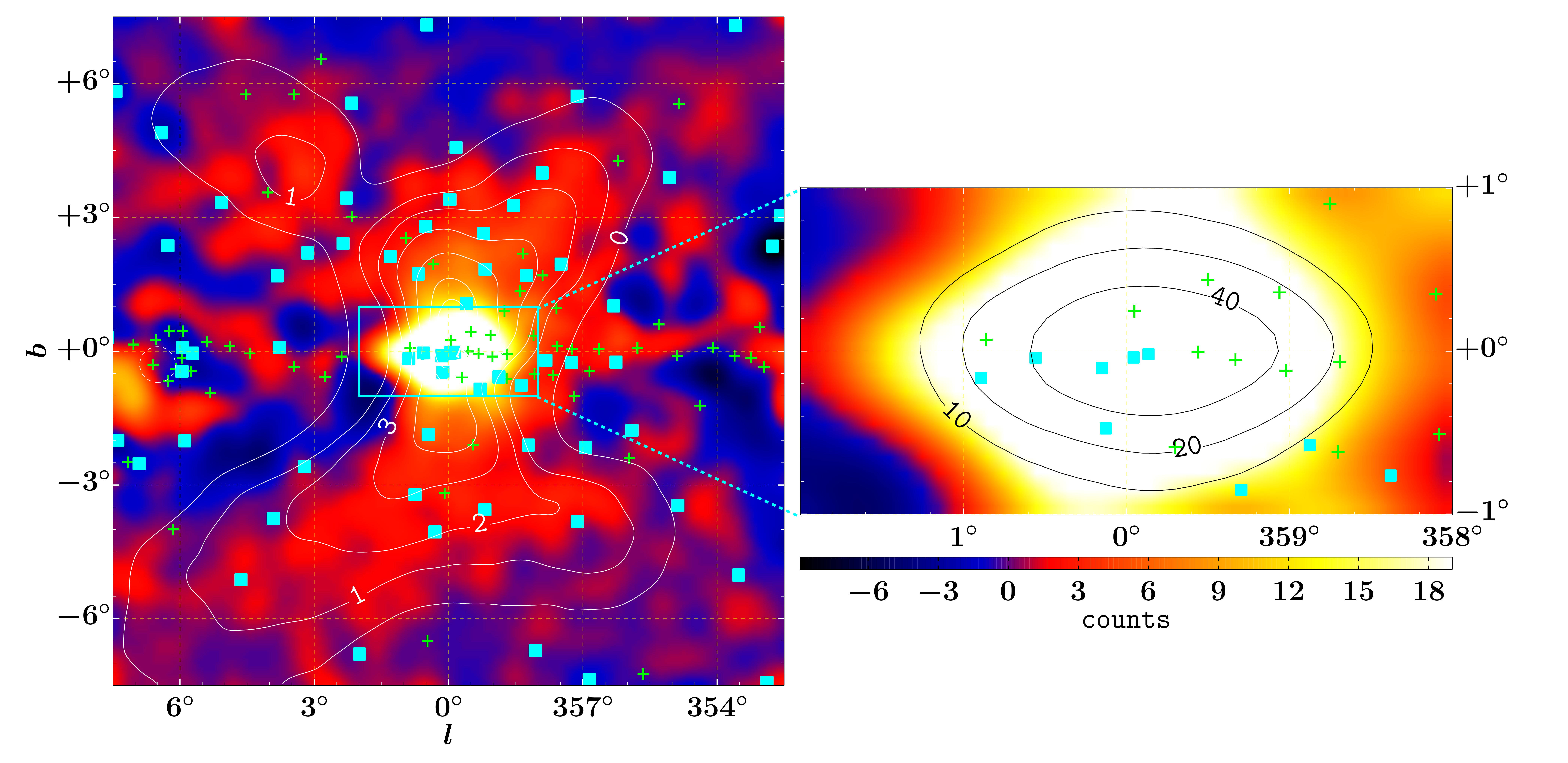}
}
\caption
{
\textbf{Residual map of the \boldmath$15^\circ \times 15^\circ$ region of interest for $E \ge 667$ MeV}. The residuals are obtained as $(\mbox{Data} - \mbox{Model})$, where the model
includes 
 previously-detected 3FGL point sources (cyan squares)\cite{3FGL}, 64 additional point source candidates (green crosses) and
 the standard diffuse Galactic emission components related to the interstellar gas and radiation field. 
 The white contours 
are the best-fit model counts from the X-bulge map obtained from analyses of WISE\cite{NessLang2016} infrared data after convolution the Fermi-LAT instrument response function.
The addition of a template based on the X-bulge 
 significantly improved the model fit to the gamma-ray data. 
The cluster of point sources on the Galactic plane at 
 $l\approx6^\circ$ may be associated with the W28 (white dashed circle) supernova remnant\cite{Ajello2016,3FGL}. The zoomed-in region on the right shows the correlation with the near-infrared stellar density nuclear bulge data\cite{Nishiyama2015}, the black contours display the best-fit model counts associated with this component after convolution with the Fermi-LAT instrument response function. 
The X-bulge and nuclear bulge templates were included when the best fit parameters for the above model were found, but not when evaluating the above residuals.  A Gaussian with radius 0.3$^\circ$ was used to smooth the images and the upper limit of the colour scale has also been clipped for display purposes.
 \label{fig:X-bulge_correlation}}
\end{figure*}

Interstellar gas map templates constructed using the results of hydrodynamical simulations were found to be a better description of the diffuse gamma-ray data than the standard interpolation-based maps with a log likelihood ratio $\approx$ 1362. 
As we have additional data compared to that used to construct the Fermi-LAT 3FGL catalogue\cite{3FGL} and since we also use a different Galactic diffuse emission model, we searched for new point sources. We found 64
 candidates (each with significance $\ge 4\sigma$) in our region of interest 
 which are shown as green crosses in Fig.~\ref{fig:X-bulge_correlation}. We  found multi-wavelength counterparts for 18 of our 64 point source candidates. 
This is similar to the 3FGL catalogue\cite{3FGL}
where some $\sim 1/3$ of the point sources do not have a multi-wavelength associations, especially in the GC region where there is high extinction and the diffuse Galactic emission model is more likely to require corrections.
Given our point source candidates have high statistical significance, including them quantitatively affects our results; however, they do not qualitatively affect our conclusions (see the Systematic Errors subsection of the Methods).


Our dark matter template for the GCE is modelled by the square of an Navarro-Frenk-White (NFW) template with an inner slope of 1.2. 
 Note that when all the uncertainties are accounted for, the dwarf spheroidals do not definitively rule out the dark-matter interpretation of the GCE\cite{Keeley2017}. Extended gamma-ray emission in the GC may also arise from unresolved sources such as millisecond pulsars (MSPs)\cite{AbazajianKaplinghat2012,MaciasGordon2014} or young pulsars\cite{OLeary2015}, both of which have GeV-peaked gamma-ray spectra. Studies have also shown detectable non-Poissonian features in photon statistics\cite{Lee_etal:2016,Bartes_etal:2016}. However, these may be due to defects in the Galactic emission model\cite{Horiuchi2016}.
The young pulsar hypothesis requires  recent star formation given the $\lsim$ few Myr $\gamma$-ray lifetimes of ordinary pulsars. Such star formation is absent from most of the bulge except in the $r \lsim 100$ pc nuclear region; a young pulsar explanation of the GCE thus requires that the bulge contain pulsars that are launched out of the nucleus. It has been claimed that this can be achieved   by the pulsar's natal kicks\cite{OLeary2015}. While MSPs can be generated  from old stellar populations\cite{Ploeg2017}.

We thus also consider Galactic bulge stellar templates. Almost half the stars\cite{Portail2015} in the Galactic bulge are on orbits that contribute to the appearance (from the Earth) of an X-shaped over-concentration (the `X-bulge')\cite{Nataf2010}. This structure has been   revealed in an  analysis\cite{NessLang2016} of 3.4 and 4.6 micron data collected by the WISE telescope\cite{WISE:instrument}. However, ref.~\citenum{JooLeeChung2017} argued that the X-shape is a processing artefact. Our aim is \textit{not} to scrutinize what is the correct bulge template, but rather, to explore the bulge as an example astrophysical template for the GCE that is alternative to dark matter.
Using standard maximum-likelihood estimation with an X-bulge template based on the average of the 3.4 and 4.6 micron data, we find that the addition of our X-bulge template improves the fit to the gamma-ray data at about the $16.1\sigma$ level (see Table~1).
 Note that we do not claim that the X-bulge provides a fit 16.1-sigma better than any other possible bulge stellar template. 
What we have done is evaluate how much the likelihood improved when an X-bulge was added relative to the likelihood without any new component being added to account for the GCE. 
 We also tried a boxy shaped bulge (see Methods section), and found that it improves the fit to the data at the 14.6$\sigma$ level. 
Importantly, the last row of Table I demonstrates that even when the boxy bulge stellar map is considered instead of the X-bulge, the dark matter template is still confidently ruled out as an explanation of GCE. As the X-bulge has a slightly higher $\sigma$ than the boxy bulge,  we used the X-bulge in the rest of the article. Similar, results would be obtained with the boxy bulge template.

A further  stellar component within the wider Galactic bulge is the so-called nuclear bulge.
This disky distribution of stars concentrated within $\sim 230$ pc radius of the Galactic nucleus has experienced on-going star formation over the life of the Galaxy and represents $\sim 10$\% of the overall bulge mass\cite{LaunhardtZylkaMezger2002}. 
To determine whether the nuclear bulge template 
improves the fit to the Fermi-LAT gamma-ray data, we used a map constructed from a near-infrared stellar density measurement\cite{Nishiyama2015} and found a $10\sigma$ improvement in the fit (see Table~1). In Fig.~\ref{fig:X-bulge_correlation} we show the residual gamma-ray map obtained after subtraction of our best-fit galactic diffuse emission and point sources model. 
As can be seen, the X-bulge and nuclear bulge are well traced by our residual gamma-ray maps. The correlation with the X-bulge is more evident away from the plane where the hard-to-model diffuse Galactic emission is no longer so dominant and the Poisson noise is much lower. We have performed a  $\pm 1^\circ$ masking of the plane, and found consistent results (see Methods section).

\begin{figure}
\centering
\centerline{
\includegraphics[width=1\linewidth,trim=4pt 10pt 5pt 5pt, clip=false ]{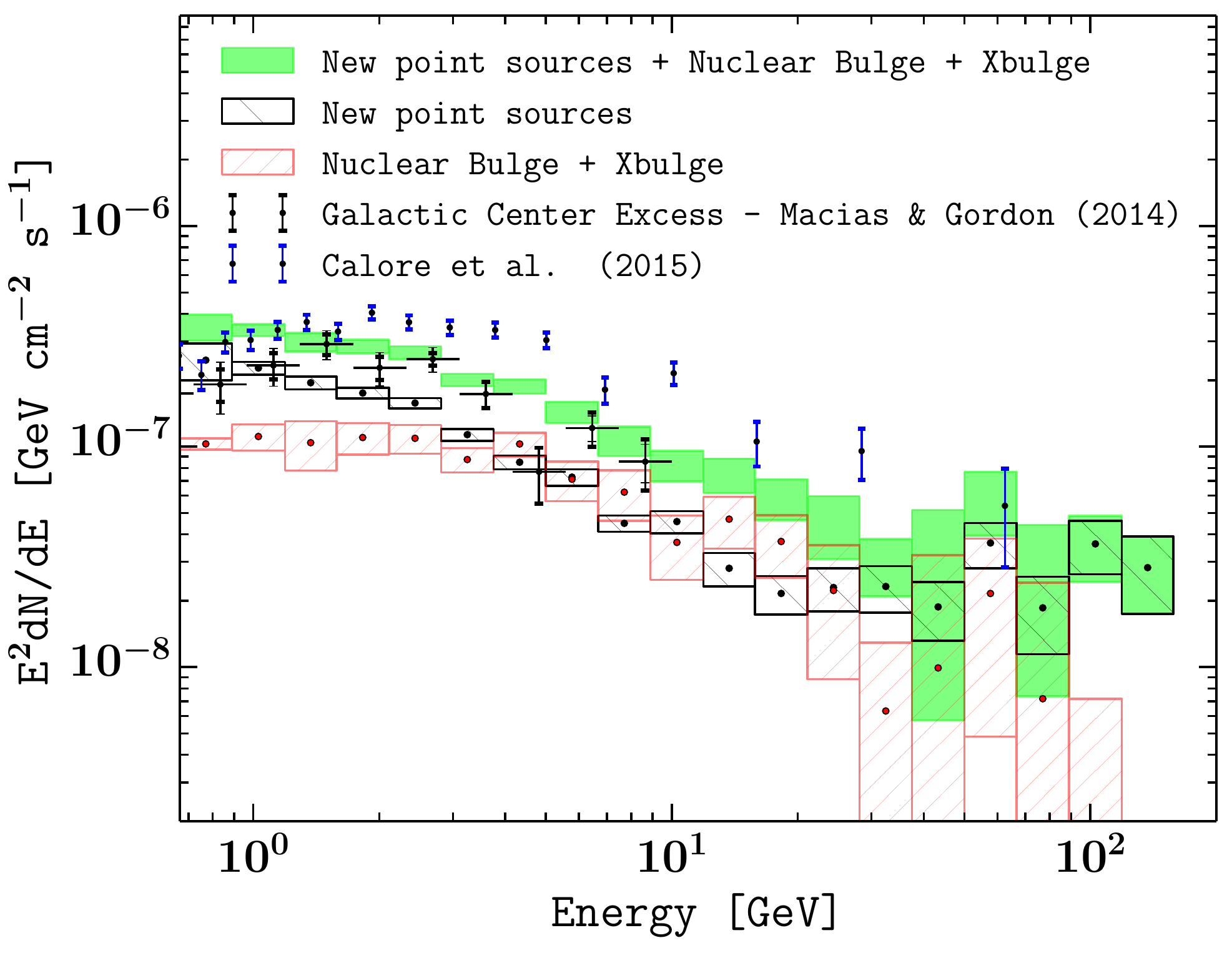}
}
\caption
{
\textbf{Differential flux of the new, statistically significant components in the Galactic Centre. }
The black boxes are the combined spectrum of the 64 new point source candidates, red boxes are the superposition of the nuclear bulge and X-bulge differential fluxes, while the green boxes display the sum of these three components and are compatible with previous results\cite{MaciasGordon2014,CaloreCholisWeniger2015},
although, the comparison should only be taken as qualitative as the previous results are for slightly different region of interest.
The box sizes encompass the 68\% confidence intervals. We found the combined nuclear bulge and X-bulge spectrum prefers an exponential cut-off to a power law model at 5$\sigma$. 
\label{fig:X-bulge+NBstars_spectra}}
\end{figure}

Best-fit spectral parameters were found using $\chi^2$ fitting to the inferred flux points for the energy bins. Relative to a power-law spectrum, the preferred spectral model (at $3.5\sigma$ and $5.1 \sigma$ respectively) for both the X-bulge and nuclear bulge templates was a power law with an exponential cut-off ($dN/dE\propto E^{-\Gamma}\exp(-E/E_{\rm cut})$, where $N$ is the photon flux). The X-bulge had a spectral slope of $\Gamma=1.9\pm 0.1$, an energy cut-off $E_{\rm cut}=10\pm 5$ GeV and a luminosity $L=(4.5\pm 0.3)\times 10^{36}$ erg/s  for $E\ge 100$ MeV. Similarly, the fit for the nuclear bulge yielded $\Gamma=1.9\pm 0.1$, $E_{\rm cut}=13\pm 4$ GeV and $L=(3.3\pm 0.3)\times 10^{36}$ erg/s. Here and in the rest of the article 68\% confidence intervals are used for error bars and we adopted 8.25 kpc for the   distance to the GC. When the X-bulge and nuclear bulge spectra were combined  (see  Fig.~\ref{fig:X-bulge+NBstars_spectra}), we found a power law exponential  model was still preferred relative to a  power law ($dN/dE\propto E^{-\alpha} $) model  at 5.0$\sigma$. The best fit spectral parameters for this case  
are shown in  Table~2.
The combined X-bulge and nuclear bulge spectral parameters are compatible with previous estimates of the GCE based on templates of dark matter as well as resolved MSPs and globular clusters containing MSPs\cite{MaciasGordon2014}. 
 
When the X-bulge and nuclear bulge templates are included in the fit, we found that a squared NFW template with an inner slope of 1.2 was not significantly detected, see Table~1. 
 The 95\% upper limit on this component's luminosity was found to be  $5\times 10^{36}$~erg/s. 
The luminosity for an NFW-squared profile from our model baseline+NP+NFW, displayed in  
Table~2, is
consistent with previous estimates\cite{MaciasGordon2014} and is about an order of magnitude larger than the limit we obtain when the X-bulge and nuclear bulge templates are included.
Similarly, a dark matter template based on the square of 
 an NFW profile with an inner slope of 1.0 was undetected.

\begin{table*}[!htbp]\caption{{\bf Summary of the Likelihood analysis results\label{Tab:likelihoods}}}
\centering
\begin{threeparttable}
\begin{tabular}{llllcrc}
\hline\hline
Base  & Source &  $\log(\mathcal{L}_{\rm Base})$  & $\log(\mathcal{L}_{{\rm Base}+{\rm  Source}})$  &  $\mbox{TS}_{\rm Source}$& $\sigma$ & Number of\\ 
           &              &                                  &                         &                                               & & source parameters\\\hline
baseline  & FB&     -172461.4          &   -172422.3    &  78 & 6.9 & 19\\ 
baseline  & NFW-s &    -172461.4       &  -172265.3       & 392 & 18.4 &  19\\  
baseline  & Boxy bulge& -172461.4         & -172238.7          &445 & 19.7   & 19\\
baseline  & X-bulge& -172461.4         &  -172224.1         & 475&  20.5  & 19\\
baseline  & NFW &    -172461.4         &  -172167.9        & 587  &23.0 &  19\\  
baseline  & NB&   -172461.4         &  -171991.8    & 939  & 29.5 & 19\\
baseline  & NP&     -172461.4          &   -169804.1    & 5315& 55.7  &  $64\times 19$\\\hline
baseline$+$NP  & FB&     -169804.1    & -169773.6      &61&  5.8   & 19\\
baseline$+$NP  & NB&     -169804.1  & -169697.2      &214  &  13.0 & 19\\
baseline$+$NP  & Boxy bulge&     -169804.1  & -169663.7      &  281& 15.3 & 19\\
baseline$+$NP  &NFW&     -169804.1   &   -169623.3         &  362& 17.6   & 19\\
baseline$+$NP  & X-bulge&     -169804.1  &   -169616.2   &376 &  18.0  & 19\\
\hline
baseline$+$NP$+$X-bulge     &NFW & -169616.2    &-169568.4 & 96& 7.9 & 19 \\ 
baseline$+$NP$+$X-bulge     & NB & -169616.2    & -169542.0 & 148&10.4& 19 \\ \hline
baseline$+$NP$+$X-bulge$+$NB  & NFW& -169542.0         &-169531.0      & 22 & 2.4  & 19\\
baseline$+$NP$+$X-bulge$+$NB  & FB& -169542.0         &-169525.5      & 33   & 3.5& 19\\ 
 \hline
baseline$+$NP$+$NB  & X-bulge&  -169697.2       &  -169542.0    & 310   & 16.1& 19\\
baseline$+$NP$+$NB  & Boxy bulge&  -169697.2       &  -169566.0    &  262  & 14.6& 19\\ 
baseline$+$NP$+$NFW  & X-bulge+NB&  -169623.3       &  -169531.0    & 185   & 10.8& $2\times 19$\\
baseline$+$NP$+$NFW$+$NB     & X-bulge & -169598.9    & -169531.0 &136 &9.9& 19 \\
baseline$+$NP$+$Boxy bulge$+$NB  & NFW& -169566.0         &-169553.3      & 25 & 2.7  & 19\\
\hline\hline
\end{tabular}
\begin{tablenotes}
      \item The \textit{baseline} model consists of all 3FGL point sources in the region of interest, Loop I, an IC template predicted by GALPROP, the hydrodynamic based gas maps, the recommended isotropic emission map, and a model for the Sun and the Moon.
 Other model templates considered are: the 64 new point sources (NP), the square of a generalised NFW profile with an inner slope $\gamma=1.2$ or the square of a ``standard NFW'' (NFW-s) with inner slope $\gamma=1$, an infrared X-bulge and a Boxy bulge template tracing old stars in the Galactic bulge, 
       a nuclear bulge (NB) template 
      and a template accounting for the Fermi Bubbles (FB). 
      The maximized likelihoods ($\mathcal{L}$) are given for the Base and Base$+$Source models and the significance of the new source is given by 
       TS$_{\rm Source}\equiv 2(\log(\mathcal{L}_{{\rm Base}+{\rm  Source}}) - \log(\mathcal{L}_{\rm Base}))$. Note that for both likelihoods all parameters are maximized and so the $\mathcal{L}_{{\rm Base}+{\rm  Source}}$ will have additional parameters whose number is given in the last column of the table. 
       The conversion between TS$_{\rm Source}$ and $\sigma$ is discussed in the Methods section.
    \end{tablenotes}
\end{threeparttable}
\end{table*}



\begin{table*}[!htbp]
\centering
\caption{\bf Exponential cut-off best-fit parameters with statistical and systematic errors to the X-bulge + nuclear bulge}
\begin{tabular}{ | r | r | r | r | r | r | r | r | r | }
\hline
	\multicolumn{1}{|c|}{Parameter}& \multicolumn{1}{|c|}{Best fit} & \multicolumn{1}{|c|}{Statistical} & 
    \multicolumn{5}{|c|}{Systematic Error} \\ \cline{4-8}
	   &  & \multicolumn{1}{|c|}{Error} & \multicolumn{1}{|c|}{Spin}& \multicolumn{1}{|c|}{Inverse} & \multicolumn{1}{|c|}{Dust} & \multicolumn{1}{|c|}{Fermi} &\multicolumn{1}{|c|}{Total}\\ 
        &  &  & \multicolumn{1}{|c|}{Temperature}      &  \multicolumn{1}{|c|}{Compton}  &      & \multicolumn{1}{|c|}{Bubbles}  &   \\ \hline 
$\Gamma$      & 2.0      & 0.1         & 0.1        & 0.1  & 0\phantom{.0}  & 0.1 & 0.2 \\
$E_{\rm cut}/$GeV & 13\phantom{.0}       & 5\phantom{.0}            & 1\phantom{.0}           & 3\phantom{.0}  & 1\phantom{.0}     & 2\phantom{.0}    & 4\phantom{.0}            \\
$L/(10^{36}$erg$/$s$)$              & 6\phantom{.0}         & 1\phantom{.0}            & 1\phantom{.0}           & 1\phantom{.0}   & 0\phantom{.0}     & 0\phantom{.0}    & 1\phantom{.0}  \\
\hline
\end{tabular}
\T
\parbox[t]{0.8\textwidth}{\label{SysErr}
The statistical errors  are 1$\sigma$. The total systematic error was obtained by adding in quadrature the individual systematic errors. 
Luminosities were computed for $E>100$ MeV.}
\end{table*}

The Fermi Bubbles\cite{Fermi:LatBubbles,Su:etal} 
are lobes that extend up to  $\sim 7$  kpc from the Galactic plane. 
The boundaries of the Fermi bubbles are described by two catenary curves\cite{Casandjian:andFermiLat2016}, see Methods section.  
However, the catenary  geometry does not match the
 excess we
see in our residual and test statistic (TS) maps in Fig.~\ref{fig:X-bulge_correlation} and Supplementary Fig.~\ref{fig:tsmaps}.  We found the addition of a catenary template did not qualitatively affect our results. As this template had a low TS value (see Table~1) we did not include it in our final results but rather treated its potential presence as a systematic effect (see Table~2, Methods section, and Supplementary Sec.~3).

A comparison between the properties of our proposed X-bulge and nuclear bulge MSPs with other MSP populations can be obtained by comparing the MSP luminosity to stellar mass ratio. Note that this comparison assumes the ratio of MSP mass to stellar mass is constant across the different considered regions which will only be approximately correct. The stellar mass\cite{Bland-Hawthorn2016} of the Galactic bulge is $1.5 \times 10^{10}$ solar masses ($M_\odot$) and the nuclear bulge has a mass of $\sim 1.4 \times 10^9 \msun$. According to ref.~\citenum{Portail2015}, instantaneous stellar mass contributing to the appearance of the X-bulge over-density represents about 25\% of the Galactic bulge.
However, as can be seen by comparing 
our white contours in Fig.~\ref{fig:X-bulge_correlation}
 to those in Fig.~18 of ref.~\citenum{Portail2015},
the method they use to determine this contribution underestimates the X-bulge we extract from the WISE data as it eliminates the central regions of the X-bulge that we have included in our template. Therefore, we estimate our X-bulge has an instantaneous stellar mass $\gtrsim 4 \times 10^{9} M_\odot$.
From these data we infer a $E\ge  100$ MeV luminosity-to-mass (LtM) ratio of  $\lesssim 2 \times 10^{27}$ erg/s/$M_\odot$ for the combined X-bulge and nuclear bulge structure.
This is similar to the LtM ratio we infer from ref.~\citenum{Winter2016} for the MSP emission from the entire Milky Way of 
$\sim 2 \times 10^{27}$ erg/s/$M_\odot$ and  less than the
  $\sim 5 \times 10^{28}$ erg/s/$M_\odot$ for globular cluster 47 Tuc\cite{FermiGlobularClusters}
(which has an MSP-dominated $>$ 100 MeV luminosity of $(4.8\pm 1.1)\times 10^{34}$ erg/s
 and stellar mass of about $10^6 M_\odot$).


We have shown that the X-bulge (or boxy bulge) plus the nuclear bulge provide a better fit than the dark matter explanation. We have not explicitly checked the burst model, but current implementations 
are highly model dependent and  fine-tuned\cite{Cholis2015}. Also, the young pulsar explanation is typically associated with spherically symmetric templates\cite{OLeary2015} and so also disfavoured compared to our MSP based X-bulge (or boxy bulge)$+$nuclear bulge explanation.
Therefore,
we have shown, in agreement with other recent results\cite{Bartels2017},
that the Galactic bulge stellar distribution is preferred over a spherically symmetric NFW-squared  excess template.



\FloatBarrier

\begin{addendum}

 \item  RMC was the recipient of an Australian Research Council Future Fellowship (FT110100108). SH is supported by the U.S. Department of Energy under award number de- sc0018327. We thank Dustin Lang for  making available code and data which helped with generating the  X-bulge template and both
 Shogo Nishiyama 
and Kazuki Yasui for providing the data for the Nuclear bulge template. 
We
acknowledge the use of public data and software from the  Fermi data archives (http://fermi.gsfc.nasa.gov/ssc/). Finally,
the authors would also like to  thank Felix Aharonian, Anthony M. Brown,   Francesca Calore, 
Jean-Marc Casandjian,
H. Thankful Cromartie, Seth Digel, Torsten En{\ss}lin, Manoj Kaplinghat,
Ken Freeman, Ortwin Gerhard, Oleg Gnedin, Xiaoyuan
Huang, Naomi McClure-Griffiths, David Nataf, Ben Roberts, Miles Winter, Richard Tuffs, and Gabrijela Zaharijas for enlightening discussions.

\item[Author contributions]  OM designed and performed the majority of the data analysis. OM also constructed the Fermi Bubbles, Sun, Moon, Inverse Compton and Loop I templates. 
CG processed the WISE data and derived the mixture distribution formulas. RMC suggested the link with the X-bulge. BC processed the hydrodynamical 3D map into annuli density maps. DP created the interpolated annuli density maps. CG and SH assisted with the PS modelling. BC and DP created the dust maps. MP created the three dimensional HI and CO maps.
All authors
  contributed to the interpretation of the results.  The text of the final manuscript was mainly written by OM and CG, but all authors did have some contribution.

 \item[Competing Interests] The authors declare that they have no
competing financial interests.

\end{addendum}

\FloatBarrier

\begin{methods} \label{Methods}

\subsubsection*{\\Observations}

We examined $\sim 7$ years of Fermi-LAT data\cite{FermiInstrument2009} (August 4, 2008$-$September 4, 2015) selecting  \textsc{Pass 8} \textsc{ULTRACLEANVETO} 
class events. The data was extracted from a square region of $15^{\circ}\times15^{\circ}$ centred at Galactic coordinates $(l,b) = (0,0)$ and made no distinction between \textit{Front} and \textit{Back} events. Furthermore, we restricted our analysis
to the 667 MeV to 158 GeV energy range and used the
P8R2$_{-}$ULTRACLEANVETO$_{-}$V6 instrument response functions. To avoid contamination from terrestrial gamma-rays, we used events with zenith angles smaller than 90$^{\circ}$. This work made use of the Fermi Science Tools \texttt{v10r0p5} software package.

Employing the \texttt{gtmktime} tool we selected the recommended data filters\\ (DATA$_{-}$QUAL$>$0)\&\&(LAT$_{-}$CONFIG==1). Spatial binning was performed with the \texttt{gtbin} utility with which we divided the LAT data into $150\times 150$ angular bins of size $0.1^{\circ}$ in a \texttt{CAR} sky projection.


\subsubsection*{Templates}
The Galactic diffuse gamma-rays resulting from the interaction of cosmic-ray electrons and protons with the interstellar gas and radiation field were modelled with a similar method used for the standard Galactic diffuse emission model\cite{Casandjian:andFermiLat2016}.
We fitted a linear combination of atomic and molecular hydrogen gas templates (Supplementary Fig.~\ref{fig:densities} and \ref{fig:dust_maps}),
 an inverse Compton (IC) 
 energy-dependent spatial template
  as obtained with GALPROP\cite{Galpropsupplementary}, specialized templates for the \textit{Sun} and the \textit{Moon}, an isotropic component (\texttt{iso$_{-}$P8R2$_{-}$ULTRACLEANVETO$_{-}$V6$_{-}$v06.txt}), and a model for the gamma-ray emission associated with Loop I (Supplementary Fig.~\ref{fig:LoopI}).

The atomic and molecular hydrogen gas column densities were each distributed within four Galactocentric annuli to account for the non-uniform cosmic-ray flux in the Galaxy. The construction of these templates is described in Supplementary Sec.~1.

\subsubsection*{Bin-by-Bin Analysis}
\label{sec:methods} 

Similar to other works\cite{Ajello2016,FermiGCE2017},  we employed a bin-by-bin analysis technique, in which we split the Fermi-LAT data into 19 logarithmically spaced energy bins. Within each energy bin we performed a separate maximum-likelihood fit\cite{3FGL} with the \textit{pyLikelihood} analysis tool.
The bin size was chosen to be larger than the LAT energy resolution, but narrow enough that the Galactic emission spectral components can be simply approximated by a power law model.
 We note that this bin-by-bin method enables us to evaluate the likelihood for a test source with an arbitrary spectral model and significantly reduces the CPU power required to reach convergence as only the flux normalization of the sources are free to vary during the fits.

Once the bin-by-bin method had converged, the inferred spectrum of each source was either fitted by a power law or an exponential cut-off model. 
When energy bins had 
 $TS<1$ or $\Delta F_i/F_i>1$ they were combined with adjacent energy bins until $TS>1$ and $\Delta F_i/F_i<1$. 

 The errors from the bin-by-bin fit were added in quadrature to the  errors caused by the uncertainties in the
effective area\cite{3FGL}. These effective area errors were taken to be $f_i^{\rm rel}$ times the predicted flux for bin $i$. Where $f_i^{\rm rel}$ is interpolated from the values given in ref.~\citenum{3FGL}.
The spectrum was modelled by an exponential cut-off if   
\begin{equation}
{\rm TS}_{\rm curvature}\equiv 2\left(\log {\cal L}(\mbox{exp.~cut-off})-\log
{\cal L}({\mbox{power~law}})\right)\ge 9
\end{equation}
where ${\cal L}(i)$ is the maximum likelihood value for model $i$.

\subsubsection*{Comparing Hydrodynamic and Interpolated Gas Templates}

Initially we fit the LAT data with a model comprised of the 3FGL\cite{3FGL}  point sources present in our region of interest plus four other spatially extended sources (HESS J1825-137, RX J1713.7-3946, W28 and W30) reported in the 3FGL\cite{3FGL}. The spatial templates used to model these extended sources correspond to Version 14.  

To identify the most suitable gas templates for our study, we performed a scan in which we evaluated the improvement of the likelihood fit to the region of interest when the gas maps used were the ones created with the interpolation method or the hydrodynamical method (see Supplementary Sec.~1 for details of their construction). Supplementary Fig.~\ref{fig:pohvsinterp} shows that the data preferred
 the hydrodynamical method.

During optimisation, the flux normalisation of the 3FGL sources were left free in each energy bin.
We also simultaneously fit the 13 diffuse components' (H{\small I} annuli, CO annuli, dust templates, Loop I, inverse Compton, and isotropic) normalisation but kept the \textit{Sun} and \textit{Moon} fluxes fixed to their nominal values.

\subsubsection*{Point Source Search}
\label{subsec:ptsrcsearch}

In order to check whether any additional point sources are required, we followed a methodology similar to that described in refs.~\citenum{2FGL} and~\citenum{3FGL}.
We started from our \textit{baseline} model, which consists of all 3FGL point sources in the region of interest, Loop I, an inverse Compton template predicted by GALPROP, the hydrodynamical based gas maps, the recommended isotropic emission map, and a model for the Sun and the Moon (see Supplementary Sec.~1). We examined the significance of a trial point source with a power-law spectrum, with a fixed slope of two, 
at the centre of each pixel. The outcome of this  was a residual TS map where the \texttt{gttsmap} utility was used for this step. 
In accordance with our bin-by-bin method, a residual TS map was computed for each energy bin and these were then added to get a total residual map for the full energy range.

From the total residual TS map we generated a list of all the pixel clusters with TS values above the detection threshold that looked reasonably isolated under visual inspection ($\sim 0.5^{\circ}$ of angular separation). The coordinates of the source candidates were calculated as the average of adjacent pixel positions weighted by their respective TS values\cite{2FGL}. A more sophisticated technique would be to fit the TS map with a two dimensional parabola. But, we found our results were not sensitive to the small difference in position that this gave.
To avoid convergence issues, at each step we added
only the ten (or fewer if less were available) brightest seeds  to our model and reran the bin-by-bin analysis routine where {\em all\/} components $-$ including the new point sources $-$ were simultaneously fitted.
Only the seeds that were found to have a TS above the detection threshold from this step were allowed to stay in the model. Note that this is an iterative procedure that goes from bright sources to faint ones. This procedure was repeated seven times in our region of interest until no seeds were found or no more point source candidates passed the thresholding step. This made the method more robust against source confusion. 
%
When doing a global analysis, as opposed to a bin-by-bin analysis, each new point source candidate has two parameters for the power law and two parameters for its position. In that case a $\mbox{TS} \geq 25$ (which corresponds to 4$\sigma$) is used as the detection condition\cite{mattox,3FGL}. However, we used a bin-by-bin analysis with 19 energy bands, where in each band the point source amplitude was not allowed to take on a negative value. As shown in  Supplementary Sec.~2, we thus have a mixture distribution given by 
\begin{equation}
p({\rm TS})= 2^{-n}\left(\delta({\rm TS}) +\sum_{i=1}^{n} \binom{n}{i} \chi_{i+2}^2({\rm TS})\right)
\label{eq:pTS}
\end{equation}
where  $\delta$ is the Dirac delta function, $\binom{19}{i}$ is a binomial coefficient, and $\chi_{i+2}^2$ is a $\chi^2$ distribution with $i+2$ degrees of freedom.
To work out the number of  $\sigma$ of a detection we evaluate the equivalent p-value for a one new parameter case\cite{wilks}:
\begin{equation}
\mbox{Number of $\sigma$}\equiv \sqrt{\rm InverseCDF\left(\chi_1^2,{\rm CDF}\left[p(\mbox{TS}),\hat{{\rm TS}}\right]\right)}
\end{equation}
where CDF and InverseCDF are the cumulative distribution and inverse cumulative distribution respectively. The first argument of each of these functions is the distribution function and the second is the value the CDF or InverseCDF is evaluated at. The observed TS value is denoted by $\hat{\rm TS}$.
It follows that 
we use a threshold of  $\mbox{TS} \ge 41.8$ to correspond to a 4$\sigma$ detection.  

The total set of new point source candidates found in this work are displayed in Supplementary Fig.~\ref{fig:tsmaps} along with the TS residual map obtained in our last iteration. Although the model including all the new point sources is a much better representation of the region of interest, a few hot spots still remain. These are, however, found to be below the detection threshold of $\mbox{TS}\ge 41.8$ in the maximum likelihood step. As can be seen from  Supplementary Fig.~\ref{fig:tsmaps}, the X-bulge morphology is clearly visible in the residual TS values. Although, along the Galactic plane, in comparison to  Fig.~\ref{fig:X-bulge_correlation} 
the X-bulge morphology is shifted to negative longitudes (a similar shift is seen in refs.~\citenum{Casandjian:andFermiLat2016}, \citenum{YangAharonian2016}, and \citenum{FermiGCE2017} with the use of different methods), this may be due to degeneracies between the X-bulge template and the amplitude of the 3FGL and new point sources  around $l=0,b=0$. A similar phenomenon can be seen in Figs.~1 and 2 of ref.~\citenum{GordonMacias2013}. The Poisson noise is also very large in this region and so the mismatch did not have much bearing on the fit.

To identify possible multi-wavelength counterparts to the gamma-ray sources we searched in the seed locations 
within the 68\% containment of the point spread function  for one of our highest energy bands $\sim 0.1^\circ -$ around each source in the ATNF pulsar\cite{ATNF:catalog}, globular cluster\cite{GBC:catalog}, supernova remnant (SNR)\cite{Green:SNRcatalog} and the Roma-BZCAT blazar\cite{Massaro2009} catalogues for potential gamma-ray emitters. We found spatial overlaps for 18 of our 64 point source candidates (see Supplementary Table~1). Note that this does not preclude the possibility that the other 46 point source candidates are real sources since high extinction towards the GC makes it difficult to have very complete  multi-wavelength source catalogs.

Our new point sources are compared to the 2FIG\cite{Ajelloetal:2017} in Supplementary Fig.~\ref{fig:ptsrcs_comparison}. As can be seen, there is a reasonable overlap considering that the two analyses used different diffuse galactic emission templates and that 2FIG was based on a wider energy range (0.3 to 500 GeV) and time interval (7.5 years). It is reassuring that the majority of new 2FIG sources that are not associated with one of our new point sources are on a hot spot of our TS map. The list of point sources found in this work are provided in Supplementary Table 1.

 Including the 3FGL sources, we had a total of 116 point source candidates in our $\left|l\right|\leq 7.5^\circ$, $\left|b\right|\leq 7.5^\circ$ region of interest. This is compatible with the 127 point source candidates found by ref.~\citenum{FermiGCE2017} for a disk like region of interest with radius 10$^\circ$.

\subsubsection*{New Templates}

\subsection*{X-bulge:}

We followed a procedure based on the method described in ref.~\citenum{NessLang2016} applied to the WISE data.
To use this template in comparisons with the Fermi-LAT data, all pixels values below zero were set to zero in each median filtered exponential subtracted map and our template was then constructed by taking the average of the two resultant maps. The resulting template is displayed 
 by the white contours in Fig.~\ref{fig:X-bulge_correlation}.

\subsection*{Boxy Bulge:}

We assumed the triaxial model for the Galactic bulge derived in ref.~\citenum{Freudenreich:1998}. This was obtained by fitting to COBE/DIRBE near-infrared ($1.25-4.9$~$\mu$m) data. We adopted the best-fitting model in that reference, which is \textit{Model S}. This consists on a $\sech^2$ function on the bar radial spatial profile. The template used in our analysis is shown in Supplementary Fig.~\ref{fig:Boxybulge}. 

\subsection*{Nuclear Bulge:}

 We used a map constructed from a near-infrared stellar density measurement of the central region of our Galaxy ($|l|\geq 3^\circ$ and $|b|\geq 1^\circ$) and subtracted a best fit Galactic disk component\cite{Nishiyama2015}. 
 In order to remove artificial sharp boundaries in the map induced by survey patches, all pixels below 15 stars/arcmin$^2$ were set to zero. The resulting template is displayed 
by the black contours in Fig.~\ref{fig:X-bulge_correlation}. 

\subsection*{Dark Matter:}

We modelled the potential annihilating dark matter signal in the GC as the square of an NFW density profile with an inner slope of 1.2, which had been shown to describe the GCE well in previous works\cite{
hooperlinden2011,
Abazajian2014,Daylan:2014
}.
%
The square of an NFW density profile 
is representative of a tentative annihilating dark matter signal in the GC. 

\subsection*{Fermi Bubbles:}
As found by  Acero et al.\cite{Casandjian:andFermiLat2016}, we use
two catenary curves of the form $10.5^\circ \times (\cosh((l - 1^\circ)/10.5^\circ) - 1^\circ)$ and $8.7^\circ \times(\cosh((l+1.7^\circ)/8.7^\circ) - 1^\circ)$ for the Northern and the Southern bubbles, respectively.



\subsubsection*{Testing Extended Emission Templates}
\label{Sec:extendedemission}

We fitted the gamma-ray emission with our bin-by-bin method to derive fluxes that are independent of the choice of spectral model. Within each bin, the spectrum of the included point and extended sources were modelled as power laws with fixed spectral index of two. Due to the small size of the bins, our results were not sensitive to the precise spectral index  used.
In each energy bin, the amplitudes of all included
point sources and all included extended templates were simultaneously fit.
This allowed us to effectively marginalise over the statistical uncertainties.
Table~1 shows the steps we took to evaluate whether a template was significantly detected.
We started with the \textit{baseline} model and then evaluated the TS of each new template. We then added the template with the highest TS to our model and repeated the procedure with this appended to the Base model. We iterated through these steps until the highest TS-value of a new template was below our $4\sigma$ threshold.
For each new template there are  $n\times 19$ new parameters, where $n$ is an integer. The probability distribution is the same as equation~(\ref{eq:pTS}) except that 
the 19 should be replaced by $n\times 19$ and 
the $\chi^2_{i+2}$ should be replaced with $\chi^2_i$
as we are not fitting the positions of the template.  This is the same formula as in case 9 of ref.~\citenum{SelfLiang87}. For one new template being considered (i.e.\ 19 new parameters),  our
 $4\sigma$  detection threshold corresponded to $TS\ge 38.4$. 


The contributions of the different components for our proposed model of the GCE are shown in Supplementary Fig.~\ref{fig:ptsrc_spectra}. 
The fractional residuals are shown in Supplementary Fig.~\ref{fig:Residuals}. As can be seen from the top left  panel, the X-bulge is needed even when the interpolated gas maps are used. Comparing the top and bottom right panels shows that the X-bulge morphology has been accounted for by our template.


\subsubsection*{Molecular-Hydrogen-To-CO Conversion Factor}
Our inferred molecular-hydrogen-to-CO conversion factor ($X_{CO}$) is shown in Supplementary Table~\ref{X_CO}. As in ref.~\citenum{Casandjian:andFermiLat2016} we evaluated $X_{CO}$ using our $2$~GeV energy bin fit results. Our values are consistent with the results from all sky fits as can be seen from comparing to Fig.~25 of ref.~\citenum{ackermannajelloatwood2012}. Although our 8 to 10 kpc best fit $X_{CO}$ is somewhat larger than the all sky and high latitude\cite{Casandjian2015} measurements, it does also have large error bars and so is still consistent with the other measurements at about  2$\sigma$ or better level even when only statistical errors are accounted for. As in ref.~\citenum{ackermannajelloatwood2012}, our $X_{CO}$ values for the outer annuli were strongly biased because the model  under-fitted the data in the outer galaxy and so we do not report them.

We found most of our results were not sensitive to changing the number of annuli to five. This was done by splitting our first annuli into two annuli with radii 0 to 1.5 kpc and 1.5 to 3.5 kpc as in ref.~\citenum{FermiGCE2017}. However,  we got unrealistic $X_{CO}$ values for the innermost annuli in that case.  Therefore, we stuck to the four annuli case shown in Supplementary Fig.~\ref{fig:densities}. Note that we did not explicitly use the $X_{CO}$ values in performing our fits and
we are only quoting their inferred values to demonstrate the fitted amplitudes for our interstellar gas maps are reasonably consistent with previous results.

\subsubsection*{Systematic Errors}\label{subSec:systerrors}
Our analysis involved making choices for the spin temperature, the cosmic-ray source distribution, the 	 $E(B-V)$ reddening  map magnitude cut,  and also whether or not to include a template for the Fermi bubbles. We evaluate the associated systematic errors by seeing how the best fit exponential cutoff parameters change when different choices were made. To do this for the  spin temperature, we used $T_S=150$~K instead of $T_S=170$~K.  For the inverse Compton model we changed the cosmic ray source distribution from ``Lorimer'' to ``OBstars''\cite{ackermannajelloatwood2012} as this was found to have the greatest effect on the inverse Compton morphology.   We found changing other parameters for the inverse Compton model had negligible effects. The dust systematic error was evaluated by changing the    $E(B-V)$ reddening  map magnitude cut from 5 mag and above to 2 mag and above.  The Fermi bubbles systematic error was obtained by including the catenary template in the fit. The resulting systematic errors are given in Table~2. 

We also checked the sensitivity of our results to the background model and new point sources. As can be seen from Supplementary Table~\ref{Tab:syslikelihoods}, the Xbulge+nuclear bulge are still needed by the data when no new point sources are included, or when the interpolated gas maps are used, or when the 2FIG point sources are used. The NFW-squared template was still needed by the data, after the X-bulge and nuclear bulge were included, only in the case of the interpolated gas maps. 

To evaluate the impact that potential mismodeling of the Galactic plane could have in our main results we masked the inner $|b|<1^{\circ}$ of the region of interest (see Supplementary  Fig.~\ref{fig:PlaneMask}) and utilized a statistical procedure similar to that in Table~1. We used the \texttt{Composite2} tool within the Fermi Science Tools and performed a composite likelihood analysis of the unmasked region of interests simultaneously for each energy bin (this method was first used in the GC region in ref.~\citenum{Horiuchi2016}). We combined the regions $b>1^{\circ}$ and $b<-1^{\circ}$ of the inner $15^{\circ}\times 15^{\circ}$ of the GC and constrained the normalization of all the extended templates to be the same throughout the two separate region of interests at each energy bin. The majority of the new point sources as well as the nuclear bulge template reside in the $|b|<1^{\circ}$ region, so this analysis pipeline considered the 22 new point sources outside the masked region, the X-bulge and the NFW-squared templates in an attempt to model the GCE. The results are shown in Supplementary Table~\ref{Tab:PlaneMask}. As it can be seen, the data still significantly favours an X-bulge template over a spherical template.

As an additional check, we also evaluated the correlation matrix around our best fit model. We found that the correlation coefficients between all extended components and our X-bulge and nuclear bulge templates were in the range $10^{-10}$ to $10^{-2}$ in all energy bins.
With such small correlations even
systematic biases in the templates several times larger than the statistical uncertainties would not substantially affect the fitted normalizations of the two bulge templates. Further sources of systematic errors may arise from the choices made in constructing the X-bulge and nuclear bulge templates. We will investigate these in future work.

\end{methods}
\vspace{0.5cm}

\noindent {\bf Data Availability \& Correspondence:} Correspondence and requests for materials
   that support the plots within this paper and other findings of this study are available from  Oscar Macias.~(email: oscar.macias@vt.edu).
 

 \clearpage

\setcounter{figure}{0}
\setcounter{table}{0}
\renewcommand{\figurename}{Supplementary Figure}
\renewcommand{\tablename}{Supplementary Table}

\section*{\centering 
Supplementary Information}

\vspace{1cm}

\subsubsection*{1. Template Construction}

We do not include the empirically-derived extended emission templates used by the Fermi-Collaboration when generating 
the publicly available version of the Galactic diffuse emission model provided for standard data analysis\cite{Casandjian:andFermiLat2016}, because they were used to flatten residuals over larger regions of the sky than we are interested in this analysis and they are specifically constructed in support of the generation of the Fermi-LAT point-source catalogues.


\subsection*{  H{\small I} and H\boldmath$_{2}$ Gas Column Density Templates:}
\label{subsec:gasmap}

We investigate two different methods for constructing the gas map templates: 
\begin{itemize}

\item { Interpolation approach:} this method has been used to produce the interstellar gas distribution models employed in most previous analyses of the GCE and is the standard approach employed in GALPROP and by the Fermi team.

In the interpolation approach, the gas column density maps are produced using the method given in Appendix B of ref.~\citenum{ackermannajelloatwood2012}. 
Atomic hydrogen column density is derived from the 21 cm LAB survey of Galactic H{\small I}\cite{kalberla2005}. Molecular hydrogen is traced by the 2.6 mm emission line of carbon monoxide (CO) from the 115 GHz centre for Astrophysics survey of CO\cite{dame2001}. 
The emission maps are decomposed into Galactocentric annuli via the relation,
  \begin{equation} \label{circular_motion}
  V_{\rm{LSR}} =  \left( V(R)  \frac{R_{\odot}}{R} - V(R_{\odot}) \right) \sin(l) \cos(b),
  \end{equation}
  where $V_{\rm{LSR}}$ is the radial velocity, relative to the local standard of rest, of the gas at Galactocentric radius $R$ with orbital velocity $V(R)$ observed in the direction $(l,b)$
  and $R_{\odot}$ is the distance of the Sun to the Galactic centre.
We  assumed a spin temperature of $T_{S} = 170 \, \rm{K}$ throughout the Galaxy and 
$R_{\odot}=
8.25
$
kpc. 
As in ref.~\citenum{ackermannajelloatwood2012}, for CO, we assigned all high-velocity emission in the innermost
annulus. For H{\small I} and the other CO annuli, the annuli decomposition interpolation method does not produce a reliable map of the gas column density as kinematic resolution is lost near $l = 0^{\circ}$. In the interpolation approach, the gas column density in the region $|l| < 10^{\circ} $ was estimated by interpolating within each annulus from the boundaries. Values at the boundaries of the interpolated region were chosen as the mean gas column density within a range of $\Delta l = 5^{\circ}$ on both sides of the boundary. Each pixel in the interpolated region was then renormalised to preserve the total gas column density in each line of sight.

\item {Hydrodynamic approach:} this method was pioneered by ref.~\citenum{Pohl2008}. See also ref.~\citenum{Timur2011} for discussion regarding the construction of the H{\small I} map.
%
%
The presence of non-circular motion in the inner Galaxy provides kinematic resolution towards the GC.
Outside the solar circle, pure circular motion with velocities described by equation (\ref{circular_motion}), was assumed with an additional correction for the motion of the Sun relative to the local standard of rest. This meant there was lack of CO and H{\small I} gas placed in the outer annuli.
We corrected for this by interpolating the outer annuli pixels in the line-of-sight maps which have $|l| < 15^{\circ} $ and then normalised each pixel to  preserve the total gas column density in each line of sight. For CO we interpolated beyond 10 kpc  and for H{\small I} we interpolated beyond 8 kpc from the GC.

\end{itemize}
The results of the two methods are compared in Supplementary Fig.~\ref{fig:densities}. Note that in the interpolated approach, 
 if the gas is placed above
a certain height above the Galactic plane, it is assumed to be
local\cite{ackermannajelloatwood2012}. This height differs between the gas distributions and was
chosen to be 1 kpc for H{\small I} and 0.2 kpc for CO. This explains the high latitude CO gas difference in the 3.5--8.0 kpc and 8--10 kpc range of Supplementary Fig.~\ref{fig:densities}.
 Somewhat broader radii were used in comparison to ref.~\citenum{ackermannajelloatwood2012} and \citenum{Casandjian:andFermiLat2016} as we were fitting to a much smaller region of interest (ROI) and so the data we used had less constraining power and so broader annuli were needed.

\subsection*{Dust Correction Templates:}
Molecular hydrogen that is not well mixed with carbon monoxide will not be traced by the CO 2.6 mm emission. Furthermore, assuming a constant atomic hydrogen spin temperature $T_{S} = 170 \, \rm{K}$ can give an incorrect estimate of column density as the spin temperature can vary along a line of sight. 
To correct for these deficiencies we included dust templates based on the methods used in ref.~\citenum{ackermannajelloatwood2012}.
  Infrared thermal emission from dust provides an alternative method of tracing hydrogen gas in the Milky Way. The correction templates are obtained by subtracting the components of the dust emission that are correlated with the gas already traced by 21 cm and 2.6 mm emission. 

We applied this method to $E(B-V)$ reddening maps\cite{schlegel1998}. Regions that are densely populated by infrared point sources (or potentially a collection of unresolved point sources) could contaminate the $E(B-V)$ reddening map, resulting in an over estimation of the dust column density. To mitigate this, we apply a magnitude cut of 5 mag and higher to the $E(B-V)$ reddening maps. After subtracting the components of the $E(B-V)$ map that were linearly correlated with the hydrogen gas maps, the residuals were separated into positive and negative components. The positive residuals physically represent hydrogen that is not traced by the relevant emission, known as the dark neutral medium, or an over estimation of the atomic hydrogen spin temperature. Negative residuals represent an underestimation of the spin temperature. The results are displayed in Supplementary Fig.~\ref{fig:dust_maps}. 

\subsection*{Inverse Compton Emission Template:}
\label{subsec:IC}

Whereas the bremsstrahlung and $\pi^0$-decay  components can be adequately described by the gas maps described above, there is no analogous empirical template  for the Galactic IC emission. We use the GALPROP package v54.1\cite{Galpropsupplementary} to generate such a template. The authors of ref.~\citenum{ackermannajelloatwood2012} did a comprehensive comparison of $128$ different GALPROP models with all sky Fermi-LAT data. They found a range of possible values for the input GALPROP parameters that are consistent with gamma-ray and local measurements of  cosmic-ray data.  In this work, we consider the IC template generated with GALDEF file\\ \texttt{galdef$_-$54$_-$Lorimer$_-$z10kpc$_-$R20kpc$_-$Ts150K$_-$EBV2mag}
as our reference model\cite{ackermannajelloatwood2012}. 
As seen from Supplementary Fig.~\ref{fig:ptsrc_spectra}, the IC component is of much smaller intensity compared to the 
interstellar gas diffuse Galactic emission components.  

\subsection*{Loop I Template:}
\label{subsec:LoopI}

Loop I is a bright, large angular scale, non-thermal structure.
Obtaining a precise template of this source is not possible since its gamma-ray emission is not well traced by radio emission. 
Following the same approach taken in recent studies by the Fermi Collaboration\cite{Fermi:LatBubbles}, in this work we used a geometrical template proposed by ref.~\citenum{Wolleben:2007} which is based on a polarization survey at 1.4 GHz (see Supplementary Fig.~\ref{fig:LoopI}).
We adopted the same morphological parameters for the shells assumed in ref.~\citenum{Fermi:LatBubbles}. \\

\subsection*{Sun and Moon emission Templates: }
\label{subsec:Sunadark matteroon}

In order to account for the diffuse emission from these objects, we construct specialized templates for our ROI by making use of the \texttt{gtsuntemp}\cite{gtsuntemp} tool. We used as input the intensity profiles  \texttt{lunar\_profile\_v2r0.fits} and \texttt{solar\_profile\_v2r0.fits} available at the \textit{Fermi Science Support Center} website.  
The templates were constructed in Galactic coordinates with a Cartesian projection method. We note that the templates obtained with the use of this tool depend on the data selection. As such, we did not use the Sun and Moon templates that come with the 3FGL\cite{3FGL} as those were constructed for analysis of only 4 years of data.

\subsubsection*{2. Mixture Distributions}

Wilks' theorem states that the asymptotic distribution of TS is given by Chi squared distribution ($\chi^2_q$) where $q$ is the number of new parameters. This can be used to evaluate  a  p-value $[1-p({\rm TS}>{\rm TS}_{\rm thresh})]$ of some threshold value of TS above which the new parameters are accepted as being statistical significant. However, this theorem does not hold if any of the null values of the new parameters are on the boundary of the allowed parameter space. This is problematic for when one wants to decide whether there is a new source in the data which can only have a non-negative amplitude. This limitation can be alleviated using the Chernoff theorem\cite{Chernoff1954} which implies that if the new parameter is the proposed source's amplitude then the TS distribution will be given by a mixture distribution:
\begin{equation}
p({\rm TS})=\frac{1}{2}(\delta({\rm TS})+\chi^2_1({\rm TS})).
\end{equation}
where $\delta$ is the Dirac delta function. This formula just states that under the null hypothesis, half the time the evaluated amplitude will be negative (in which case ${\rm TS}$ is assigned) and the other half of the time (when the amplitude is non-negative) the distribution will follow from Wilks' theorem. This method was shown to work well for simulations of EGRET data by ref.~\citenum{mattox}. They also proposed that an extension which includes parameters for the two dimensional position of the source has a distribution of 
\begin{equation}
p({\rm TS})=\frac{1}{2}(\delta({\rm TS})+\chi^2_3({\rm TS}))
\label{eq:3param}
\end{equation}
which they checked is correct by comparing to simulated EGRET data. 
In some cases sources are found to significantly prefer a curved spectrum\cite{3FGL}. For example, pulsars generally have an exponential cut-off spectrum.
 Therefore, it may be preferable to test for some new source with a non-parametric spectrum so as not too cause any bias by using a spectrum different from the spectrum the source has. This can be done by doing a bin by bin analysis where the energy range is broken up into a number of bins. If the bins are made sufficiently small, then a power law with $\Gamma$ set to some fixed value (such as 2) can be used in each bin without loss of generality. Although, the bins shouldn't be too small otherwise one starts to get correlations between the different energy bins due to the finite energy dispersion of the Fermi-LAT instrument\cite{FermiInstrument}. For the case of a testing for a new source at a fixed position
we can use ${\rm TS}=\sum_{i=1}^n {\rm TS}_i$ where ${\rm TS}_i$ is the ${\rm TS}$ for bin $i$ and there are $n$ bins.
We can then utilize the formula given in case 9 of ref.~\citenum{SelfLiang87}:
\begin{equation}
p({\rm TS})= 2^{-n}\left(\delta({\rm TS}) +\sum_{i=1}^{n} \binom{n}{i} \chi_{i}^2({\rm TS})\right)
\label{eq:pTS1}
\end{equation}
where the amplitude in each energy bin needs to be non-negative. This equation has a simple interpretation in terms of mixture distributions. The $2^{-n}$ term is equal to the number of distinct ways $n$ bins could have a non-negative or negative best fit amplitude. As there is only one way they could all have a negative amplitude, that is the weight of the $\delta$ function. While if the there are $i$ non-negative amplitudes they would have $\binom{n}{i}$ distinct configurations and each of these configurations would have a $\chi^2_i$ distribution. As discussed in the Methods section, we used equation~(\ref{eq:pTS1}) in testing for new extended sources.

When testing for new point sources, the two position variables of the proposed new point source should also be included. They are not on the boundary of the allowed range under the null hypothesis. Case 9 of ref.~\citenum{SelfLiang87} also covers the case where $n$ of the new variable are restricted and two are not:
\begin{equation}
p({\rm TS})= 2^{-n}\left(\sum_{i=0}^{n} \binom{n}{i} \chi_{i+2}^2({\rm TS})\right).
\label{eq:pTS2}
\end{equation}
However, this formula is not quite the one we are looking for as when all the bins have non-positive best fit amplitudes, we need to  have ${\rm TS}=0$ regardless of the values of the position variables. Therefore, in order to modify equation~(\ref{eq:pTS2}) for our case of interest, we need to subtract off $2^{-n} \binom{n}{0}\chi^2_2=2^{-n} \chi^2_2$ and add on a $2^{-n}\delta({\rm TS})$ term. This gives equation (\ref{eq:pTS}) of the Methods section.

\subsubsection*{3. Fermi Bubbles}
In ref.~\citenum{Fermi:LatBubbles}, the spectrum of the Fermi bubbles was determined
from the residuals in the energy range between 0.7 GeV and 10 GeV. They examined an area near the southern edge of the Fermi bubbles
($-55^\circ < b < -40^\circ$ and $-15^\circ < l < 15^\circ$). A power-law fit
in this region had a spectral index index $1.9\pm0.1$. In contrast, we found our X-bulge and nuclear bulge, for the same energy range, had a power-law fit with a 
softer spectral index of $2.20 \pm 0.05$.
Also, we estimate the X-bulge plus nuclear bulge has a solid angle of 0.03~sr, and therefore a luminosity per solid angle of  $(1.0\pm 0.1)\times10^{38}$~erg/s/sr. By contrast, the Fermi bubbles, for $|b|\ge 10^\circ$,  have a solid angle of about 0.7~sr and a luminosity\cite{Fermi:LatBubbles} per solid angle of $(6.3\pm 0.1)\times 10^{37}$~erg/s/sr. Even taking a wide range of systematic effects into account, the highest luminosity\cite{Fermi:LatBubbles} per solid angle  for the Fermi bubbles was only $8.6\times 10^{37}$~erg/s/sr. 

Ref.~\citenum{FermiGCE2017} used a spectral decomposition technique to determine the low latitude Fermi bubble contribution. However, their residuals for their GCE show the characteristic X-shape in their Fig.~3. Also, they found the addition of their Fermi bubbles template did not remove the need for an extra template to explain the GCE. In summary, current spatial and spectral evidence seems to disfavour a Fermi-bubbles explanation for all of the X-bulge and nuclear bulge emission.

\subsubsection*{Supplementary References}


\begin{figure*}[ht]
\begin{center}
\includegraphics[width=0.7\linewidth]{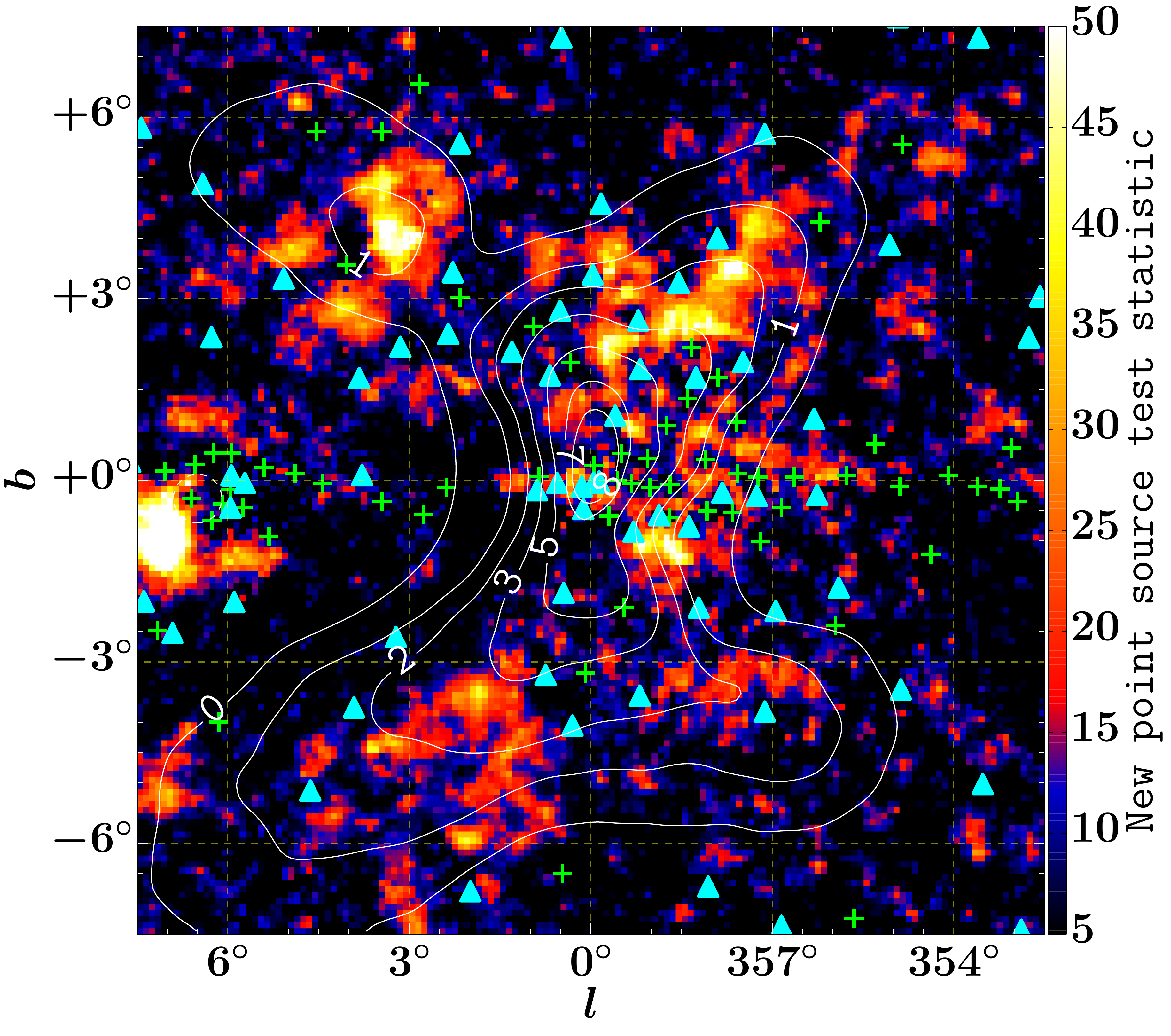}
\caption{{\bf   Significance map of the \boldmath$15^\circ \times 15^\circ$ region about the GC output from the \texttt{gttsmap} tool after the new point sources have been identified.}
The energy range shown is 667 MeV$-$158 GeV. The   3FGL  point sources  (cyan triangles) and the new point sources (green crosses) are displayed. Supplementary  Table~1  summarizes the basic properties of the new point sources.
The white contours 
are the best-fit model counts from the X-bulge map obtained from analyses of WISE\cite{NessLang2016} infrared data after convolution the Fermi-LAT instrument response function.
 \label{fig:tsmaps}}
\end{center} 
\end{figure*}

\begin{figure*}[t!]
\begin{center}
\begin{tabular}{cc}
\includegraphics[scale=0.5]{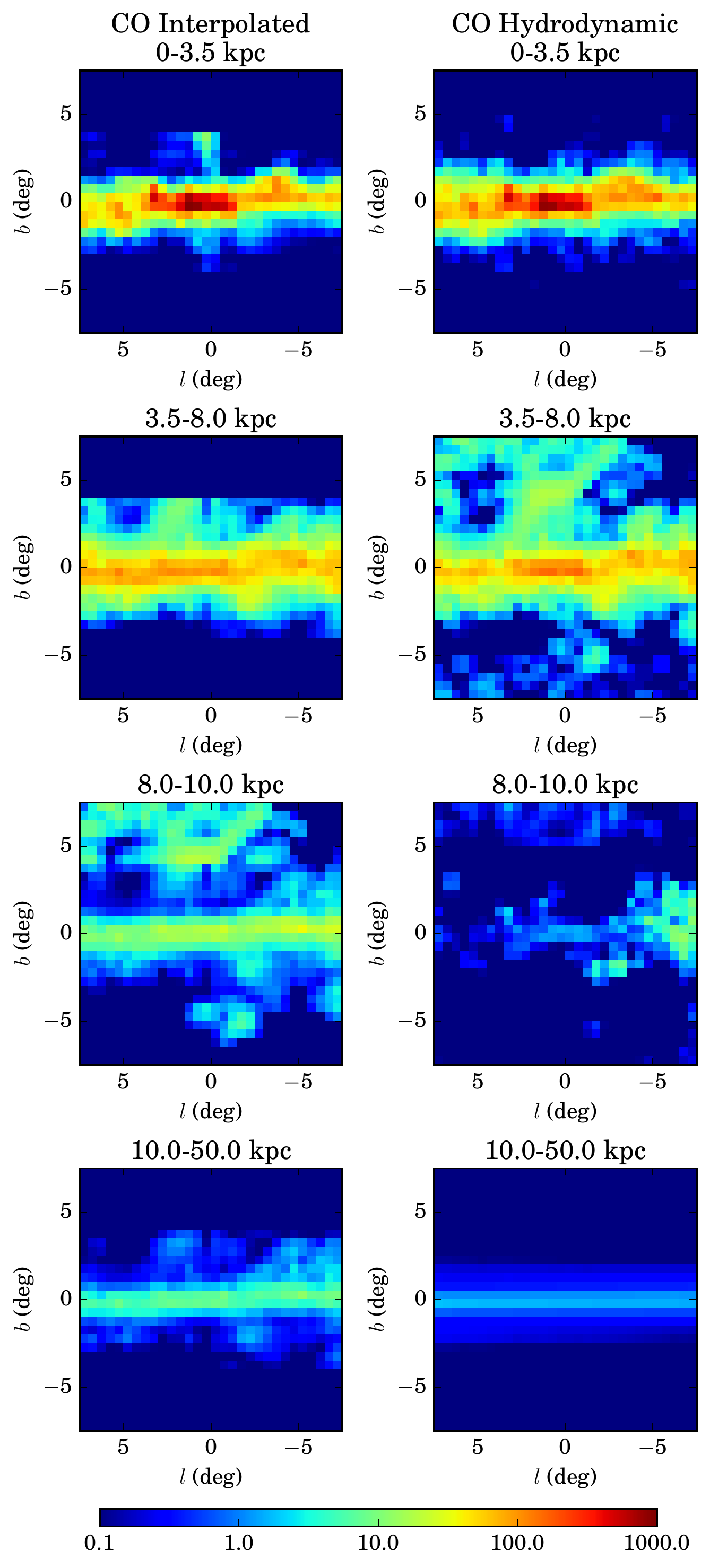} & \includegraphics[scale=0.5]{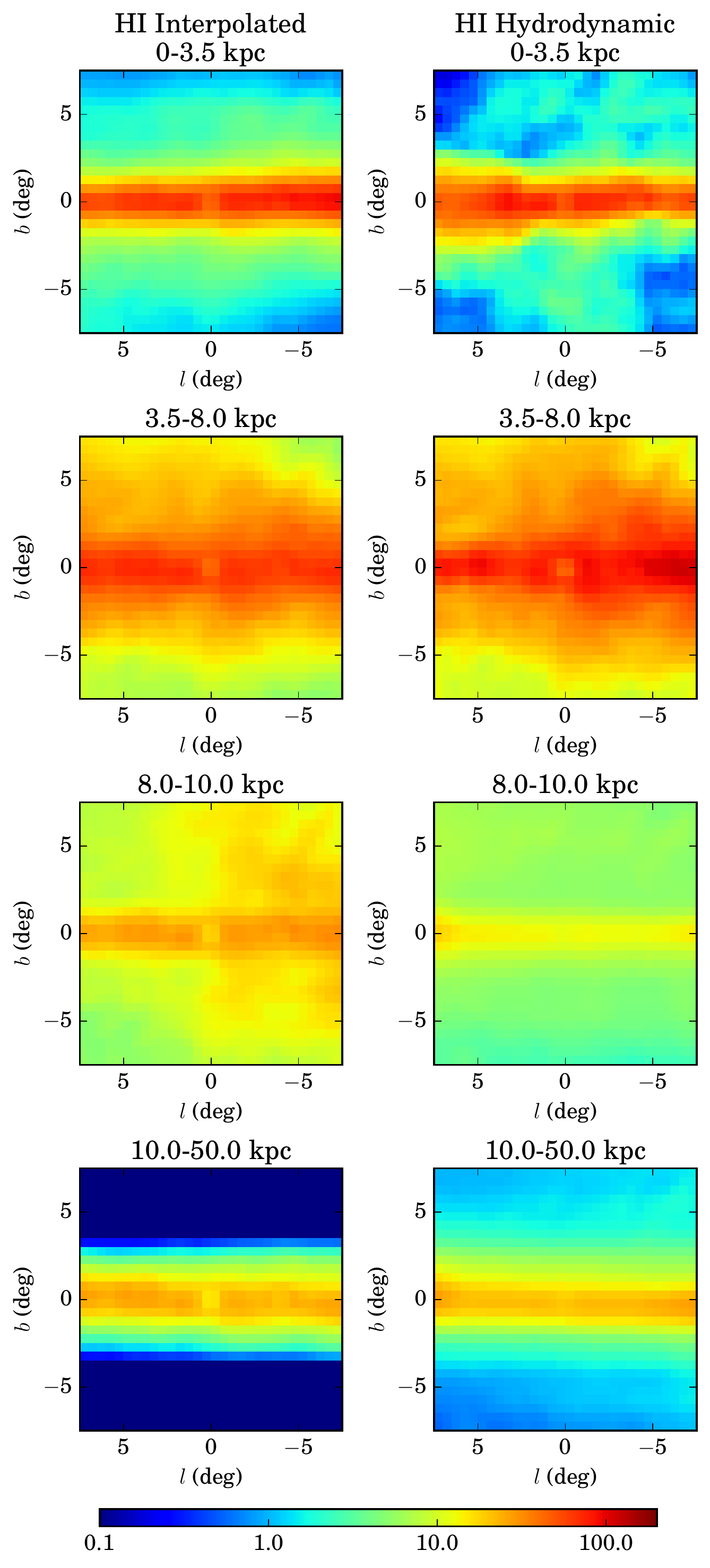}
\end{tabular}
\end{center}
\caption{
{\bf  Column density maps for the
interpolated and hydrodynamic methods.}
The minimum and maximum radii of each annulus is listed.
The units for the H$_2$ proportional CO maps are K$\cdot$km$/$s. The units for the H{\small I} maps are $10^{20}$cm$^{-2}$.
 \label{fig:densities}}
\end{figure*}

\begin{figure*}[t!]
\begin{center}

\begin{tabular}{cc}
\centering
\includegraphics[scale=0.2]{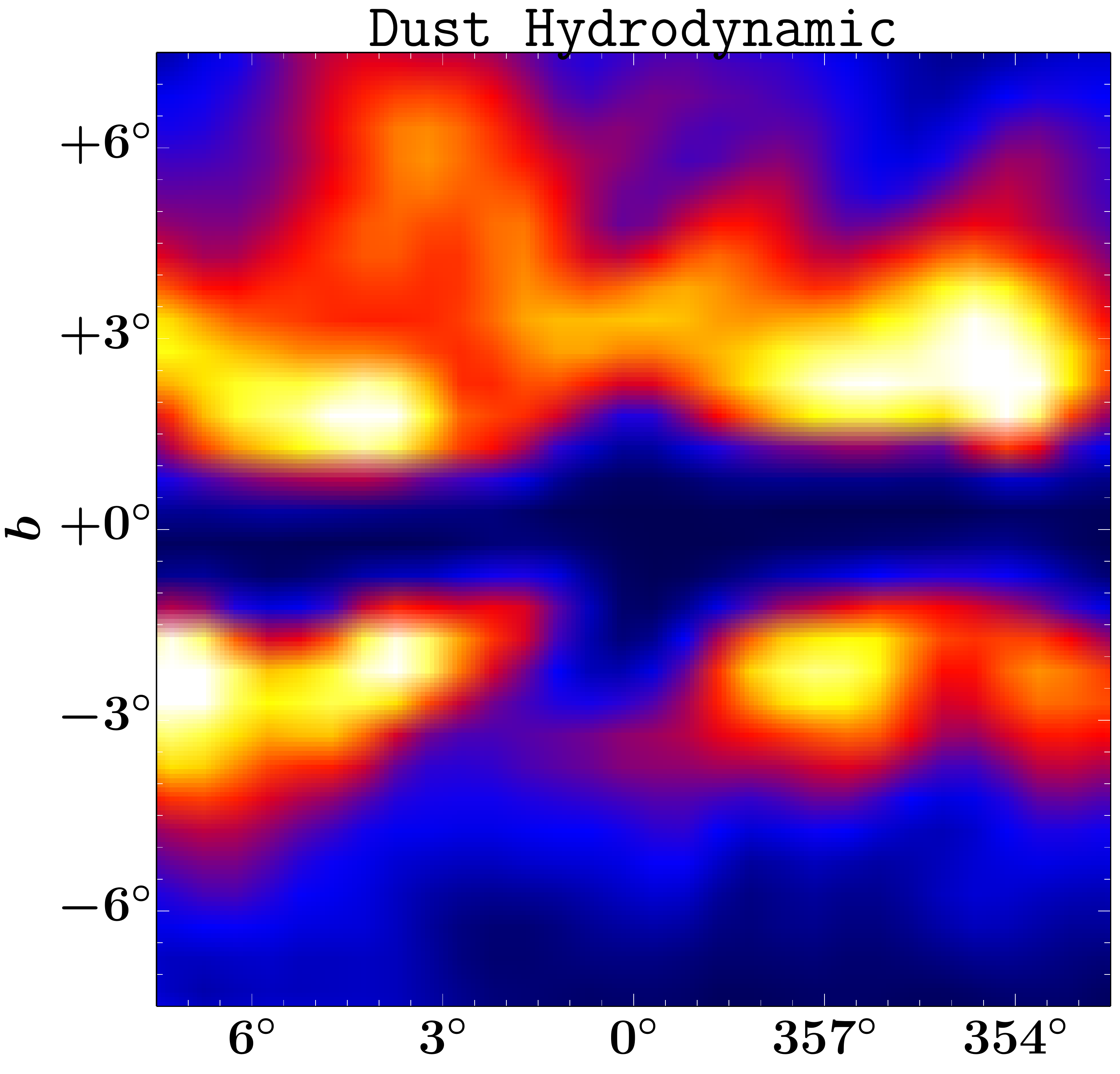} &  \includegraphics[scale=0.2]{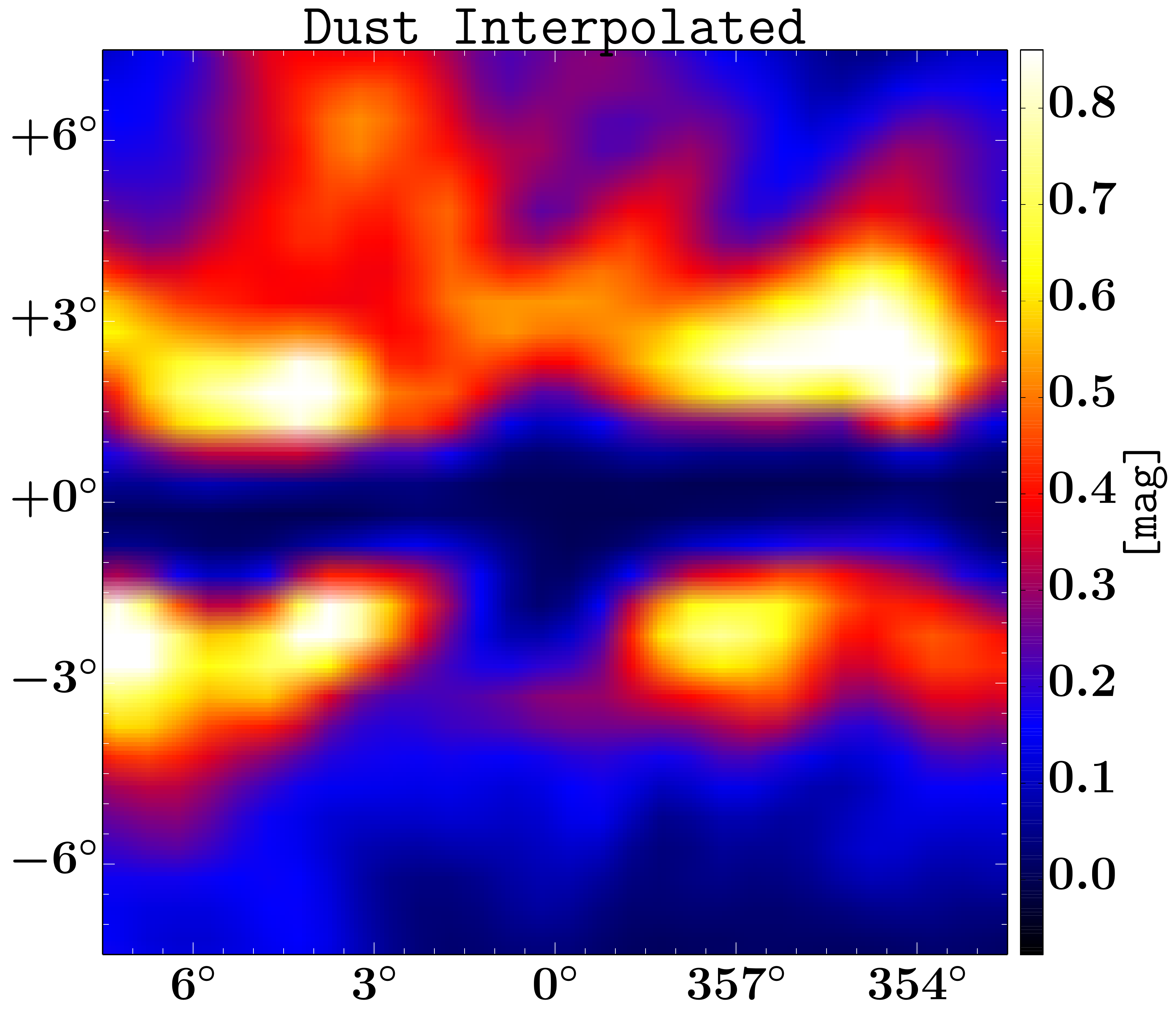}\\
\includegraphics[scale=0.2]{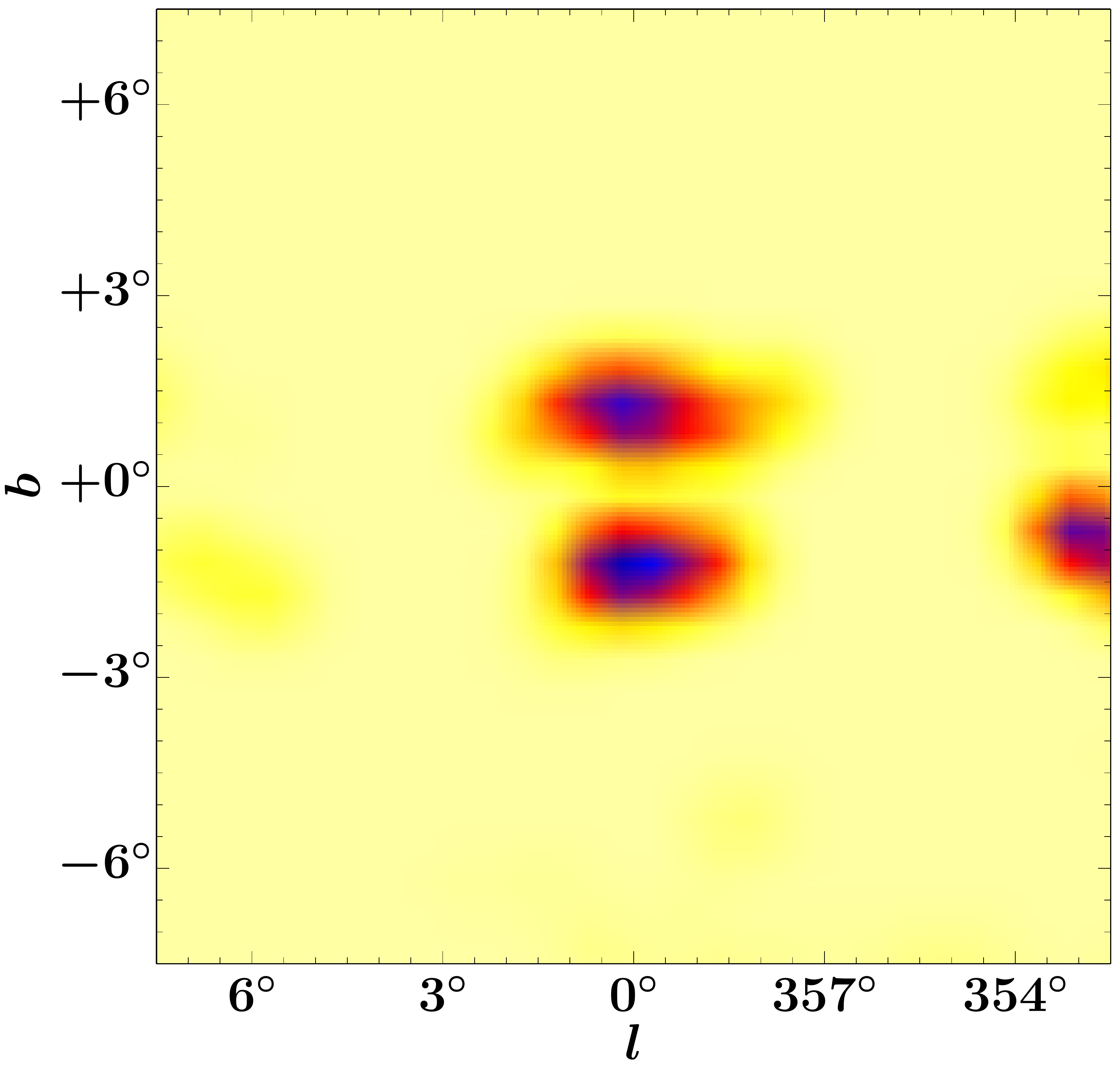} &  \includegraphics[scale=0.2]{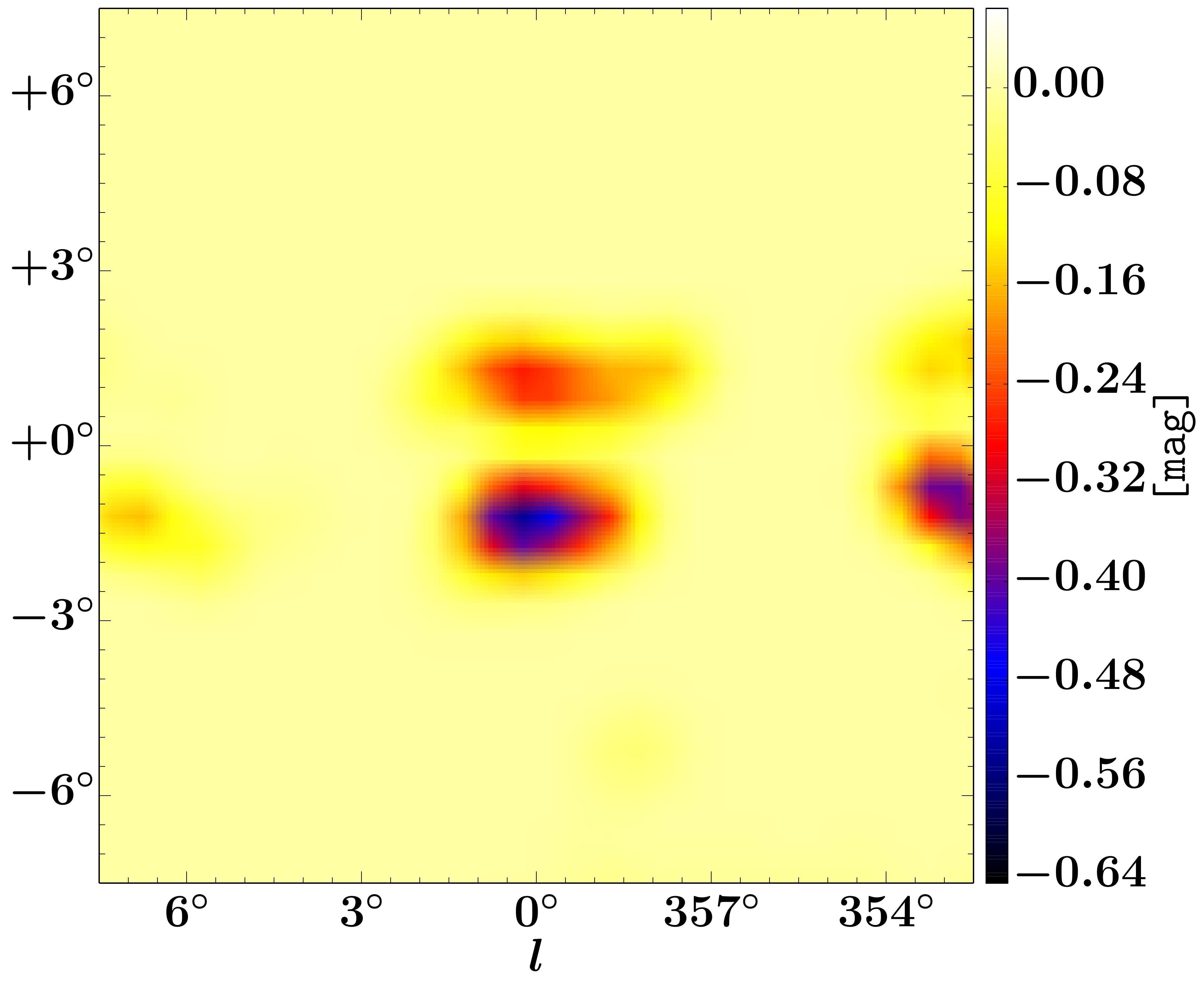}
\end{tabular}

\caption{
{\bf  Dust reddening E(B-V) positive (top) and negative (bottom) residual maps.}
The left (right) hand side maps were based on fitting the hydrodynamic (interpolated) HI and CO maps to the E(B-V) reddening map which had be a magnitude cut of 5 mag and higher. For display purposes, Gaussian smoothing  with a radius of $0.5^\circ$ was performed.
\label{fig:dust_maps}}
\end{center}
\end{figure*}

\begin{figure*}[t]
\begin{center}

\begin{tabular}{cc}
\centering
\includegraphics[scale=0.4]{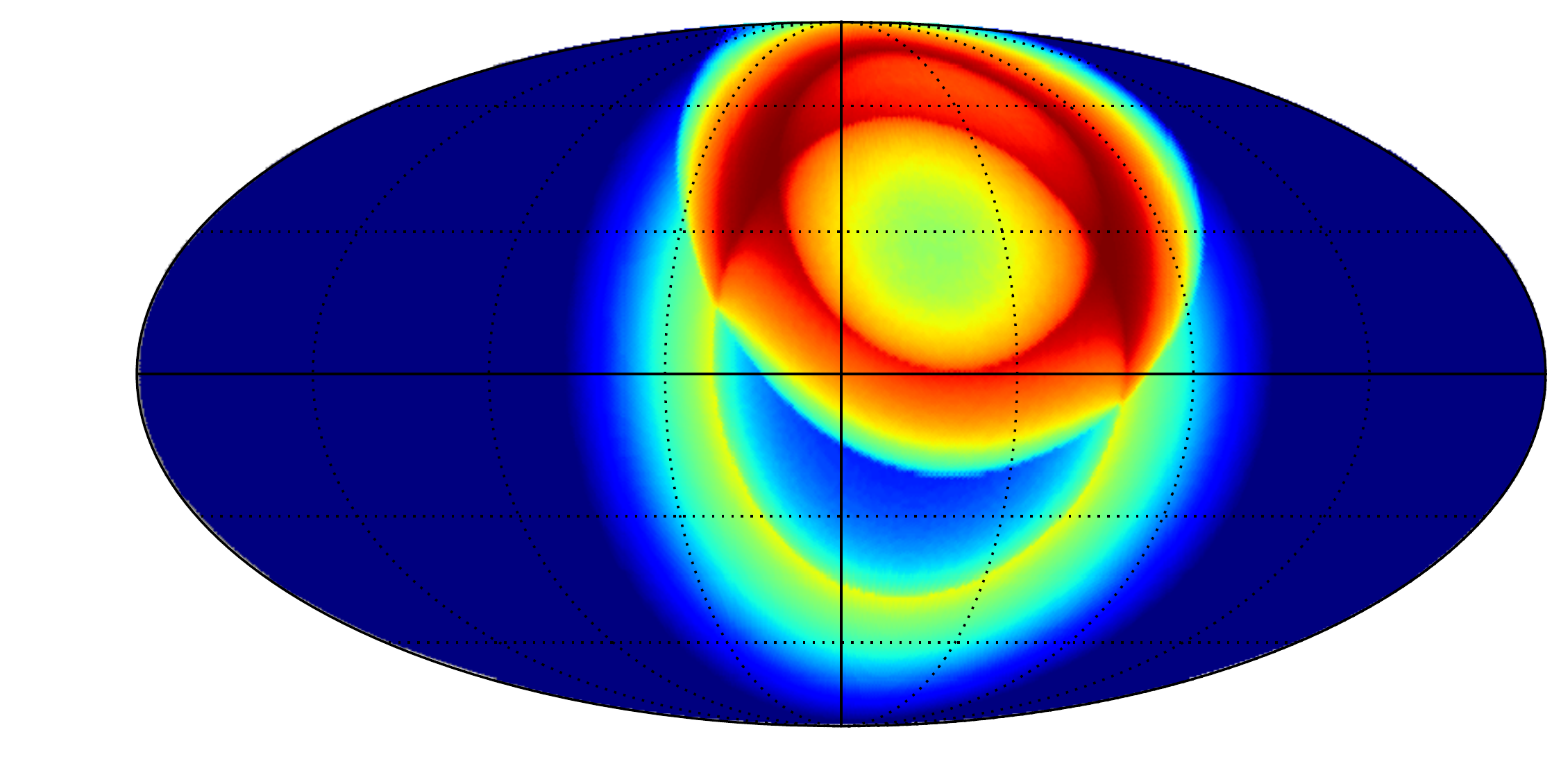} & \includegraphics[scale=0.25]{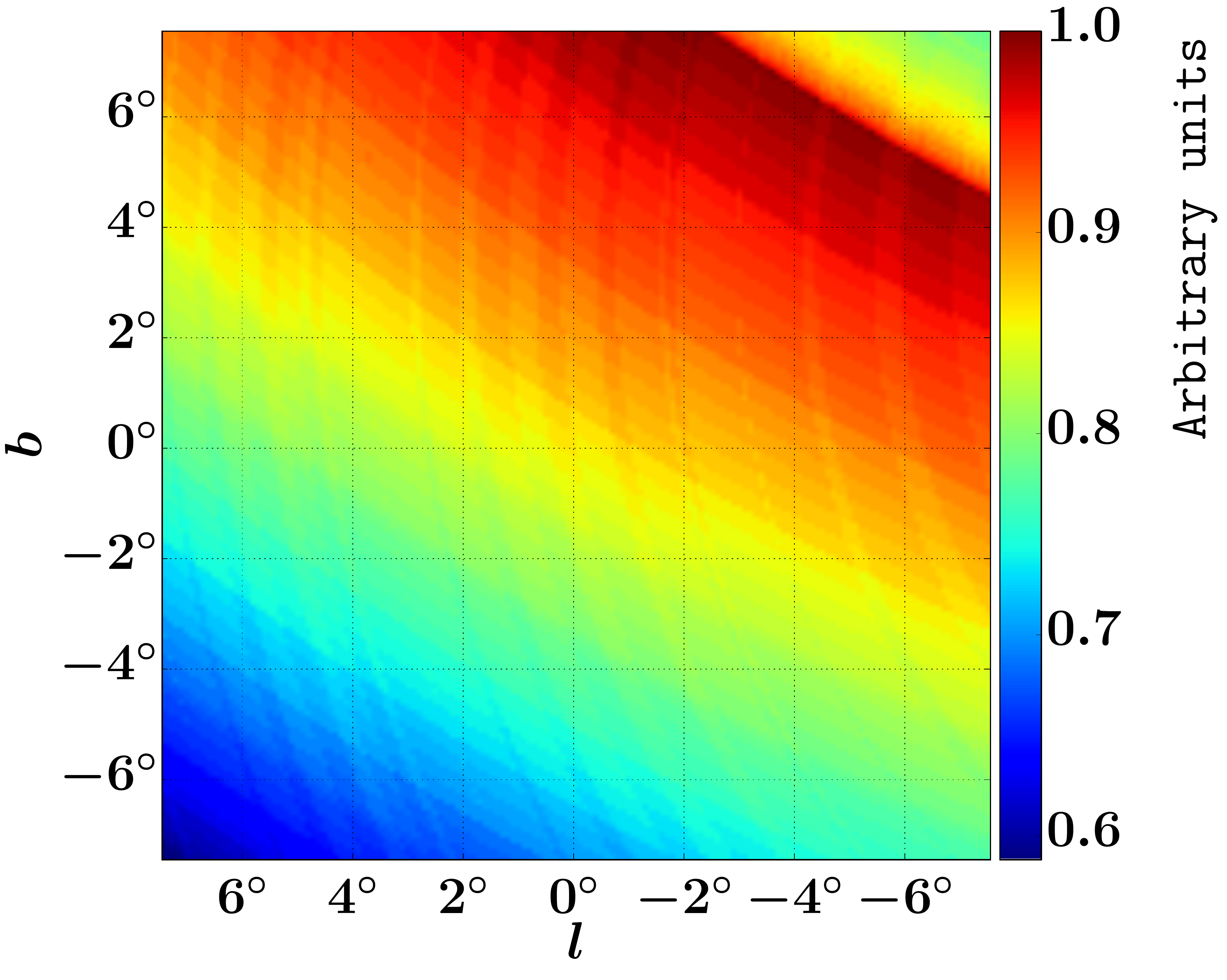}

\end{tabular}

\caption{ {\bf  Loop I template adapted from ref.~\citenum{Wolleben:2007}.} A histogram equalized colour mapping is used for display. The map is appropriately normalized, in our region of interest, for analysis with the Fermi Science Tools. On the left we show a full sky version of the map in the Mollweide  projection and on the right we have zoomed in to our region of interest.
\label{fig:LoopI}}
\end{center}
\end{figure*}

\begin{figure*}[t]
\begin{center}
\includegraphics[scale=0.45]{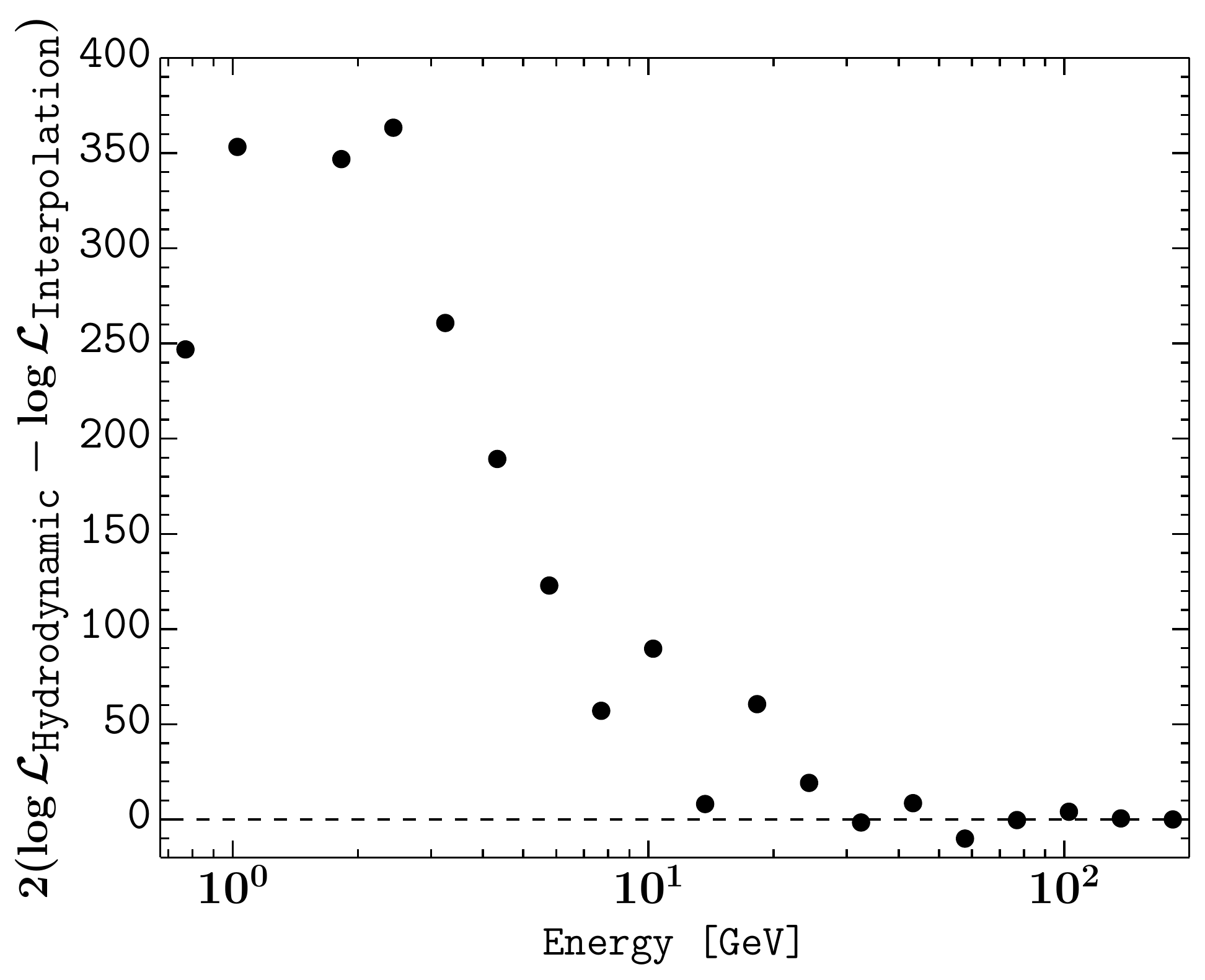}

\caption{{\bf   Comparison of the log-likelihood obtained for two different interstellar gas models.} The likelihood is compared for hydrodynamic\cite{Pohl2008} gas maps vs the interpolation ones used in the standard Galactic diffuse emission model. Summing over the energy bins gives ${\rm TS}_{\rm Hydrodynamic}=2\times 1362$. 
 \label{fig:pohvsinterp}}
\end{center} 
\end{figure*}

\begin{figure*}[ht!]
\begin{center}
\begin{tabular}{c}
\centering
\includegraphics[scale=0.35]{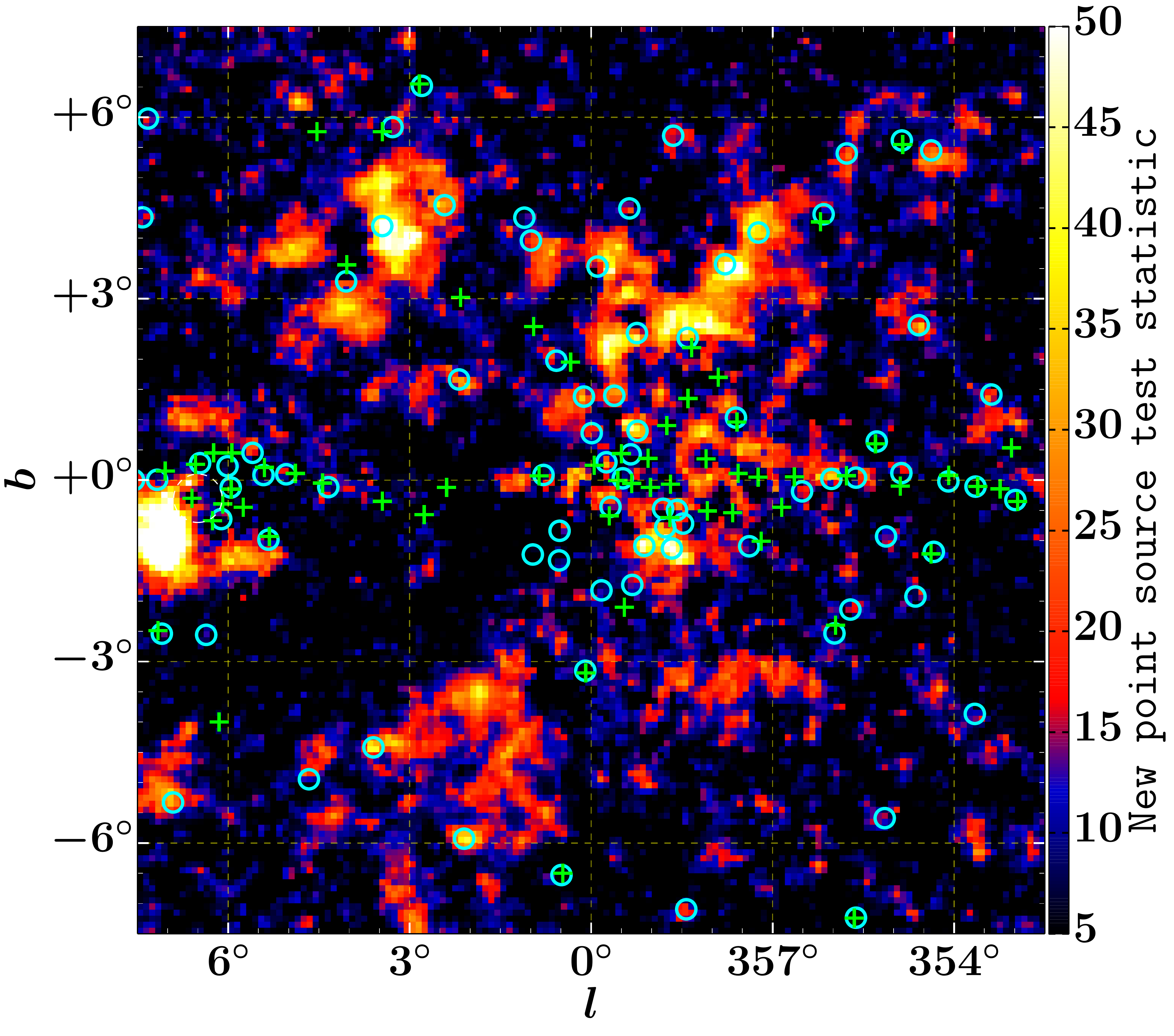}
\end{tabular}
\caption{
{\bf  Comparison of new gamma-ray point sources in this work vs new point sources in the 2FIG catalog\cite{Ajelloetal:2017}.}
See caption of Supplementary Fig.~\ref{fig:tsmaps} for a description of this residual TS map. Cyan circles display the 81 new gamma-ray point sources (not already present in the 3FGL\cite{3FGL}) found in the 2FIG catalog\cite{Ajelloetal:2017}. Green crosses correspond to the 64 new point sources found in this work. Despite the different data sets and 
background models used in both studies, our analysis confirms the existence of 31 of the new point sources found in 2FIG (see Supplementary Table~1). Note that many undetected 2FIG point source candidates are placed in hot spots in this residual map. It is very likely that by augmenting the time and the photon energy range utilized in this analysis to the same levels used in the 2FIG analysis, such hot spots would be found with greater TS-values and would therefore constitute positive point source detections.
\label{fig:ptsrcs_comparison}}
\end{center}
\end{figure*}

\begin{figure*}[ht]
\begin{center}
\includegraphics[width=0.7\linewidth]{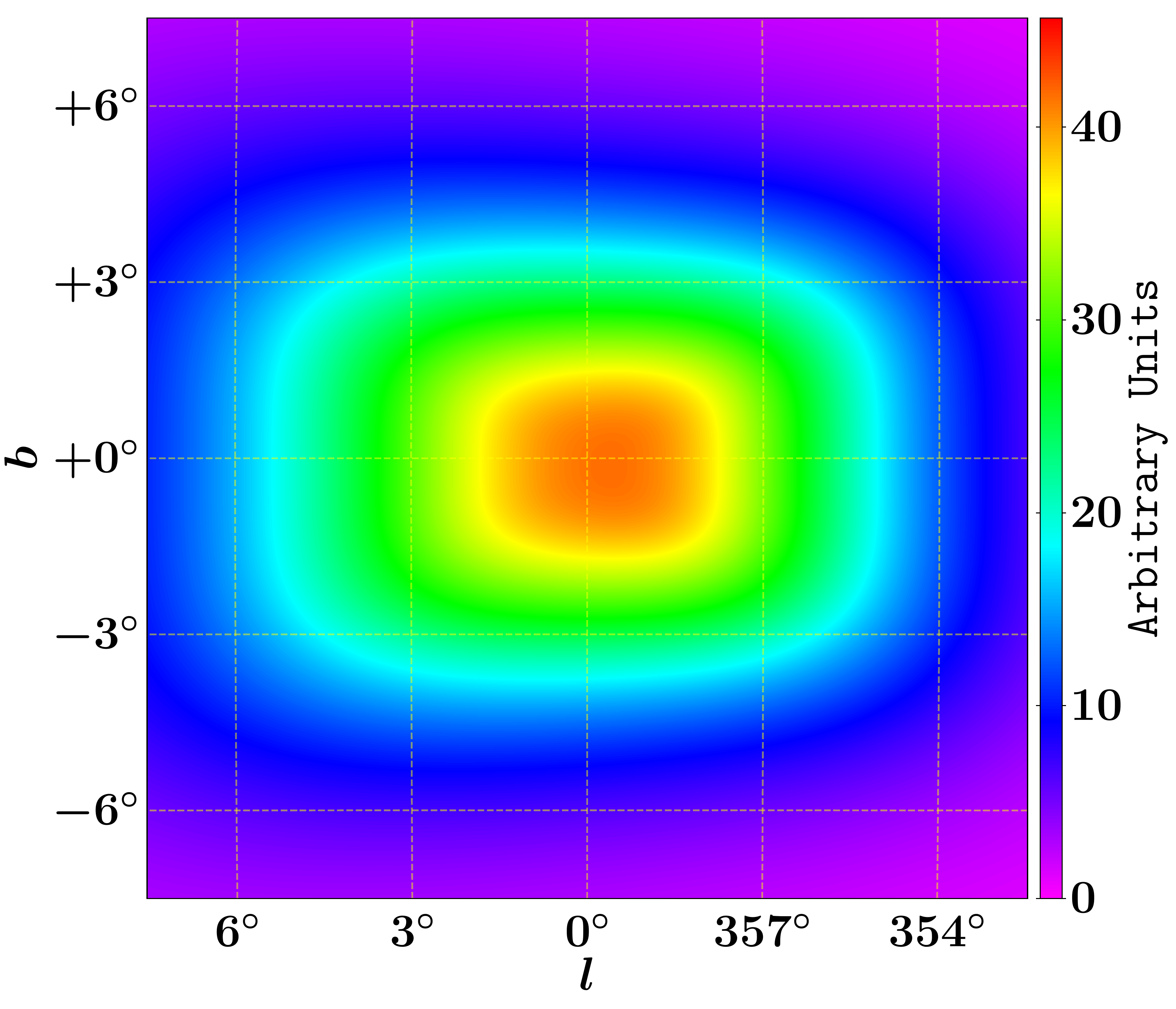}
\caption{{\bf   The Boxy bulge template obtained from a parametric fit to diffuse infrared DIRBE data\cite{Freudenreich:1998}.}
 \label{fig:Boxybulge}}
\end{center} 
\end{figure*}

\begin{figure*}[ht!]
\begin{center}
\begin{tabular}{c}
\centering
\includegraphics[scale=0.5]{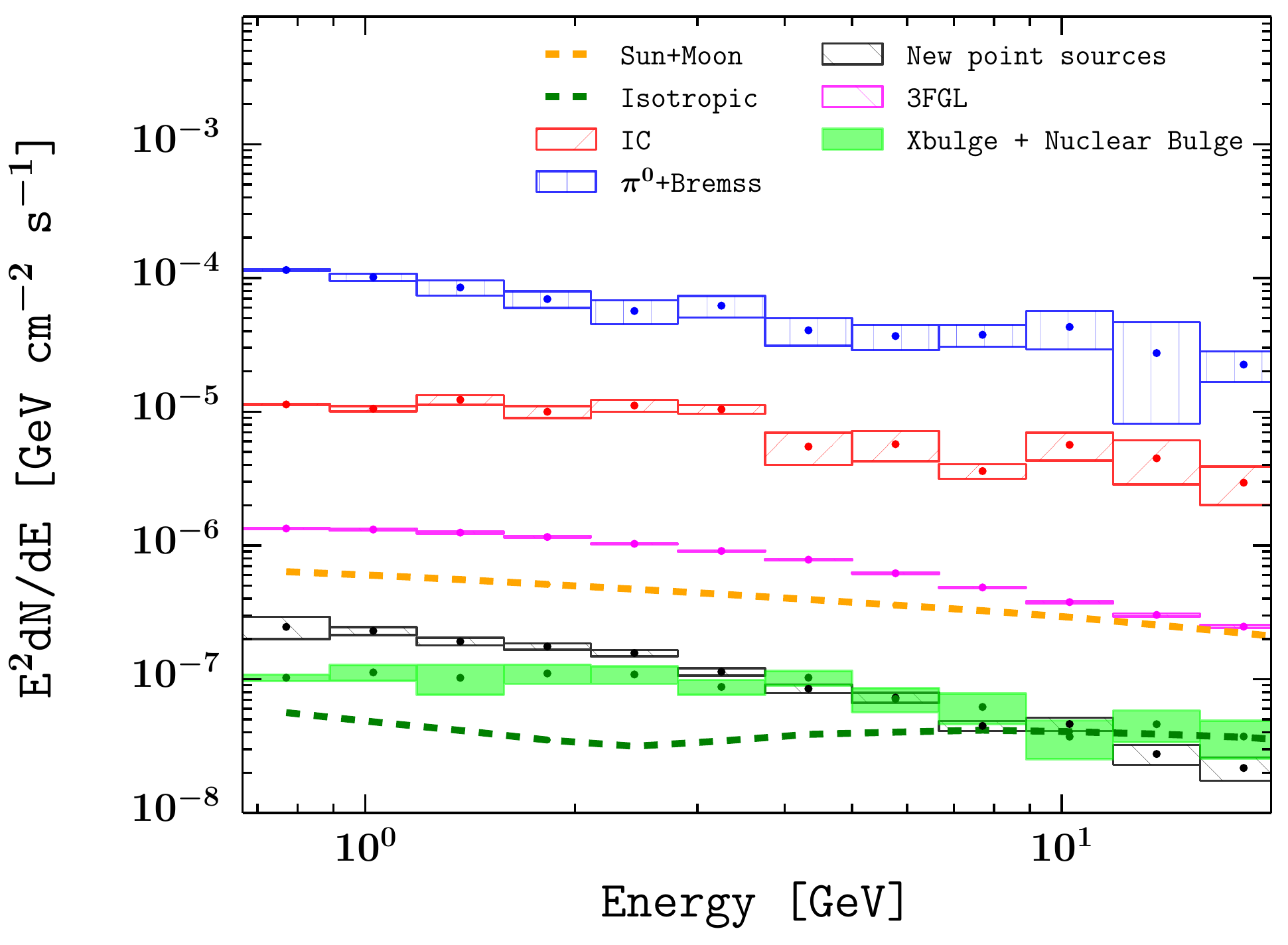}
\end{tabular}
\caption{
{\bf  Model components.}
Differential flux of the different components in the $15^{\circ}\times 15^{\circ}$ region about the Galactic Center. The normalisations are for the full sky. The model considered in the fit is the \textit{baseline$+$NP$+$nuclear bulge$+$X-bulge} model. To reduce clutter the spectrum of the combined $\rm H_I$, CO, and dust maps maps are shown as  $\pi^0+$Bremss. The Loop I spectrum had a low signal to noise for the majority of the bins and so has not been included in the above plot.
The box heights represent  the 68\% confidence interval regions.
\label{fig:ptsrc_spectra}}
\end{center}
\end{figure*}

\begin{figure*}[!htbp]
\begin{center}

\centering
\begin{tabular}{cc}

\includegraphics[scale=0.2]{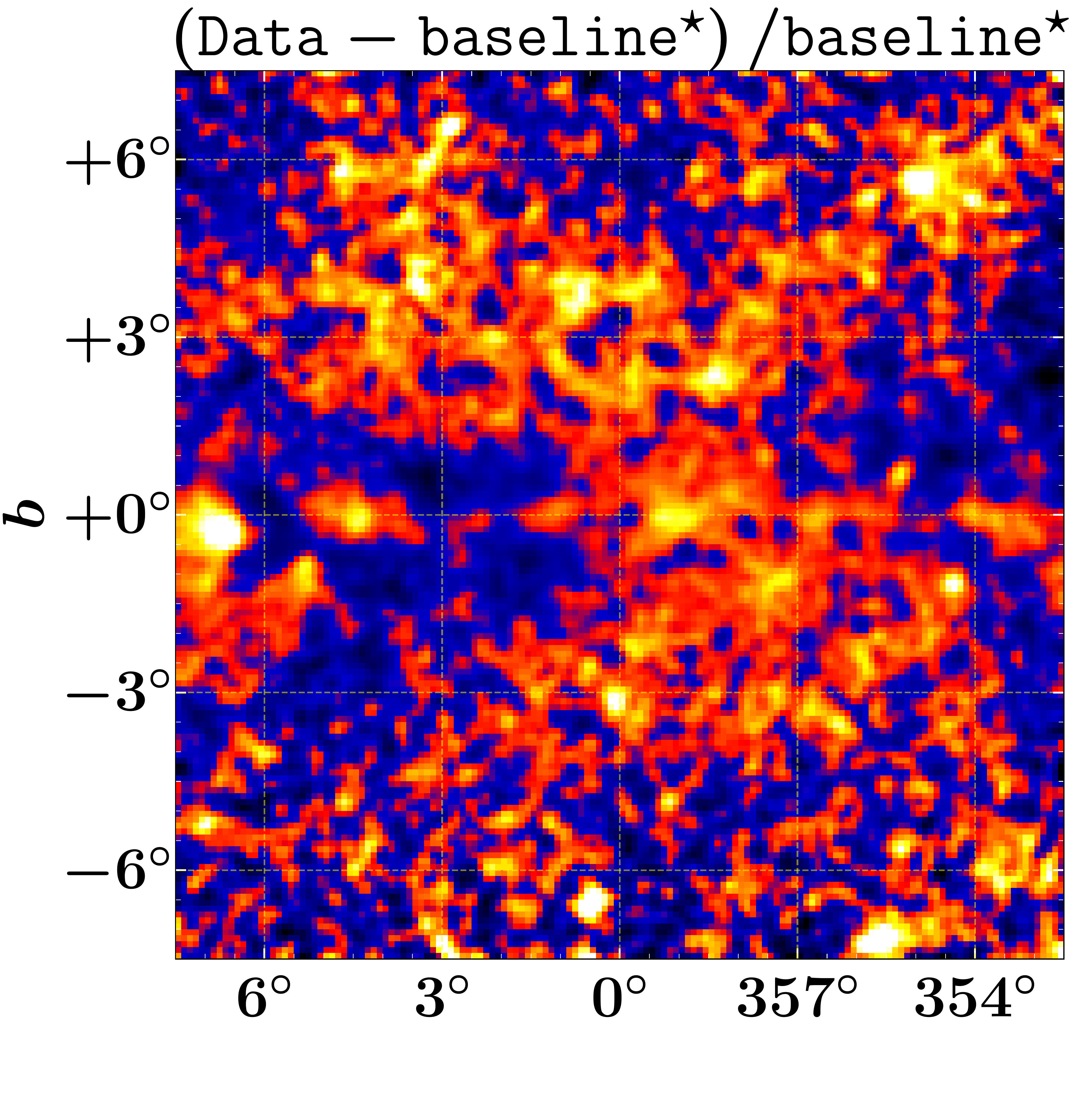} &
\includegraphics[scale=0.2]{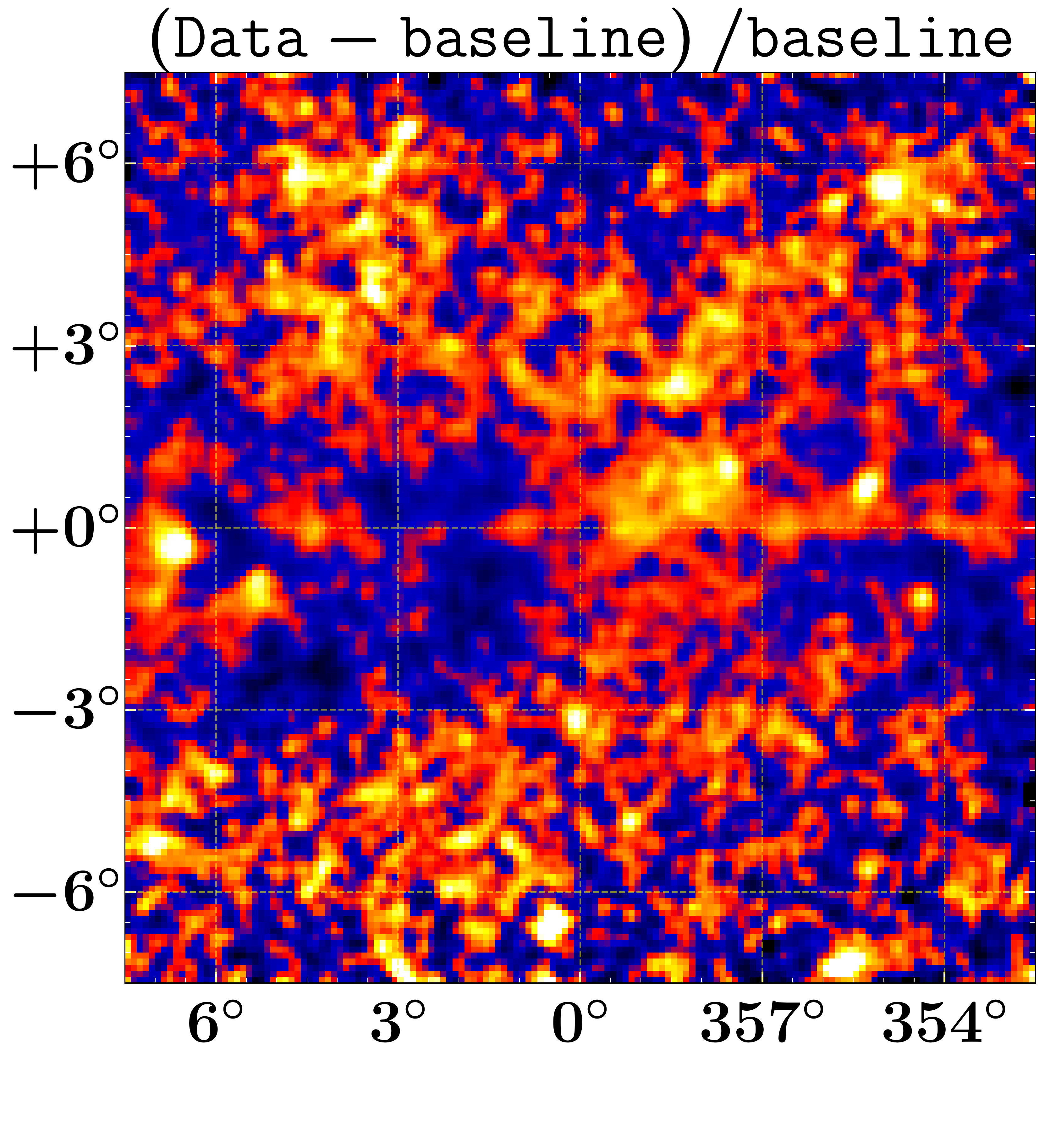}\\

\includegraphics[scale=0.2]{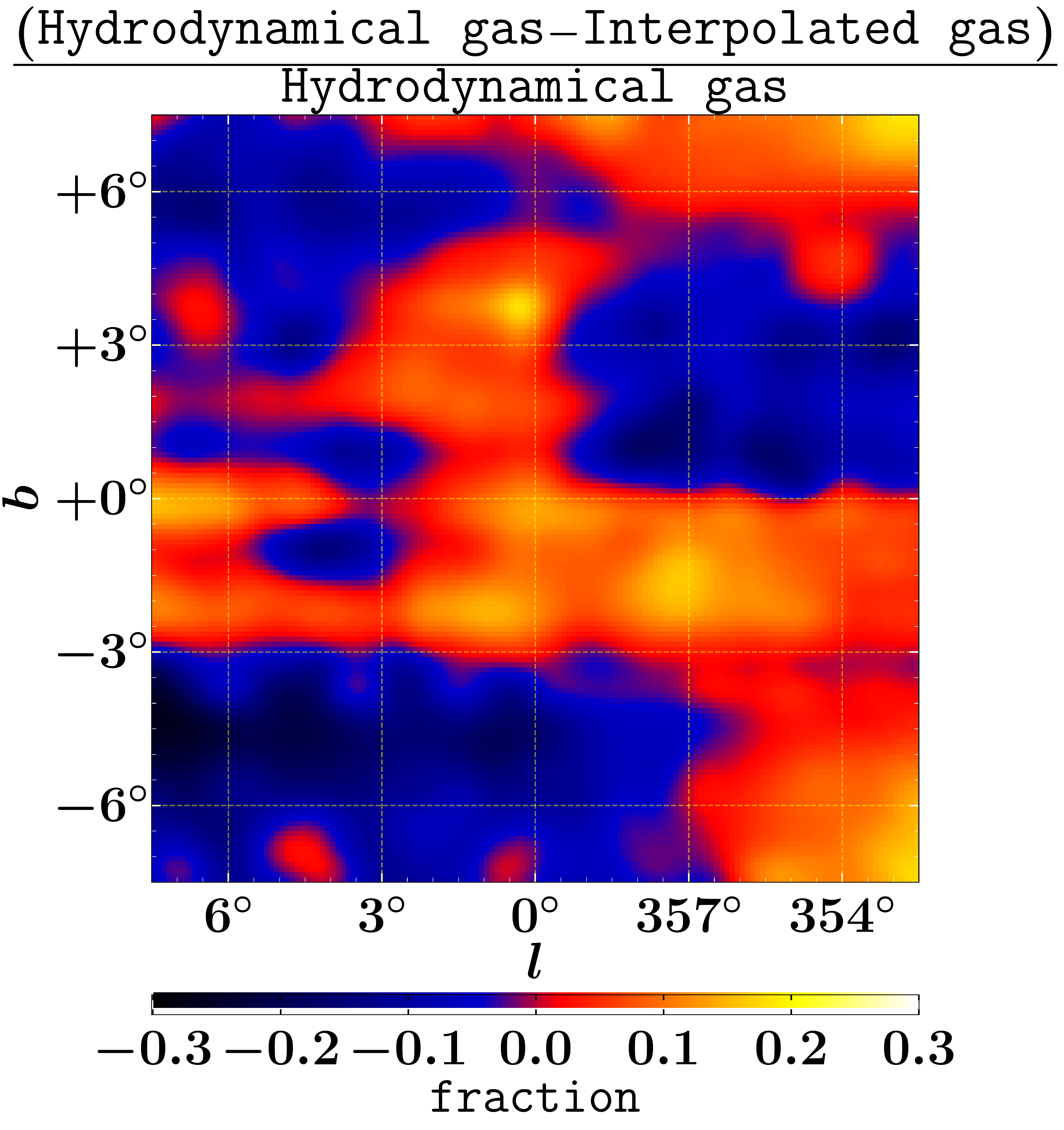} & \includegraphics[scale=0.2]{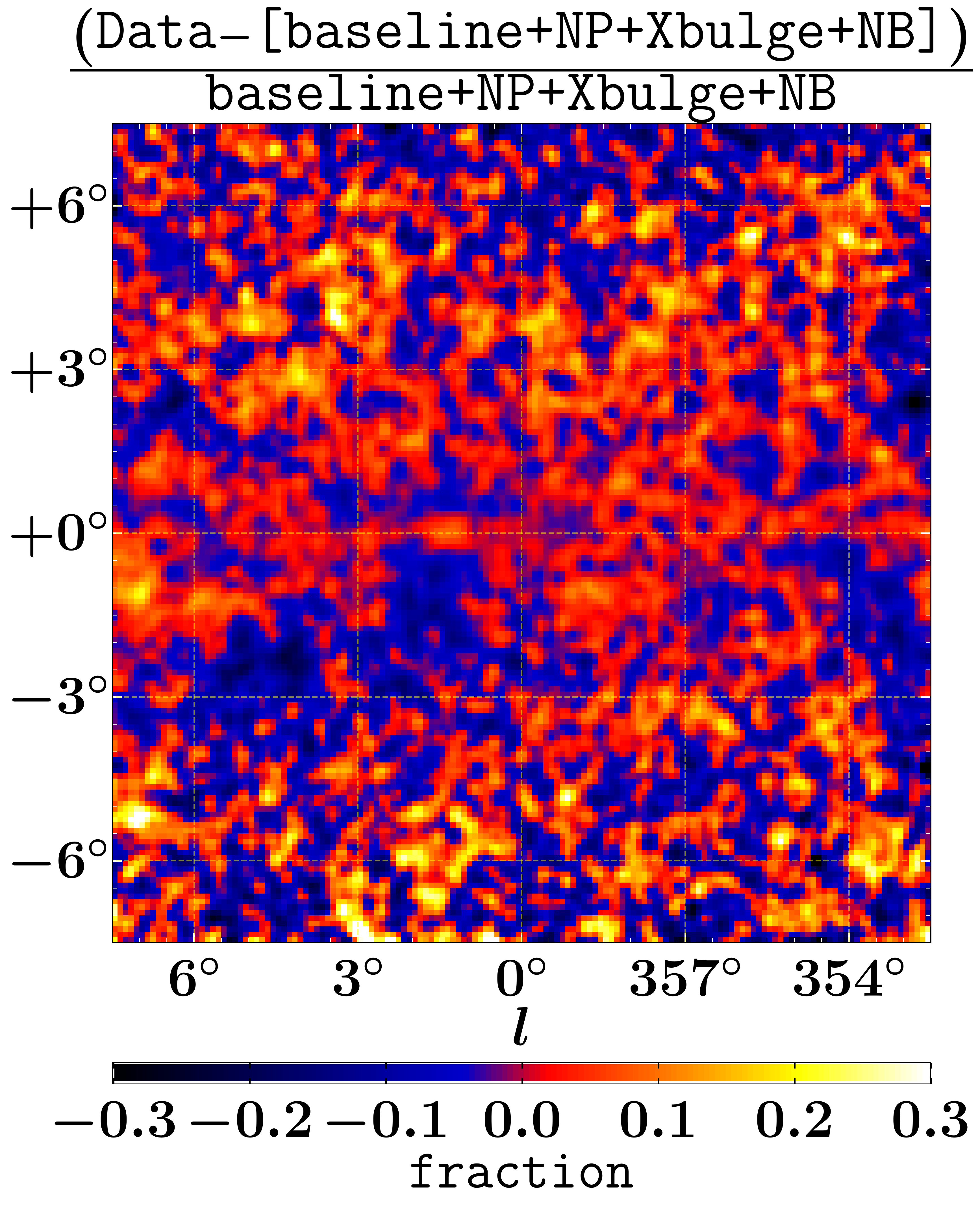} 
\end{tabular}

\caption{
{\bf   Fractional residuals for $E\ge 667$~MeV: }
Top left: The \textit{baseline}$^{\star}$ model consists of all 3FGL point sources in the ROI, Loop I, an IC template predicted by GALPROP, the interpolation based gas maps, the recommended isotropic emission map, and a model for the Sun and the Moon. Top right: The \textit{baseline} model is same as the top left, except that it uses the hydrodynamical based  gas maps. Bottom left: the fractional difference between the best fit interpolation based and the hydrodynamical gas maps. Bottom right: Our full model (\textit{baseline$+$NP$+$Xbulge$+$NB}) consisting of the baseline plus the 64 new point sources (NP),  an infrared X-bulge template tracing old stars in the Galactic bulge, 
        and a Nuclear Bulge (NB) template. For display purposes, we smoothed these images with a 0.1$^\circ$ Gaussian filter.
 \label{fig:Residuals}}
\end{center}
\end{figure*}

\begin{figure*}[ht!]
\begin{center}
\begin{tabular}{c}
\centering
\includegraphics[scale=0.35]{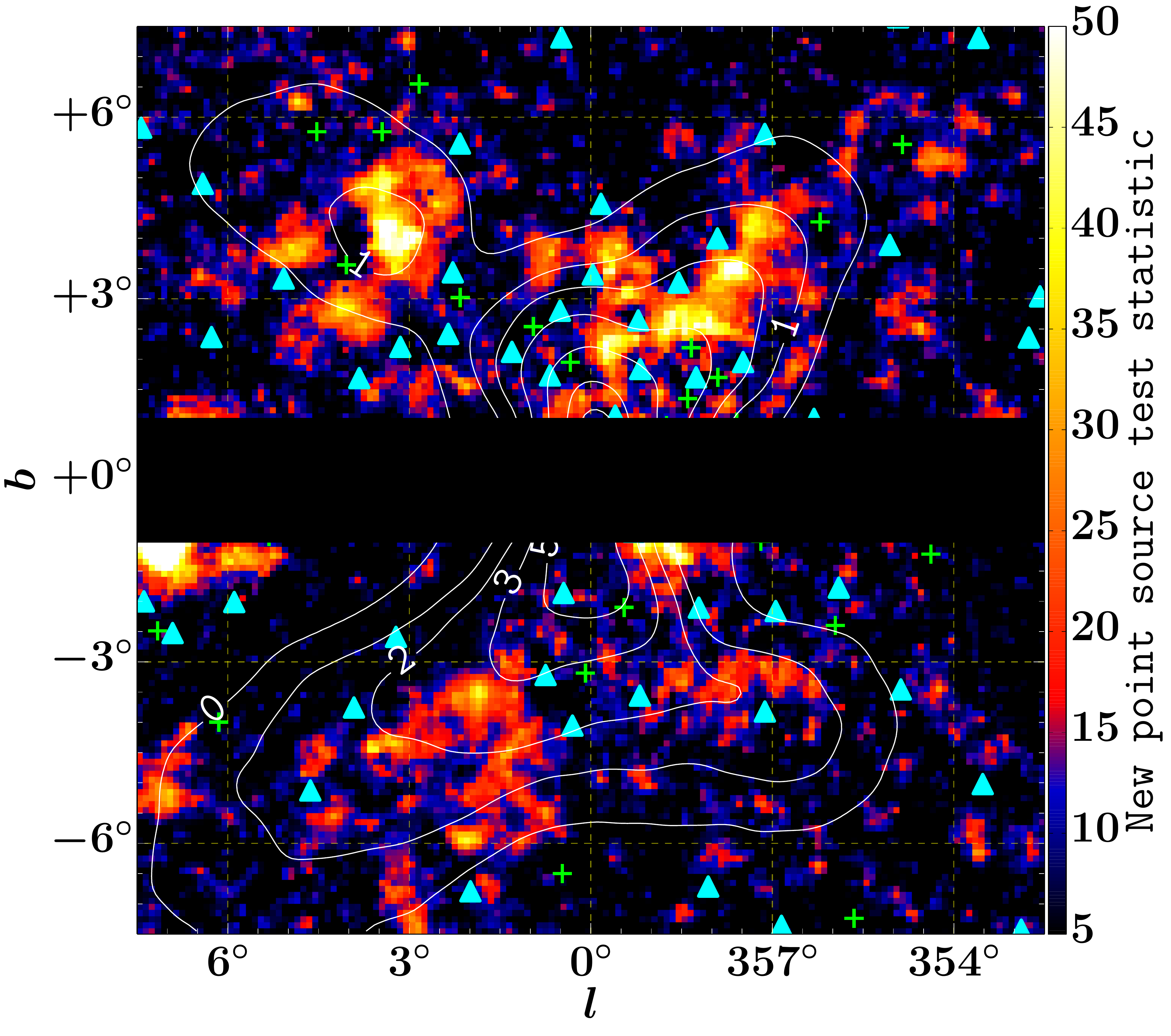}
\end{tabular}
\caption{
{\bf  Masking of the  Galactic plane ($|b|<1^{\circ}$).} This image displays the ROI for one of our systematic error analyses. The unmasked region contains 22 new point sources. More details of this pipeline are given in the Methods Section and Supplementary Table~\ref{Tab:PlaneMask}.
\label{fig:PlaneMask}}
\end{center}
\end{figure*}

\FloatBarrier
\newpage

\begin{center}
\begin{longtable}{ccccccc}
\caption{
{\bf  Point Sources detected with $TS \ge 41.8$ for the \boldmath$15^\circ \times 15^\circ$ region about the Galactic Center. } 
	\label{tab:Sources_xbulge1}
    }\\  
	\hline\hline
	Name &   $l$    &   $b$     &    Association     & TS    &   $F_{1-100\,{\rm GeV}}$                   &Spectrum \\
	     &   [deg]  &   [deg]   & or spatial overlap &       &   $\times 10^{-9}$ [ph cm$^{-2}$ s$^{-1}$] &                 \\ \hline
\endfirsthead
\multicolumn{7}{c}%
{{\bfseries \tablename\ \thetable{} -- continued from previous page}} \\
\hline\hline
	Name &   $l$    &   $b$     &    Association & TS    &   $F_{1-100\,{\rm GeV}}$                      &Spectrum \\
	     &  [deg]   &   [deg]   & or spatial overlap &      &   $\times 10^{-9}$ [ph cm$^{-2}$ s$^{-1}$] &                 \\ \hline
\endhead
\hline  \multicolumn{7}{r}{{Continued on next page}} \\ 
\endfoot
\hline \multicolumn{7}{l}{\parbox[t]{1.0\textwidth}{\T The full version of this table is also provided as a FITS file in the supplementary online material. The third column displays 1FIG\cite{Ajello2016} and 2FIG\cite{Ajelloetal:2017} associations as well as spatial overlaps with tentative multi-wavelength counterparts. We use the globular cluster catalog\cite{GBC:catalog}, ATNF pulsar catalog\cite{ATNF:catalog},  Green's SNR catalog\cite{Green:SNRcatalog},
and the Roma-BZCAT blazar catalog\cite{Massaro2009}. Best-fit fluxes were obtained in the $1-100$ GeV energy range are denoted by $F_{1-100\,{\rm GeV}}$. The spectrum column gives the spectral slope for a power law fit, except in the case where an exponential cutoff model was preferred at $\ge 4\sigma$ level where the $E_{\rm cut}$ in GeV units is also given. Similarly to ref.~\citenum{3FGL}, we combined our energy bands into four larger bands before fitting the spectrum. The errors are $1\sigma$.}
}\\
\endlastfoot
FGC	1711.3	-3008	&	354.9	&	5.6	&	2FIG J1711.0-3005	&	78.2	&	  $ 	2.5	\pm	0.3	 $	 & 	$ 	1.9	 \pm	0.2	 $  \\ \hline				
FGC	1719.7	-2947	&	356.2	&	4.3	&	2FIG J1719.1-2945	&	47.7	&	  $ 	1.9	\pm	0.3	 $	 & 	$ 	2.2	 \pm	0.4	 $  \\ \hline				
FGC	1725.9	-3429	&	353.1	&	0.5	&		&	55.7	&	  $ 	5.7	\pm	0.8	 $	 & 	$ 	2.6	 \pm	0.2	 $  \\ \hline				
FGC	1727.7	-2302	&	2.8	&	6.6	&	2FIG J1727.7-2305	&	71.5	&	  $ 	1.3	\pm	0.2	 $	 & 	$ 	1.9	 \pm	0.1	 $  \\ \hline				
FGC	1729.1	-3443	&	353.2	&	-0.1	&	1FIGJ1728.6-3433	&	109.2	&	  $ 	5.4	\pm	3.4	 $	 & 	$ 	2.7	 \pm	0.3	 $  \\ \hline				
FGC	1729.2	-3504	&	353.0	&	-0.4	&	2FIG J1729.1-3501, 	&	52.7	&	  $ 	2.6	\pm	16.9	 $	 & 	$ 	1.9	 \pm	0.4	 $  \\ 				
			&		&		&	1FIGJ1729.1-3502	&		&						 & 					 \\ \hline				
FGC	1729.9	-3423	&	353.6	&	-0.1	&	2FIG J1730.0-3421	&	57.2	&	  $ 	4.9	\pm	0.9	 $	 & 	$ 	2.7	 \pm	2.8	 $  \\ \hline				
FGC	1730.5	-3353	&	354.1	&	0.1	&	2FIG J1730.8-3356, 	&	113.2	&	  $ 	9.8	\pm	0.8	 $	 & 	$ 	3.2	 \pm	0.2	 $  \\ 				
			&		&		&	1FIGJ1730.2-3351,	&		&						 & 					\\				
			&		&		&	 G354.1+0.1~\cite{Green:SNRcatalog}, 	&		&						 & 					\\				
			&		&		&	PSR J1730-3350~\cite{ATNF:catalog}	&		&						 & 					\\\hline				
FGC	1731.6	-3235	&	355.3	&	0.6	&	2FIG J1731.3-3235, 	&	143.3	&	  $ 	15.1	\pm	0.8	 $	 & 	$ 	0.0	 \pm	0.0	 $  \\ 				
			&		&		&	1FIGJ1731.3-3235, 	&		&						 & 					\\				
			&		&		&	G355.4+0.7~\cite{Green:SNRcatalog}	&		&						 & 					\\\hline				
FGC	1732.1	-2257	&	3.5	&	5.8	&	2FIG J1731.3-2303	&	49.6	&	  $ 	1.7	\pm	0.3	 $	 & 	$ 	2.3	 \pm	0.7	 $  \\ \hline				
FGC	1733.1	-2910	&	358.3	&	2.2	&	2FIG J1732.6-2901	&	71.3	&	  $ 	3.4	\pm	0.5	 $	 & 	$ 	2.2	 \pm	0.4	 $  \\ \hline				
FGC	1733.3	-3318	&	354.9	&	-0.1	&	Liller 1~\cite{GBC:catalog}	&	57.4	&	  $ 	4.6	\pm	0.8	 $	 & 	$ 	2.8	 \pm	0.2	 $  \\ \hline				
FGC	1733.9	-2948	&	357.9	&	1.7	&		&	55.8	&	  $ 	1.3	\pm	0.4	 $	 & 	$ 	2.3	 \pm	0.2	 $  \\ \hline				
FGC	1734.6	-2202	&	4.5	&	5.8	&		&	73.7	&	  $ 	2.1	\pm	0.3	 $	 & 	$ 	1.9	 \pm	0.5	 $  \\ \hline				
FGC	1734.9	-3228	&	355.8	&	0.1	&	2FIG J1734.6-3237,	&	91.3	&	  $ 	8.8	\pm	0.8	 $	 & 	$ 	2.6	 \pm	0.3	 $  \\ 				
			&		&		&	1FIGJ1734.6-3228,	&		&						 & 					\\				
			&		&		&	G355.6−0.0~\cite{Green:SNRcatalog}	&		&						 & 					\\\hline				
FGC	1735.1	-3028	&	357.6	&	1.0	&	2FIG J1735.7-3025,	&	163.3	&	  $ 	8.0	\pm	0.7	 $	 & 	$ 	1.9	\pm	0.2	,\;	5.9	\pm	1.3	 $  \\
			&		&		&	1FIGJ1735.4-3030,  	&		&						 & 									\\
			&		&		&	Terzan 1~\cite{GBC:catalog}	&		&						 & 									\\\hline
FGC	1736.5	-2934	&	358.4	&	1.4	&		&	44.2	&	  $ 	3.3	\pm	0.6	 $	 & 	$ 	2.4	\pm	0.2	 $  \\ \hline				
FGC	1736.5	-3420	&	354.4	&	-1.2	&	2FIG J1736.2-3422, 	&	101.9	&	  $ 	3.3	\pm	0.5	 $	 & 	$ 	1.9	\pm	0.1	,\;	5.7	\pm	1.1	 $  \\ 
			&		&		&	1FIGJ1736.1-3422	&		&						 & 									\\\hline
FGC	1737.2	-3145	&	356.6	&	0.1	&	1FIGJ1737.4-3144	&	177.7	&	  $ 	12.0	\pm	0.9	 $	 & 	$ 	2.4	 \pm	0.4	 $  \\ \hline				
FGC	1738.1	-2647	&	1.0	&	2.5	&	PSR J1801-2451~\cite{ATNF:catalog}	&	43.1	&	  $ 	1.7	\pm	0.5	 $	 & 	$ 	2.4	 \pm	0.6	 $  \\ \hline				
FGC	1738.7	-3115	&	357.2	&	0.1	&		&	49.6	&	  $ 	3.3	\pm	0.9	 $	 & 	$ 	2.0	 \pm	1.8	 $  \\ \hline				
FGC	1738.9	-2737	&	0.3	&	2.0	&		&	50.4	&	  $ 	1.7	\pm	0.5	 $	 & 	$ 	2.1	 \pm	0.7	 $  \\ \hline				
FGC	1739.1	-2931	&	358.8	&	0.9	&		&	84.5	&	  $ 	5.1	\pm	0.7	 $	 & 	$ 	1.9	 \pm	0.4	 $  \\ \hline				
FGC	1739.2	-2530	&	2.2	&	3.0	&		&	56.2	&	  $ 	2.7	\pm	0.5	 $	 & 	$ 	2.3	 \pm	0.4	 $  \\ \hline				
FGC	1739.3	-3056	&	357.6	&	0.1	&	1FIGJ1740.1-3057	&	111.3	&	  $ 	9.5	\pm	1.0	 $	 & 	$ 	2.6	 \pm	0.4	 $  \\ \hline				
FGC	1739.6	-3022	&	358.1	&	0.4	&	1FIGJ1739.4-3010, 	&	363.3	&	  $ 	12.5	\pm	0.9	 $	 & 	$ 	1.9	 \pm	2.3	 $  \\ 				
			&		&		&	PSR J1739-3023~\cite{ATNF:catalog}	&		&						 & 					\\\hline				
FGC	1739.7	-3151	&	356.9	&	-0.5	&		&	48.0	&	  $ 	2.9	\pm	0.8	 $	 & 	$ 	2.9	 \pm	0.5	 $  \\ \hline				
FGC	1741.1	-2932	&	359.1	&	0.4	&		&	50.0	&	  $ 	3.0	\pm	1.1	 $	 & 	$ 	2.0	 \pm	0.9	 $  \\ \hline				
FGC	1741.5	-2337	&	4.0	&	3.6	&		&	50.3	&	  $ 	2.2	\pm	1.7	 $	 & 	$ 	1.9	\pm	0.6	,\;	1.5	\pm	0.3	 $  \\ \hline
FGC	1742.1	-3050	&	358.1	&	-0.5	&		&	66.3	&	  $ 	5.6	\pm	1.0	 $	 & 	$ 	2.7	 \pm	0.2	 $  \\ \hline				
FGC	1742.1	-3112	&	357.7	&	-0.5	&		&	46.7	&	  $ 	3.0	\pm	0.8	 $	 & 	$ 	1.9	\pm	0.2	,\;	1.8	\pm	0.2	 $  \\ \hline
FGC	1742.7	-2907	&	359.5	&	0.4	&	2FIG J1742.3-2916	&	90.4	&	  $ 	4.7	\pm	1.2	 $	 & 	$ 	2.4	 \pm	0.8	 $  \\ \hline				
\hline		
\end{longtable}
\end{center}

\begin{table}[h]
\centering
\caption{\bf  Radial distribution of $X_{CO}$ as obtained from our maximum likelihood estimation using the hydrodynamic gas template models. }
\begin{tabular}{r|c|c|c}
	\hline\hline
Annulus&  $0-3.5$ kpc &  $3.5-8.0$ kpc  &  $8.0-10.0$ kpc\\ \hline
$X_{CO}$&$0.4 \pm 0.1$ & $1.0 \pm 0.2$ & $3.9 \pm 1.3$ \\ 	\hline\hline
\end{tabular}

\parbox[t]{0.8\textwidth}{\label{X_CO}
 The quoted values are for the first three annuli and are in units of $ 10^{20}[$ cm$^{-2}$ (K km s$^{-1}$)$^{-1}]$. The errors are 1$\sigma$.}
\end{table}

\begin{table*}[!htbp]\caption{{\bf  Systematic likelihood analysis results\label{Tab:syslikelihoods}}}
\centering
\begin{threeparttable}
\begin{tabular}{llllcrc}
\hline\hline
Base  & Source &  $\log(\mathcal{L}_{\rm Base})$  & $\log(\mathcal{L}_{{\rm Base}+{\rm  Source}})$  &  $\mbox{TS}_{\rm Source}$& $\sigma$ & Number of\\ 
           &              &                                  &                         &                                               & & source parameters\\\hline

baseline$+$NB$+$X-bulge      & NFW& -171956.4         &-171948.7  & 15 & 1.5& 19 \\
baseline$+$NFW      &  NB$+$X-bulge& -172167.9         &-171948.7  & 438 & 18.6 & $2\times 19$ \\
\hline
baseline$^{\star}$      &  NFW& -173565.0         &-172929.2  & 1272 & 34.6 & 19 \\
baseline$^{\star}+$NFW      &  NB$+$X-bulge& -172929.2         &-172592.0  & 674 & 23.8& $2\times 19$ \\
baseline$^{\star}+$NB$+$X-bulge      &  NFW& -172631.5         &-172592.0  &79& 6.9 & 19 \\
\hline
baseline      & 2FIG& -172461.4         &-170710.5  & 3501  &37.3 & $81\times 19$ \\ \hline
baseline$+$2FIG      &  X-bulge&-170710.5 &-170487.3  & 446 & 19.8 &  19 \\
baseline$+$2FIG      &  NFW&-170710.5          &-170484.6  & 452 &  19.9&  19 \\
baseline$+$2FIG      &  NB&-170710.5          &-170470.5  & 480  &  20.6&  19 \\ \hline
baseline$+$2FIG$+$NB  & NFW&-170470.5         & -170387.8     & 165   & 11.1& 19\\ 
baseline$+$2FIG$+$NB  & X-bulge&-170470.5         &-170307.6      & 326   & 16.6& 19\\\hline
baseline$+$2FIG$+$NB$+$Xbulge  & NFW& -170307.6         &-170301.8      & 12   &1.1 & 19\\ \hline\hline

\end{tabular}
\begin{tablenotes}
      \item 
      See the caption of Table~1 for definitions.  The \textit{baseline}$^{\star}$ is same as the \textit{baseline} model, except that it uses the interpolated gas maps. 2FIG are the 81 new point sources found in our ROI by ref.~\citenum{Ajelloetal:2017}.
    \end{tablenotes}
\end{threeparttable}
\end{table*}

\begin{table*}[!htbp]\caption{{\bf Likelihood analysis results with a Galactic plane mask\label{Tab:PlaneMask}}}
\centering
\begin{threeparttable}
\begin{tabular}{llllcrc}
\hline\hline
Base  & Source &  $\log(\mathcal{L}_{\rm Base})$  & $\log(\mathcal{L}_{{\rm Base}+{\rm  Source}})$  &  $\mbox{TS}_{\rm Source}$& $\sigma$ & Number of\\ 
           &              &                                  &                         &                                               & & source parameters\\\hline

baseline      & NFW& -430289.1       &-430155.5  &134  & 9.8& $19$ \\
baseline      & X-bulge& -430289.1   &-430089.2  & 200 &12.5 & $19$ \\
baseline      & NP& -430289.1        &-429657.8  & 631 & 12.9& $22\times 19$ \\
\hline
baseline$+$NP      & NFW&-429657.8      &-429559.9  & 98 &8.0 & $19$ \\
baseline$+$NP      & X-bulge&-429657.8  &-429496.6  & 322 & 16.5& $19$ \\
\hline
baseline$+$NP$+$X-bulge      & NFW&-429496.6     &-429487.2
  & 19 & 2& 19 \\
\hline
\end{tabular}
\begin{tablenotes}
      \item 
      See the caption of Table~1 for definitions. After masking the Galactic plane ($|b|<1^{\circ}$), the number of new point sources (NP) added to the fit were 22 (see also Supplementary Fig.~\ref{fig:PlaneMask}).   
    \end{tablenotes}
\end{threeparttable}
\end{table*}

\end{document}


\noindent
 {\large {\bf {\fontfamily{phv}\selectfont Discovery of Gamma-Ray Emission from the X-shaped Bulge of the Milky Way
}}}

\noindent
Oscar~Macias$^{1}$, 
Chris~Gordon$^{2}$,
Brendan~Coleman$^{2}$, 
Dylan~Paterson$^{2}$,
Shunsaku~Horiuchi$^{1}$,\\
Roland~M.~Crocker$^{3}$, 
Martin Pohl$^{4,5}$\&
Shogo Nishiyama$^{6}$

\begin{affiliations}
 \item Center for Neutrino Physics, Department of Physics, Virginia Tech, Blacksburg, VA 24061, USA
\item Department of Physics and Astronomy, Rutherford Building, University of Canterbury, Private Bag 4800, Christchurch 8140, New Zealand
\item Research School of Astronomy and Astrophysics, Australian National University, Canberra, Australia
\item Institute of Physics and Astronomy, University of Potsdam, 14476 Potsdam-Golm, Germany
\item DESY, Platanenallee 6, 15738 Zeuthen, Germany
\item Miyagi University of Education, Aoba-ku, Sendai, Miyagi 980-0845, Japan.
\end{affiliations}

\begin{abstract}
An anomalous, apparently diffuse, $\gamma$-ray signal not readily attributable to known Galactic sources has been found\cite{Goodenough2009gk,Hooper:2010mq,hooperlinden2011,AbazajianKaplinghat2013,GordonMacias2013,MaciasGordon2014,hooperslatyer2013,HuangUrbanoXue2013,Abazajian2014,Daylan:2014,CaloreCholisWeniger2015,Ajello2016} in  {\it Fermi} Gamma-Ray Space Telescope (Fermi-LAT)   data covering the central $\sim 10$ degrees of the Galaxy.
This `Galactic Center Gamma-Ray Excess' (GCE) signal has a spectral peak at $\sim 2$ GeV and reaches its maximum intensity at the Galactic Centre (GC) from where it falls off as a radial power law $\propto r^{-2.4}$.
Given its morphological and spectral characteristics, the GCE is readily ascribable to self-annihilation of dark matter particles governed by an Navarro-Frenk-White (NFW) like density profile. 
However, it could also be composed of many dim, unresolved point sources for which millisecond pulsars\cite{AbazajianKaplinghat2012,MaciasGordon2014} (MSPs) or pulsars\cite{OLeary2015} would be natural candidates given their GeV-peaked spectra.
Statistical evidence that many sub-threshold point sources contribute up to 100\% of the GCE signal~\cite{Lee_etal:2016,Bartes_etal:2016} has recently been claimed.
We have developed a novel analysis that exploits hydrodynamical modeling\cite{Pohl2008} to better register the position of $\gamma$-ray emitting gas in the Inner Galaxy.
We also nonparametrically model the diffuse Galactic background and incorporate this uncertainty in our analysis.
%
%
These improvements allow us to demonstrate that the GCE is spatially correlated with the previously-known\cite{Nataf2010,NessLang2016} X-shaped stellar over-density in the Galactic bulge (`X-bulge'). 
We also find a strong correlation with 20 cm continuum emission associated with bremsstrahlung [RMC: WHY BREMSSTRAHLUNG?] emission in the galactic ridge. 
Our analysis also reveals 43 new point sources in the GC region and confirms 14 non-3FGL ones from 1FIG. 
Our findings imply that the gamma-ray GCE is associated with the $>$ 5 Gyr old stellar population of the X-bulge.
It is, therefore,
not a dark matter phenomenon nor associated with young pulsars but is very plausibly attributable to MSPs (which have multibillion year lifetimes) and, sub-dominantly, the `secondary' diffuse $\gamma$-emission from the electron and positrons injected by this MSP population.
\end{abstract}

We examine $\sim 7$ years of \textit{Fermi}-LAT data (August 4, 2008$-$September 4, 2015) selecting  \textsc{Pass 8} \textsc{ULTRACLEANVETO}\footnote{\url{http://fermi.gsfc.nasa.gov/ssc/data/analysis/documentation/Pass8_usage.html} {\bf RMC: this should go in SI references}} class events with energies between 500 MeV$-$500 GeV. Compared to previous LAT data releases, \textsc{Pass 8} provides greater acceptance and improved energy measurements. The selected events class has the most robust cuts on residual cosmic ray (CR) background contamination.

We extract the LAT data within a square region of $15^{\circ}\times15^{\circ}$ centred at Galactic coordinates $(l,b) = (0,0)$ and make no distinction between \textit{Front} and \textit{Back} events. To avoid contamination from terrestrial $\gamma$-rays, only events with zenith angles smaller than 100$^{\circ}$ are used. Time intervals when the rocking angle was more than 52$^{\circ}$ and when the Fermi satellite was within the South Atlantic Anomaly are also excluded. The \textit{Fermi} Science Tools \texttt{v10r0p5} is used for this analysis.


The Galactic diffuse $\gamma$-rays resulting from the interaction of CR $e^{\pm}$ and $p^+$ with the interstellar gas and radiation field is modeled with a similar method as for the 3FGL catalog~\cite{3FGL}. We fit a linear combination of annular gas templates divided up into 5 different rings assuming a uniform CR distribution within each ring, an IC map cube as obtained with GALPROP, specialized templates for the \textit{Sun} and the \textit{Moon} together with an isotropic component \footnote{\texttt{iso$_{-}$P8R2$_{-}$ULTRACLEANVETO$_{-}$V6$_{-}$v06.txt}}, and a model for the $\gamma$-ray emission associated with Loop I. The gas column densities are distributed within galactocentric annuli to account for the non-uniform CR flux in the Galaxy. In the following, we describe the construction of these templates tracing the $\gamma$-ray emission processes.

\begin{addendum}
\item[Supplementary Information] is linked to the online version of the paper at www.nature.com/nature. 
 \item[Acknowledgements]
 RMC was the recipient of an Australian Research Council Future Fellowship (FT110100108). OM would like to thank Dustin Lang for gracefully providing the X-bulge templates. The authors would like to sincerely thank Francesca Calore, Torsten En{\ss}lin, Ken Freeman, Oleg Gnedin, Xiaoyuan
Huang, David Nataf for enlightening discussions.

\item[Author Contributions]

 \item[Author Information] The authors declare that they have no
competing financial interests. Correspondence and requests for materials
should be addressed to O.M. (oscar.macias@vt.edu). 
Reprints and permissions information is available at npg.nature.com/reprintsandpermissions.
\end{addendum}






\clearpage




\section*{Supplementary Information}


In the main article we extended the approach used by Gordon and Macias\cite{GordonMacias2013} to cover a wider ROI, account for uncertainties in molecular and column densities, account for the low latitude Fermi-bubbles, and account for the possibility of an extra source of electrons and positrons.

\subsection{ $\rm{H}_{\rm{I}}$ and $\rm{H}_{2}$ Gas Column Density Maps}
\label{subsec:gasmap}

We use two different methods for constructing the gas map templates; one that reproduces the gas models used in most previous analyses of the GCE (called hereon ``standard model'') and that comes with the newest GALPROP distribution, and a novel approach based on hydrodynamical simulations of interstellar material in the Galaxy which will be part of the ``base model''.
 
In our standard model, the gas column density maps are produced based on the method given in Appendix B of ref.~\cite{ackermannajelloatwood2012}. Atomic hydrogen column density is derived from the 21cm LAB survey of galactic H$_{\rm{I}}$[\cite{kalberla2005}]. Molecular hydrogen is traced by the 2.6mm emission line of carbon monoxide (CO) from the 115 GHz Center for Astrophysics survey of CO[\cite{dame2001}]. 
 The emission maps are decomposed into Galactocentric annuli via the relation
 \begin{equation} \label{circular_motion}
 v_{\rm{LSR}} = R_{\odot} \left( \frac{V(R)}{R} - \frac{V_{\odot}}{R_{\odot}} \right) \sin(l) \cos(b)
 \end{equation}
 where $v_{\rm{LSR}}$ is the radial velocity of the gas at Galactocentric radius $R$ with orbital velocity $V(R)$ observed in the direction $(l,b)$.
Our standard model assumes a spin temperature of $T_{S} = 125 \, \rm{K}$ throughout the galaxy.

In our region of interest the annulus decomposition method does not produce a reliable map of the gas column density, as kinematic resolution is lost near $l = 0^{\circ}$. The gas column density in the region $|l| < 10^{\circ} $ is estimated by interpolating within each annulus from the boundaries. Values at the boundaries of the interpolated region are chosen as the mean gas column density within a range of $\Delta l = 5^{\circ}$ on both sides of the boundary. Each pixel in the interpolated region is renormalized to preserve the total gas column density in each line of sight.

An alternative model in which the emission map deconvolution is improved upon by Pohl et al.~\cite{Pohl2008} is also considered.
Pohl et al. specify a probabilistic deconvolution method for assigning gas clouds to distances via simulated gas flow model predictions. 
Their gas flow model follows hydrodynamic simulations inside the solar circle outlined in ref.~\cite{Bissantz2003}, taking the spiral arms and Galactic Bar into account.
The presence of non-circular motion in the inner Galaxy provides kinematic resolution towards the Galactic Centre.
Outside the solar circle pure circular motion with velocities described by equation (\ref{circular_motion}) is assumed.

The line intensity data resulting from the work of Pohl et al.~\cite{Pohl2008} are utilized here.
These models are derived using the same H$_{\rm{I}}$ and CO surveys as for the standard model.
A global spin temperature of 170K is assumed in the H$_{\rm{I}}$ deconvolution.
Distances are scaled to account for the 8 kpc solar circle in the simulation and line intensities converted to Galactocentric annuli column density maps.

\begin{figure*}[t!]
\includegraphics[scale=0.5]{H2__figure_horizontal.pdf}
\includegraphics[scale=0.5]{HI__figure_horizontal.pdf}
\caption{Column density maps for the CO (top two rows) and HI (bottom two rows). The first and third rows use the interpolated approach and the 2nd and 4th rows use the Pohl et al.~\cite{Pohl2008}\ approach. The units for the CO maps are K$\cdot$km$/$s. The units for the HI maps are $10^{20}$cm$^{-2}$.
 \label{fig:densities}}
\end{figure*}

\begin{figure}[t]
\begin{center}
\includegraphics[scale=0.45]{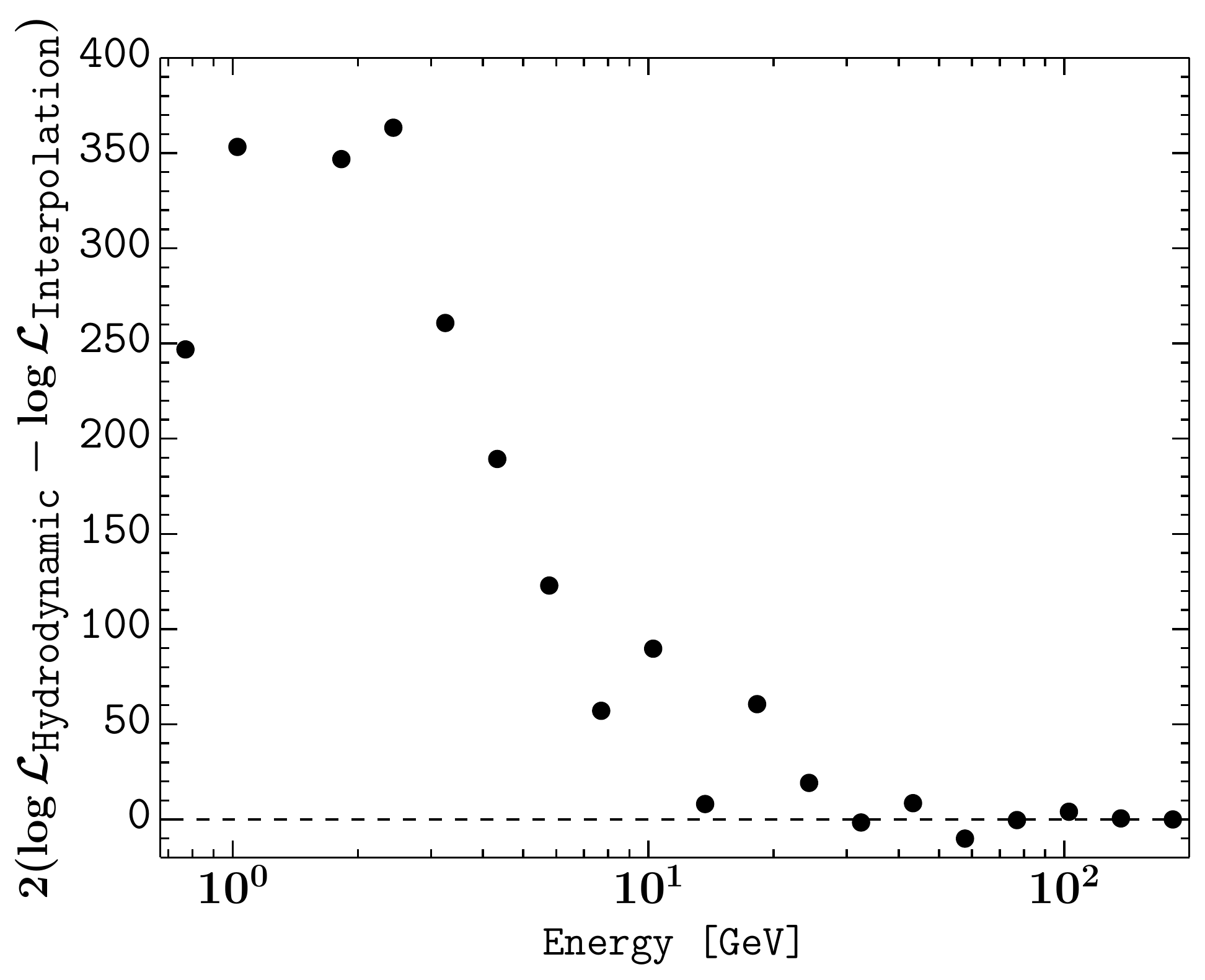}

\caption{ Comparison of the log-likelihood obtained for two different interstellar gas models: Pohl et al.~\cite{Pohl2008} gas maps vs the interpolation ones used in the standard Galactic diffuse emission model case. Summing over the energy bins gives $TS_{\mbox{\tiny Pohl et al.}}=354$. Note that this comparison was done with just the 3FGL sources and no X-bulge, 20-cm nor FB templates were added to the fit.
 \label{fig:pohvsinterp}}
\end{center} 
\end{figure}

\subsection{Dust Correction to Account for Dark Neutral Medium}
The distribution of $\rm{H}_{\rm{I}}$ and $\rm{H}_{2}$ is not completely mapped by the method in Section \ref{subsec:gasmap}. Molecular hydrogen that is not well mixed with carbon monoxide will not be traced by 2.6mm emission. Furthermore, assuming a constant atomic hydrogen spin temperature $T_{S} = 125 \, \rm{K}$ can give an incorrect estimate of column density as the spin temperature can vary along a line of sight. These potential biases can be minimized by including a template for the dark neutral medium (DNM).

Our dark neutral medium template is based on the templates used in refs.~\cite{ackermannajelloatwood2012} and \cite{casandjian2014}. Infrared thermal emission from dust provides an alternative method of tracing hydrogen gas in the Milky Way provided they are well mixed. Subtracting the components of the dust emission that is correlated with the gas already traced by 21cm and 2.66mm emission gives a template for the DNM. 

We applied this method to E(B-V) maps of ref.~\cite{schlegel1998}. Regions that are densely populated by infrared point sources (or potentially a collection of unresolved point sources) could contaminate the E(B-V) map, resulting in an over estimation of the dust column density. To mitigate this, we apply a magnitude cut of 2mag to the E(B-V) maps, which has a significant effect in our region of interest. After subtracting the components of the E(B-V) map that were linearly correlated with the hydrogen gas maps, the residuals were separated into positive and negative components. The positive residuals physically represent hydrogen that is untraced by the relevant emission or an over estimation of the atomic hydrogen spin temperature. Negative residuals represent an underestimation of the spin temperature.

\subsection{Inverse Compton emission model}
\label{subsec:IC}

Whereas the bremsstrahlung and $\pi^0$-decay  components can be adequately described by the gas maps above, there is no analogous empirical template  for the Galactic IC emission. We use the GALPROP package v54.1 to generate such a template. The authors of ref.~\cite{ackermannajelloatwood2012} did a comprehensive comparison of $128$ different GALPROP models with all sky \textit{Fermi}-LAT data. They found a range of possible values for the input GALPROP parameters that are consistent with $\gamma$-ray and local measurements of  CR data.  In this work, we consider the IC template generated with GALDEF file \texttt{galdef$_-$54$_-$Lorimer$_-$z10kpc$_-$ R20kpc$_-$Ts150K$_-$EBV2mag}~\footnote{\url{http://galprop.stanford.edu/PaperIISuppMaterial/} {\bf[RMC: put in SI REFERENCES]}} as our reference model. However, we do not view this template as preferred over other possible models. As will be shown in following sections, the IC component is in general negligible compared to the other DFGB components and so our results are insensitive to using alternative IC templates from  ref.~\cite{ackermannajelloatwood2012}.


\subsection{Loop I Template}
\label{subsec:LoopI}

\begin{figure}[t]
\begin{center}
{
\includegraphics[scale=0.43]{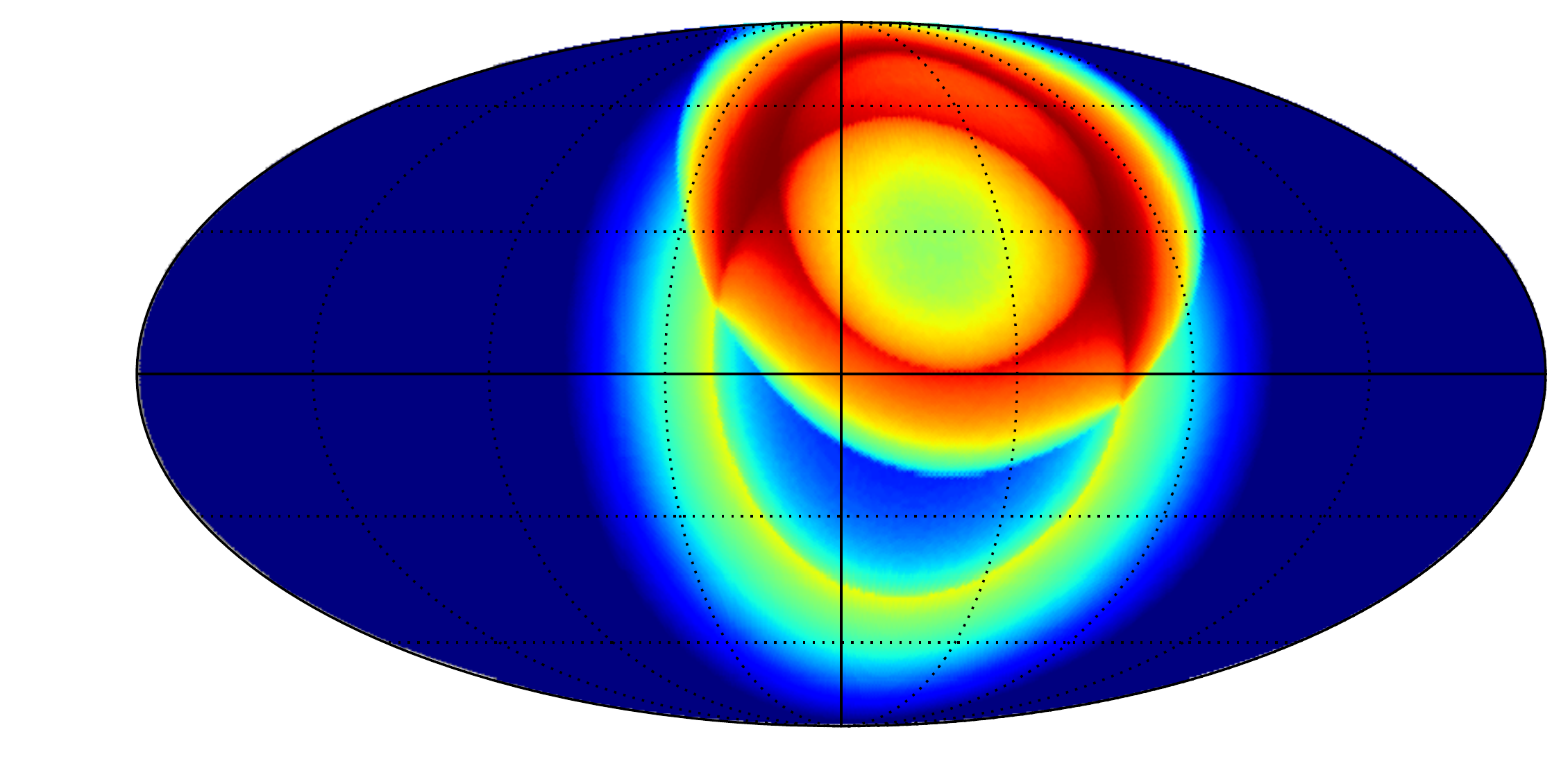}

\caption{Loop I template adapted from ref.~\cite{Wolleben:2007}. A histogram equalized color mapping is used for display. The map is appropriately normalized for analysis with the \textit{Fermi} Science Tools. }
 \label{fig:LoopI}}
\end{center}
\end{figure}

Loop I is a bright, large angular scale, non-thermal structure.
Obtaining a precise template of this source is not possible since its $\gamma$-ray emission is not well traced by radio emission. We note that the standard diffuse emission model used in our previous investigations of the GCE~\cite{GordonMacias2013, MaciasGordon2014,MacGorCroProf2015} employs a uniform-patch Loop I template whose shape was derived by visual inspection of the $\gamma$-ray residuals. Following the same approach taken in recent studies by the \textit{Fermi} team\cite{Fermi:LatBubbles}, in this work we use a geometrical template proposed by ref.~\cite{Wolleben:2007} which is based on a polarization survey at 1.4 GHz  (see Fig.~\ref{fig:LoopI}). We adopt the same morphological parameters for the shells assumed in ref.~\cite{Fermi:LatBubbles}. As will be explained in detail in Sec.~\ref{sec:methods}, our fitting procedure uses these templates as a starting point for our maximum likelihood analysis.

\subsection{Sun and Moon emission Templates }
\label{subsec:SunadMoon}

CR $e^{\pm}$ interacting through IC scattering with the solar radiation field and solar atmosphere produce extended $\gamma$-ray emission from around the Sun~\cite{Fermi-Sun}. Since the Moon moves through a large fraction of the sky, CR interactions taking place in the lunar lithosphere make the Moon appear as a reasonably bright extended $\gamma$-ray source. In order to account for the diffuse emission from these objects, we construct specialized templates for our ROI by making use of the \textit{gtsuntemp}~\cite{gtsuntemp} utility.

\subsection{Optional Templates for use in the Systematic Uncertainties Analysis}

\subsubsection{\textit{Fermi} Bubbles spatial map}
\label{subsubsec:FermiBubbules}

The \textit{Fermi} Bubbles~\cite{Su:etal,Fermi:LatBubbles} (FB)  are giant lobes that extend up to  $\sim 7$  kpc from the Galactic plane into both the North and South Galactic  hemispheres. Recently, Acero et al.~\cite{Casandjian:andFermiLat2016} performed a morphological study of these structures in the inner $20^{\circ}$ of the GC using several flat templates of different shapes. They found that the boundaries of the \textit{Fermi} bubbles are well described by two catenary curves of the form $10.5\degree \times (\cosh((l - 1\degree)/10.5\degree) - 1\degree)$ and $8.7\degree \times(\cosh((l+1.7\degree)/8.7\degree) - 1\degree)$ for the Northern and the Southern bubble respectively. 

\subsubsection{20-cm map of the Galactic Ridge}
\label{subsubsec:20cm}

A recent study~\cite{MaciasGordon2014} carried out a \textit{Fermi}-LAT analysis of the Galactic  ridge region ($2\degree \times 1\degree$ around the GC) using templates obtained from measurements at other wavelengths. Either a 20-cm  map \cite{Yusef-Zadeh2013} or a HESS residuals \cite{Aharonian:2006}  map were used. 
The 20-cm template was based on  Green Bank Telescope continuum emission data  which measures non-thermal and thermal plasma distributions \cite{law2008}.  Note,  this is  distinct from the first ring in our gas models (see Sec.~\ref{subsec:gasmap}) as the later encompass a much wider section of the Galaxy (from $0$ to $1.5$ kpc). Both the 20-cm and the HESS residuals templates were found to improve significantly the fit and in approximately the same proportion.

\section{Methods}
\label{sec:methods}





Following previous works,  we employ a bin-by-bin analysis technique, in which we split the {\it Fermi}-LAT data into 24 energy windows. Within each energy bin we perform a separate maximum-likelihood fit~\cite{3FGL} with the \textit{pyLikelihood} analysis tool~\footnote{\url{http://fermi.gsfc.nasa.gov/ssc/data/analysis/documentation/}{\bf[RMC: put in reference section]}}. The window size is chosen to be larger than the LAT energy resolution, but narrow enough that the Galactic background spectral components can be simply approximated by Power-Law (PL: $dN/dE = N_0 E^{-\alpha} $ with norm $N_0$ and slope $\alpha$) formulas with fixed slopes. We note that this bin-by-bin method makes the results independent of the spectrum of the sources and significantly decreases the number of iterations required to reach convergence in this very challenging ROI.

\subsection{Point Source Search}
\label{subsec:ptsrcsearch}

Initially we fit the LAT data with a model comprised of the 3FGL catalogue~\cite{3FGL} point sources present in our ROI plus four other spatially resolved sources (HESS J1825-137, RX J1713.7-3946, W28 and W30) reported by the {\it Fermi}-LAT team. The diffuse $\gamma$-ray background is modeled using the spatial templates described in Sec.~\ref{sec:backgroundmodel}, except for the catenary map describing the Fermi Bubbles, which is included at a later stage in the analysis.

To identify the most suitable gas templates for our study we perform a scan in which we evaluate the improvement of the likelihood fit to the ROI when the gas maps used are the ones created with the interpolation method or the Pohl et al.\cite{Pohl2008} method. Figure~\ref{fig:pohvsinterp} shows that the data prefers the Pohl et al. templates for most parts of the energy range of interest and these are, therefore, selected for our baseline model.

During optimization the flux normalization of the 3FGL sources mentioned above are left free while all the spectral shape parameters are fixed to their catalogue values\cite{3FGL}. We also varied the extended components normalization but kept the \textit{Sun} and \textit{Moon} fluxes fixed to their nominal values. These steps provided a reference log-likelihood estimate along with a best-fit baseline model.

In order to search for missing point sources we divide the ROI in $0.1^{\circ}\times 0.1^{\circ}$ grid positions and examine the significance of a trial point source with a PL spectrum and fixed slope $\alpha=2.0$ at the center of each pixel. 
The outcome of this pass is a residual test statistic (TS) map (the \texttt{gttsmap} utility is used for this step). The TS at each pixel is computed as $\mbox{TS} = 2(\log\mathcal{L} (\mbox{source}) - \log \mathcal{L} (\mbox{no-source}))$, where $\mathcal{L}$ is the likelihood of the data given the model with or without a source present at a given grid position. In accordance with our bin-by-bin method, a residual TS map is computed for each energy bin and these are then added to get a total residual map for the full energy range.

From the total residual TS map we generate a list of all the pixel clusters with TS values above the detection threshold that look reasonably isolated under visual inspection. The coordinates of the source candidates are calculated as the average of adjacent pixel positions weighted by their respective TS values~\cite{2FGL}. We then add the seeds to our model and rerun the bin-by-bin analysis routine.
Note that in the thresholding pass, convergence is achieved by adding the point source seeds to the model consecutively in order of decreasing TS's. This also makes the method more robust against source confusion. 
We do not use \textit{gtfindsrc} to refine the seeds' positions because it is based on an unbinned likelihood method\cite{2FGL}. When doing a global analysis, each new PS candidate has 2 degrees of freedom for the PL and two degress of freedom for the position. In that case a $\mbox{TS} \geq 25$ is used as detection condition. However, since we have 19 energy bands in which the PS amplitude is allowed to vary, we use a threshold of  $\mbox{TS} \geq 55$ which has the same p-value as the global analysis case.  All sources that are found to have a TS above the threshold  are allowed to remain in our baseline model. We iterate through these steps until no more seeds are found or accepted.

Figure~\ref{fig:tsmaps}-(a) shows the TS residual map obtained in the first iteration with the reference model; many pixel clumps with $\mbox{TS} \geq 55$ are readily evident. Each consecutive panel in Fig.~\ref{fig:tsmaps} shows the TS residual map obtained after augmenting  the reference model with the point source candidates satisfying the threshold condition. Remarkably, our procedure confirms the existence of 43 new point source candidates in the ROI. 
The total set of new point sources found in this work are displayed in Fig.~\ref{fig:tsmaps}-(d) along with the TS residual map obtained in our last iteration. Although the model including all the new point sources is a much better representation of the ROI, a few hot spots still remain. These are, however, found to be below the detection threshold of $\mbox{TS}\ge 55$ by a maximum-likelihood iteration.  

To identify possible multi-wavelength counterparts to the gamma-ray sources we searched in the seed locations $-$ within the 68\% containment of the point spread function (PSF) for one of our highest energy bands $\sim 0.1\degree -$ around each source in the ATNF pulsar~\cite{ATNF:catalog}, globular cluster~\cite{GBC:catalog} and SNR~\cite{Green:SNRcatalog} catalogs for potential gamma-ray emitters. We find spatial overlaps for 7 of our 43 point source candidates (see Tab.~\ref{tab:Sources_3fgl}). Note that this does not preclude the possibility that the other 36 seeds are real sources since a considerable fraction of the \textit{Fermi}-LAT sources~\cite{3FGL} have no  multi-wavelength associations.

\begin{figure*}[ht]
\begin{center}
\begin{tabular}{cc}
\centering
\includegraphics[scale=0.25]{TS_Residuals_1.pdf} &  \includegraphics[scale=0.25]{TS_Residuals_2.pdf}
\end{tabular}

\begin{tabular}{cc}
\centering
\includegraphics[scale=0.25]{TS_Residuals_3.pdf} & \includegraphics[scale=0.25]{TS_Residuals_8.pdf}
\end{tabular}

\caption{\small Significance map of the $15^\circ \times 15^\circ$ region about the GC output from \texttt{gttsmap} (the results of the point source search at each iteration are shown consecutively from [a]-[d]). The energy range shown is $1-150$ GeV. The 3FGL (white circles) point sources are displayed in all panels. Point source candidates to be passed to the thresholding step at each iteration are represented by green angled crosses, and these are then shown as white crosses in the next panel if they satisfy the threshold condition. The last panel (d) displays the 43 additional point source candidates found in this study. Table~\ref{tab:Sources_3fgl} summarizes their basic properties. 
 \label{fig:tsmaps}}
\end{center} 
\end{figure*}



\section{Search for Extended Emission}
\label{Sec:extendedemission}

\begin{figure}[t!]
\begin{center}
\begin{tabular}{cc}
\centering
\includegraphics[scale=0.4]{GCE_new_vs_old_analysis.pdf} &  \includegraphics[scale=0.4]{TS_GCE_no_new_ptsrcs.pdf}
\end{tabular}

\begin{tabular}{cc}
\centering
\includegraphics[scale=0.4]{GCE_new_ptsrcs.pdf} &  \includegraphics[scale=0.4]{TS_GCE_new_ptsrcs.pdf}
\end{tabular}

\small 
\caption{From left to right: NFW$^2$ energy spectrum as extracted from the $15\degree \times 15\degree$ around the GC assuming a generalized NFW$^2$ profile with an inner slope $\gamma=1.2$ and the TS values obtained at each energy bin for the extended NFW$^2$ source. The first row assumes the standard background model (see Tab.~\ref{Tab:likelihoods})  and  a total $\mbox{TS}_{\mbox{\tiny NFW$^2$}}=1207$ is obtained for the NFW$^2$ source. While the second row uses our baseline background and includes the set of 43 new point sources (baseline$+$NP) to the bin-by-bin fit giving a $\mbox{TS}_{\mbox{\tiny NFW$^2$}}=337$. The GCE spectra obtained by e.g. \cite{MaciasGordon2014,CaloreCholisWeniger2015} are overlaid for comparison. 
\label{fig:GCE_no_ptsrcs}}
\normalsize
\end{center}
\end{figure}

\begin{figure}[t!]
\begin{center}

\begin{tabular}{cc}
\centering
\includegraphics[scale=0.4]{GCE_new_ptsrcs_xbulge.pdf} &  \includegraphics[scale=0.4]{TS_GCE_new_ptsrcs_xbulge.pdf}
\end{tabular}
\begin{tabular}{cc}
\centering
\includegraphics[scale=0.4]{GCE_new_ptsrcs_xbulge_NBstars.pdf} &  \includegraphics[scale=0.4]{TS_GCE_new_ptsrcs_xbulge_NBstars.pdf}
\end{tabular}

\small
\caption{Same descriptions as in Fig.~\ref{fig:GCE_no_ptsrcs}. Top: NFW$^2$ spectrum when the new point sources plus the X-bulge template are added to the fit (baseline$+$NP$+$X-bulge in Tab.~\ref{Tab:likelihoods}). This gives a total $\mbox{TS}_{\mbox{\tiny NFW$^2$}}=130$. Bottom: in this case the background model is further augmented by the inclusion of the NBstars extended source. The significance of the NFW$^2$ model is downgraded to  $\mbox{TS}_{\mbox{\tiny NFW$^2$}}=29$ for 19 degrees of freedom.
\label{fig:GCE_syst_uncert}}
\normalsize
\end{center}
\end{figure}

We search for extended emission from the inner Galaxy with a physical model that has been shown by previous works to describe well the GCE, namely an NFW density profile: 
\begin{equation}
\rho (r) = \frac{\rho_\odot}{ \left(\frac{r} { R_\odot}\right)^\gamma  \left(\frac{1 + r/R_s}{1 + R_\odot/R_s}\right)^{3-\gamma} }, 
\end{equation}

\noindent
where $R_\odot$ is the Sun's distance from the GC, $\rho_\odot$ is the DM density at $R_\odot$, $R_s$ is the scale radius of the Galaxy's DM halo, and $\gamma$ is the profile slope. We set some of these parameters to $R_s = 23.1$ kpc, $R_\odot = 8.25$ kpc and $\rho_\odot = 0.36$ GeV/cm$^3$.

The recent analysis of the GCE made by the \textit{Fermi} team~\cite{TroyandSimona:2015} considered two  different profile slopes; $\gamma=1,1.2$. We refer to these alternative templates as ``Standard NFW'' (NFW-s) and ``Cuspy NFW'' (NFW) for $\gamma=1$ and $\gamma=1.2$ respectively. The square of such profiles (e.g.~\cite{GordonMacias2013}) are representative of a tentative annihilating DM signal or an unresolved population of MSPs in the GC. Throughout the rest of this article we will use the notation NFW$^2$ or NFW$^2$-s to allude to a DM annihilation or MSPs template.

We first examine the case in which the set of new point sources (Tab.~\ref{tab:Sources_3fgl}) is not included in the model. We fit for extended excess $\gamma$-ray emission with our bin-by-bin method to derive fluxes that are independent of the choice of spectral model (see Sec.~\ref{sec:methods} for a detailed description of the procedure). Within each bin, the spectrum of the NFW$^2$ source is modeled with a PL formula with fixed spectral index $\alpha=2.0$.
Due to the small size of our bins, our results are not sensitive to the precise slope used.
Figure~\ref{fig:GCE_no_ptsrcs} shows that with our improved methods we are able to reproduce the results of previous studies~\cite{Goodenough2009gk, AbazajianKaplinghat2012,GordonMacias2013, Hooper2013, Daylan:2014,CaloreCholisWeniger2015,TroyandSimona:2015}, finding a $\mbox{TS}_{\mbox{\tiny NFW$^2$}}=1207$ over the whole energy range. However, when performing the same analysis, but this time using the fiducial Pohl et al gas maps along with the set of 43 new point sources to the model (baseline$+$NP), the TS-value drops to $\mbox{TS}_{\mbox{\tiny NFW$^2$}}=337$ (see Tab.~\ref{Tab:likelihoods}). This provides an indication that the GCE may be due to a combination of incomplete modeling of the interstellar emission and faint point sources not accounted for in the ROI.


\begin{figure}[t!]
\begin{center}

\begin{tabular}{cc}
\centering
\includegraphics[scale=0.4]{spectra_xbulge.pdf} &  \includegraphics[scale=0.4]{TS_xbulge.pdf}\\

\includegraphics[scale=0.4]{spectra_xbulge+NP.pdf} &  \includegraphics[scale=0.4]{TS_xbulge+NP.pdf}

\end{tabular}

\small
\caption{Differential flux of the X-bulge template (left) and corresponding band TS-values (right). Top: The baseline model is assumed (see Tab.~\ref{Tab:likelihoods}). The spectrum peaks at $\sim 2$ GeV and has a total $\mbox{TS}_{\mbox{\tiny X-bulge}}=475$. Bottom: the background model used corresponds to baseline$+$NP. When the new point source candidates are added to the model we get $\mbox{TS}_{\mbox{\tiny X-bulge}}=316$. 
\label{fig:Xbulge_spectra}}
\normalsize
\end{center}
\end{figure}

These findings are consistent with previous studies of the Virgo Cluster~\cite{macias-ramirezgordonbrown2012, Han:2012uw, Ackermann:virgo}, which have shown that the degeneracy between a potential extended source and overlapping faint point sources decrease the significance of the former. This is also in line with the results of refs.~\cite{Lee_etal:2016,Bartes_etal:2016} which demonstrated that a population of unresolved point sources could explain the GCE.

Not only DM is theoretically motivated to produce extended gamma-ray emission in the GC, but also stellar evolution models predict the existence of an X-shaped structure in the Milky Way bulge that could contain unresolved gamma-ray emitters. As introduced in Sec.~\ref{subsubsec:Xbulge}, such formation has recently been found~\cite{Lang:2016} in WISE data. Remarkably, the residual infrared X-bulge is well traced by our residual gamma-ray TS maps (see Fig.~\ref{fig:Xbulge_correlation}). The correlation is much more evident away from the plane and for positive latitudes. 

We investigated further this possibility by performing a maximum-likelihood analysis with a flux normalized X-bulge template. We find a $\mbox{TS}_{\mbox{\tiny X-bulge}}=475$ and  $\mbox{TS}_{\mbox{\tiny X-bulge}}=316$ before and after including the set of new point sources, respectively. There is a remarkable similarity between the significance obtained for the annihilating NFW$^2$ and the X-bulge sources in each case under consideration, pointing to a high degeneracy between the two hypothesis. 

The differential flux and band TS-values for the X-bulge map are shown in Fig.~\ref{fig:Xbulge_spectra}. Similarly to the GCE, the X-bulge spectrum also peaks at $\sim 2$ GeV and is well fit by an power-law spectrum of the form $\propto E^{-2.2\pm0.1}$.
 The NFW$^2$ spectrum found with the standard model is still a factor of $\sim 6$ higher than that of the X-bulge. Note however, that a large fraction of the new point sources lay in the plane. This explains why both, the X-bulge and new point sources are needed to account for the majority of the GCE.   


\begin{figure*}[ht]
\begin{center}
\includegraphics[scale=0.5]{Xbulge_correlation.pdf}
\caption{
Significance (TS) map of the $15^\circ \times 15^\circ$ region before adding the additional point sources.  The 3FGL (white circles) point sources and the 43 additional point source candidates found in this study are displayed. The white contours are extracted from Fig.3 of ref.~\cite{Lang:2016} and correspond to the X-bulge structure as seen by WISE~\cite{WISE} at the infrared.  
 \label{fig:Xbulge_correlation}}
\end{center}
\end{figure*}

\begin{figure}[ht!]
\begin{center}
\begin{tabular}{c}
\centering
\includegraphics[scale=0.5]{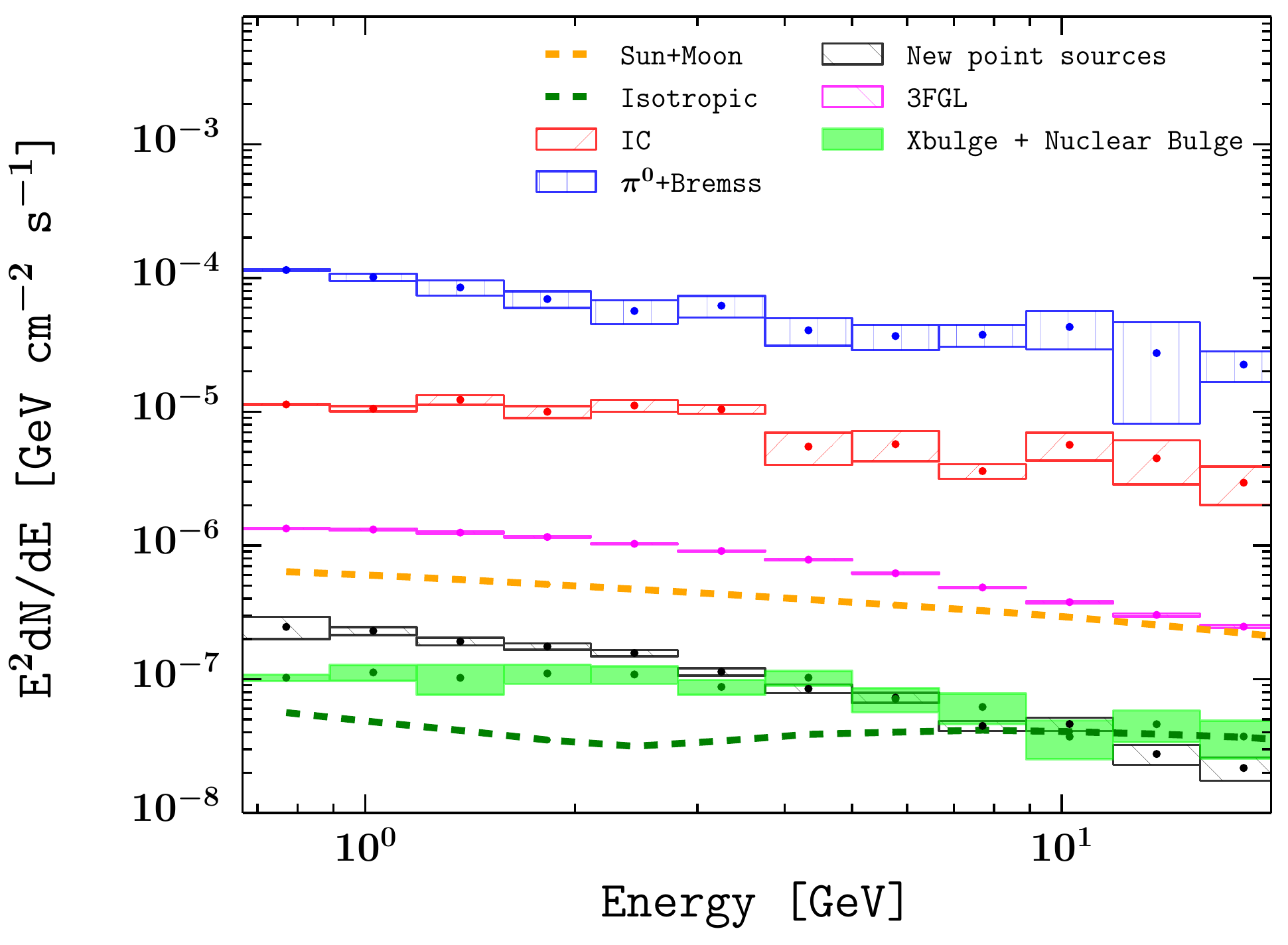}
\end{tabular}
\caption{\small Top: Differential flux of the different components in the $15^{\circ}\times 15^{\circ}$ region about the GC. The model considered in the fit is the \textit{baseline$+$NP$+$NB-stars$+$X-bulge} model (see Tab.~\ref{Tab:likelihoods}). The box heights represent the the 68\% CI regions.
Bottom: spectral residuals associated to the fit and determined from the full ROI.\label{fig:ptsrc_spectra}}
\end{center}
\end{figure}


\begin{table*}[t!]\caption{Summary of the Likelihood analysis results\label{Tab:likelihoods}}
\centering
\begin{threeparttable}
\scriptsize 
\begin{tabular}{llllcc}
\toprule
Base Model & Source Model &  $\log(\mathcal{L}_{\mbox{\tiny Base}})$  & $\log(\mathcal{L}_{\mbox{\tiny Source}})$  &  $\mbox{TS}_{\mbox{\tiny Source}}$ & Number of\\ 
           &              &                                  &                         &                                               & free parameters\\\hline
standard   &standard$+$NFW$^2$ &    -172638.1          & -172034.8        & 1207   &  19\\
baseline  & baseline$+$NFW$^2$ &    -172461.4         &  -172167.9        & 587   &  19\\
baseline  & baseline$+$NFW$^2$-s &    -172461.4       &  -172265.3       & 392   &  19\\    
baseline  & baseline$+$X-bulge& -172461.4         &  -172224.1         & 475    & 19\\
baseline  & baseline$+$NB-stars&   -172461.4         &  -171991.8         & 939    & 19\\
baseline  & baseline$+$FB&     -172461.4          &   -172422.3    &  78   & 19\\
baseline  & baseline$+$NP&     -172461.4          &   -170305.5    & 4312   &  $43\times21$\\\hline
baseline$+$NP  & baseline$+$NP$+$NFW$^2$&     -170305.5          &   -170137.3    &  337   & 19\\
baseline$+$NP  & baseline$+$NP$+$X-bulge&     -170305.5          &   -170147.5   & 316     & 19\\
baseline$+$NP  & baseline$+$NP$+$NB-stars&     -170305.5          &   -170179.1   &253     & 19\\
baseline$+$NP  & baseline$+$NP$+$FB&     -170305.5          &  -170292.8    &25     & 19\\\hline
baseline$+$NP$+$NB-stars  & baseline$+$NP$+$NB-stars$+$X-bulge& -170179.1        &-170049.3      & 260    & 19\\

baseline$+$NP$+$NB-stars$+$X-bulge  & baseline$+$NP$+$NB-stars$+$X-bulge$+$NFW$^2$& -170049.3         &-170034.7
      & 29    & 19\\ \hline
baseline$+$NP$+$X-bulge  & baseline$+$NP$+$X-bulge$+$NFW$^2$& -170147.5         &-170082.7      & 130    & 19\\
baseline$+$NP$+$NB-stars  & baseline$+$NP$+$NB-stars$+$NFW$^2$& -170179.1          & -170106.2     & 146   & 19\\
baseline$+$NP$+$NB-stars$+$NFW$^2$  & baseline$+$NP$+$NB-stars$+$NFW$^2$$+$X-bulge    & -170106.2          & -170034.7     & 143   & 19\\\bottomrule

\end{tabular}
\begin{tablenotes}
      
      \item \textbf{Notes.} The \textit{baseline} model consists of all 3FGL point sources in the ROI, Loop I, an IC template predicted by GALPROP, the Pohl et al. gas maps, the recommended isotropic emission map and a model for the Sun and the Moon. For our \textit{standard} model, the gas maps used correspond to the ones in the standard diffuse emission model~\cite{3FGL}. Other model templates considered are: the 43 new point sources (NP), a generalized NFW$^2$ profile with a mild slope $\gamma=1.2$ or a ``standard NFW$^2$'' (NFW$^2$-s) with slope $\gamma=1$, an infrared X-bulge template tracing old stars in the Galactic bulge, an NB-stars map of the Galactic ridge and a template accounting for the Fermi Bubbles (FB). See Sec.~\ref{sec:backgroundmodel} for detailed descriptions of the model templates. 
    \end{tablenotes}
\end{threeparttable}
\end{table*}
\normalsize

\section{Systematic Uncertainties}
\label{Sec:systematic_uncertainties}

We applied the bin-by-bin likelihood method to alternative versions of the background model introduced in Sec.~\ref{sec:methods}. Table~\ref{Tab:likelihoods} summarizes the TS values obtained when using optional background model templates like the X-bulge, the \textit{Fermi} Bubbles or the 20cm map.
The general shape of the NFW$^2$ spectrum and respective band TS values are appreciably affected by the inclusion of the X-bulge and/or 20cm models. This is shown in Fig.~\ref{fig:GCE_syst_uncert}.

An exception occurs when the gas maps used are the ones in the standard diffuse emission model. For this single case the NFW$^2$ source remains significant (TS$_{\mbox{\tiny NFW$^2$}}$=325). Note however that the set of point sources found in this work were singled out using our fiducial gas maps (Pohl et al gas templates). As discussed in Sec.~\ref{Sec:discussion}, a fraction of the new point sources are interpreted as corrections to the diffuse background model. In this sense, the same point source search technique should be rigorously applied to the ROI using the standard gas templates. We do not attempt this possibility in this work as the Pohl et al gas maps are a better description to the bremsstrahlung and hadronic gamma-ray emission in the ROI.   

The deduced NFW$^2$ integral flux in the energy range $1-150$ GeV is $1.9\times 10^{-7}$ ph cm$^{-2}$ s$^{-1}$ for the standard background model. Modeling the ROI with the baseline$+$NP yields a flux that is only slightly lower, $1.2\times 10^{-7}$ ph cm$^{-2}$ s$^{-1}$, whereas for the baseline$+$NP$+$X-bulge we obtain a  best-fit flux $9.6\times 10^{-8}$ ph cm$^{-2}$ s$^{-1}$. However, when our best-fitting background model baseline$+$NP$+$X-bulge$+$20-cm is used we get a much lower NFW$^2$ flux $4.6\times 10^{-8}$ ph cm$^{-2}$ s$^{-1}$.  We note that the flux detection significance for this last case is only  $\mbox{TS}_{\mbox{\tiny NFW$^2$}}=39$ for 19 degrees of freedom.

\section{Discussion}
\label{Sec:discussion}


Our analysis of seven years of LAT Pass 8 data for the GC region reveals excess emission that can be adequately described by point-like sources. About 80\% of the new point source candidates  found in this work lie in the Galactic plane ($\vert b \vert \leq 2\degree $). We identify two small regions with high concentrations of new point sources: close to the W28 supernova remnant and the GC  region.
As is apparent in Fig.~\ref{fig:tsmaps}-(f), the set of sources around the center of the image are not distributed in space with spherical symmetry. Instead, some of them seem to be correlated with the gas templates in Fig.~\ref{fig:densities}.

It is possible that a number of the point sources found in the ROI can be attributed to mis-identified diffuse Galactic emission. One example is the emission from gas clumps unseen by the radio observations used in the construction of our gas maps. We note that, based on the examination of the sources near Cygnus, Orion and molecular clouds, ref.~\cite{3FGL} established a background-dependent threshold TS to determine which sources are unlikely to be mere corrections to the diffuse emission model. We computed the mean photon count per pixel ($N_{\rm bkgd}$) integrated from 666.7 MeV to 11.8 GeV in our the diffuse model cube (that does not include the isotropic contribution) obtaining $N_{\rm bkgd} =33.5 $. According to ref.~\cite{3FGL} only sources with $\mbox{TS}>80$ (CG: THIS NEEDS TO BE CORRECTED FOR OUR LARGER DOF) are unlikely to be associated with diffuse features in our ROI. Also, small-scale self- absorption features may be caused by relatively cold (?50 K) gas clouds that are located in front of much warmer gas and sources of continuum emission \cite{Federici2015}. For example absorption features are seen in the $(l,b)=(0,0)$ pixel of the LAB survey \cite{LAB, Gibson2005}.

Figure~\ref{fig:tsmaps}-(a) indicates residual localized extended emission from an area close to the Sgr C position. In the past, the \textit{Fermi} team have modeled similar diffuse excesses with the use of several point sources close together. In our approach, we conservatively do not attempt to model the tentative localized diffuse emission with additional extended templates.

Our analysis confirms the existence of most of the 1FIG point sources found by~\cite{TroyandSimona:2015} that are not present in the 3FGL catalog. Table~\ref{tab:Sources_3fgl} shows these associations. As can be seen the set of sources with 1FIG correspondence are among the sources with the highest TS calculated values. The significantly larger number of new point source candidates found in this work could be due to a number of differences in the analysis methods and data considered: (i) this work uses higher statistics and an a higher quality data set (\texttt{Pass8} compared to the superseded \texttt{Pass7Rep}) and (ii) we use the \texttt{gttsmap} tool more efficiently as we break the energy range in a larger number of bins - which makes the assumption of a PL spectra with index $\alpha=2.0$ (in \texttt{gttsmap}) a less biased assumption. In principle, this way \texttt{gttsmap} is able to pick up sources with arbitrary spectra.

\begin{figure}[ht!]
\begin{center}
\begin{tabular}{c}
\centering
\includegraphics[scale=0.5]{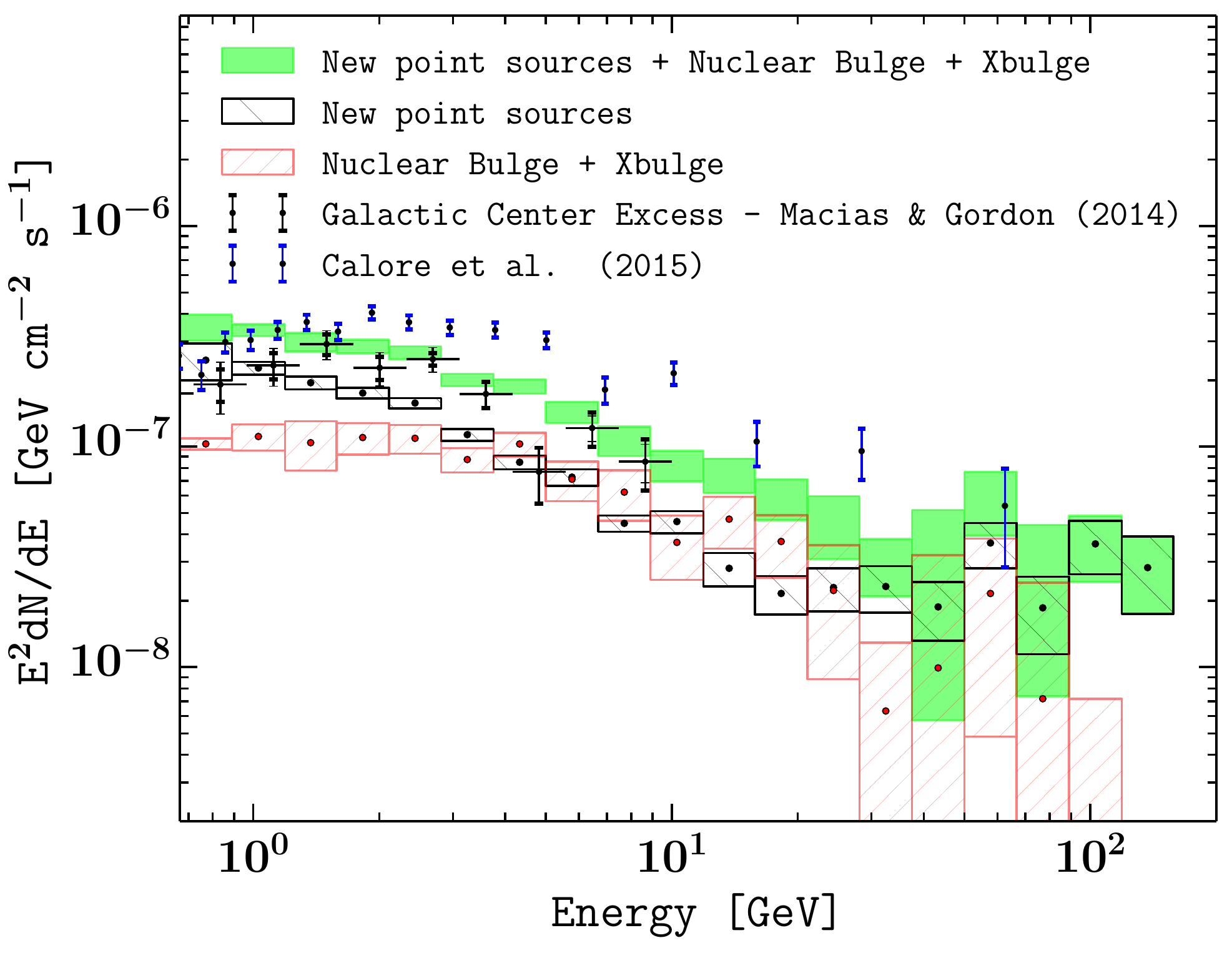}
\end{tabular}
\caption{\small Differential flux of the new statistically significant components in the GC. The model considered in the fit is the \textit{baseline$+$NP$+$NB-stars$+$X-bulge} model (see Tab.~\ref{Tab:likelihoods}). The black boxes are the combined spectrum of the 43 new point source candidates, red boxes are the superposition of the NB-stars and X-bulge differential fluxes, while the green boxes display the sum of these three components.   \label{fig:Xbulge+NBstars_spectra}}
\end{center}
\end{figure}

Most of the point sources found here can be interpreted as mere corrections to the diffuse Galactic emission model. Our analysis uses an IC emission model that is different to the one used in ref.~\cite{TroyandSimona:2015}. However, statistical analysis demonstrates that point-like rather than truly diffuse extended emission is a better hypothesis for the GCE in the GC.

We also note that ref.~\cite{TroyandSimona:2015} use a broad-band method to fit the ROI. This presents significant convergence issues for the very large number of parameters in the ROI. Our bin-by-bin method converges reliably and in a simple fashion.

\onecolumn
\scriptsize
\begin{longtable}{cccccccc}
\caption{Point Sources Detected with $TS > 55$ for the $15^\circ \times 15^\circ$ region about the GC. }\\
\hline\hline
Name &   $l$    &   $b$     &    TS     &  $F_{1-100\,{\rm GeV}}$                  & Type & Spectrum &  1FIG~\cite{TroyandSimona:2015} Association\\
     &   [deg]  &   [deg]   &           & $\times 10^{-9}$ [ph cm$^{-2}$ s$^{-1}$] &      &          & or spatial overlap \\ \hline
FGC	J1756.0	-2508	&	4.44	&	-0.049	&	225.5	&	$	13.2	\pm	1.0	$	&	PL	&	$	3.1	\pm	0.1								$	&	1FIGJ1755.5-2511	\\
FGC	J1801.3	-2449	&	5.32	&	-0.93	&	124.4	&	$	10.3	\pm	0.9	$	&	PL	&	$	2.7	\pm	0.1								$	&	1FIGJ1801.2-2451,PSR J1801-2451~\cite{ATNF:catalog}	\\
FGC	J1744.2	-2932	&	359.33	&	-0.054	&	391.7	&	$	1.1	\pm	1.4	$	&	PL	&	$	2.4	\pm	0.4								$	&	1FIGJ1744.2-2930	\\
FGC	J1739.6	-3022	&	358.1	&	0.35	&	363.3	&	$	14.0	\pm	1.0	$	&	PL	&	$	3.0	\pm	0.1								$	&	1FIGJ1739.4-3010,PSR J1739-3023~\cite{ATNF:catalog}	\\
FGC	J1735.9	-3028	&	357.59	&	0.96	&	163.3	&	$	8.0	\pm	0.8	$	&	PLE	&	$				1.9	\pm	0.4	,\;	5.0	\pm	3.3	$	&	1FIGJ1735.4-3030, Terzan 1~\cite{GBC:catalog}	\\
FGC	J1737.2	-3145	&	356.64	&	0.053	&	177.7	&	$	14.3	\pm	1.0	$	&	PL	&	$	2.9	\pm	0.1								$	&	1FIGJ1737.4-3144	\\
FGC	J1734.9	-3228	&	355.78	&	0.073	&	91.3	&	$	10.0	\pm	1.0	$	&	PL	&	$	2.8	\pm	0.1								$	&	1FIGJ1734.6-3228	\\
FGC	J1731.6	-3235	&	355.3	&	0.6	&	143.3	&	$	11.5	\pm	0.9	$	&	PLE	&	$				2.2	\pm	0.4	,\;	4.1	\pm	2.4	$	&	1FIGJ1731.3-3235	\\
FGC	J1729.1	-3443	&	353.24	&	-0.144	&	109.2	&	$	8.1	\pm	1.2	$	&	PL	&	$	2.9	\pm	0.1								$	&	1FIGJ1728.6-3433	\\
FGC	J1736.5	-3420	&	354.38	&	-1.22	&	101.9	&	$	5.1	\pm	0.6	$	&	PLE	&	$				1.5	\pm	0.5	,\;	3.5	\pm	1.8	$	&	1FIGJ1736.1-3422	\\
FGC	J1759.7	-2353	&	5.95	&	-0.15	&	100.4	&	$	7.0	\pm	2.9	$	&	PL	&	$	2.4	\pm	0.1								$	&		\\
FGC	J1758.1	-2320	&	6.24	&	0.45	&	59.7	&	$	3.8	\pm	1.1	$	&	PL	&	$	2.5	\pm	0.2								$	&		\\
FGC	J1727.7	-2302	&	2.84	&	6.55	&	71.5	&	$	1.7	\pm	0.3	$	&	PL	&	$	2.3	\pm	0.1								$	&		\\
FGC	J1711.3	-3008	&	354.85	&	5.55	&	78.2	&	$	2.3	\pm	0.3	$	&	PL	&	$	2.8	\pm	0.1								$	&		\\
FGC	J1733.1	-2910	&	358.34	&	2.19	&	71.3	&	$	2.8	\pm	0.6	$	&	PL	&	$	2.6	\pm	0.2								$	&		\\
FGC	J1805.1	-3619	&	355.65	&	-7.24	&	64.9	&	$	1.6	\pm	0.2	$	&	PLE	&	$				0.4	\pm	0.7	,\;	1.1	\pm	0.3	$	&		\\
FGC	J1734.6	-2202	&	4.531	&	5.76	&	73.7	&	$	2.9	\pm	0.3	$	&	PL	&	$	4.1	\pm	0.3								$	&		\\
FGC	J1739.2	-2530	&	2.16	&	3.02	&	56.2	&	$	2.6	\pm	0.6	$	&	PL	&	$	2.6	\pm	0.2								$	&		\\
FGC	J1754.9	-2609	&	3.45	&	-0.35	&	72.9	&	$	8.8	\pm	0.9	$	&	PL	&	$	3.0	\pm	0.2								$	&		\\
FGC	J1802.4	-2352	&	6.26	&	-0.67	&	77	&	$	7.6	\pm	1.6	$	&	PL	&	$	2.4	\pm	0.1								$	&		\\
FGC	J1739.3	-3056	&	357.57	&	0.111	&	111.3	&	$	11.4	\pm	1.2	$	&	PL	&	$	3.0	\pm	0.1								$	&	1FIGJ1740.1-3057	\\
FGC	J1742.7	-3005	&	358.69	&	-0.066	&	161.2	&	$	7.1	\pm	1.7	$	&	PL	&	$	2.9	\pm	0.2								$	&		\\
FGC	J1742.7	-2907	&	359.5	&	0.438	&	90.4	&	$	6.2	\pm	1.4	$	&	PL	&	$	3.0	\pm	0.2								$	&		\\
FGC	J1739.1	-2931	&	358.75	&	0.903	&	84.5	&	$	6.6	\pm	0.8	$	&	PL	&	$	3.3	\pm	0.2								$	&		\\
FGC	J1730.5	-3353	&	354.09	&	0.078	&	113.2	&	$	8.6	\pm	1.0	$	&	PL	&	$	3.0	\pm	0.2								$	&	1FIGJ1730.2-3351,G354.1+00.1~\cite{Green:SNRcatalog}	\\
FGC	J1752.6	-3029	&	359.45	&	-2.101	&	62.8	&	$	2.1	\pm	0.5	$	&	PL	&	$	3.1	\pm	0.3								$	&		\\
FGC	J1758.5	-3028	&	0.089	&	-3.187	&	63.9	&	$	1.2	\pm	0.3	$	&	PL	&	$	2.2	\pm	0.2								$	&		\\
FGC	J1812.9	-3145	&	0.47	&	-6.5	&	61.7	&	$	1.3	\pm	0.2	$	&	PL	&	$	2.2	\pm	0.1								$	&	NGC 6569~\cite{GBC:catalog}	\\
FGC	J1756.4	-2440	&	4.89	&	0.11	&	59.7	&	$	6.4	\pm	1.1	$	&	PL	&	$	2.6	\pm	0.2								$	&		\\
FGC	J1742.9	-3050	&	358.08	&	-0.51	&	66.3	&	$	8.1	\pm	1.5	$	&	PL	&	$	2.7	\pm	0.2								$	&		\\
FGC	J1744.6	-2918	&	359.56	&	-0.0058	&	84.5	&	$	0.9	\pm	0.4	$	&	PL	&	$	1.7	\pm	0.3								$	&	1FIGJ1745.0-2905	\\
FGC	J1744.5	-2851	&	359.95	&	0.245	&	65.7	&	$	3.5	\pm	2.2	$	&	PL	&	$	2.8	\pm	0.4								$	&		\\
FGC	J1733.9	-2948	&	357.9	&	1.7	&	55.8	&	$	1.4	\pm	0.4	$	&	PL	&	$	2.1	\pm	0.2								$	&		\\
FGC	J1801.7	-2324	&	6.6	&	-0.3	&	131.2	&	$	16.3	\pm	4.1	$	&	PL	&	$	2.6	\pm	0.1								$	&	1FIGJ1801.4-2330	\\
FGC	J1759.5	-2310	&	6.54	&	0.26	&	55	&	$	4.6	\pm	1.9	$	&	PL	&	$	2.5	\pm	0.2								$	&		\\
FGC	J1757.2	-2411	&	5.4	&	0.21	&	59.7	&	$	5.2	\pm	1.3	$	&	PL	&	$	2.2	\pm	0.3								$	&		\\
FGC	J1751.7	-2657	&	2.39	&	-0.12	&	101.1	&	$	10.8	\pm	1.0	$	&	PL	&	$	3.1	\pm	0.2								$	&		\\
FGC	J1747.4	-2809	&	0.86	&	0.07	&	56	&	$	9.6	\pm	2.1	$	&	PL	&	$	2.7	\pm	0.2								$	&	G00.9+00.1~\cite{Green:SNRcatalog},PSR J1747-2809~\cite{ATNF:catalog}	\\
FGC	J1743.7	-2950	&	359.02	&	-0.12	&	56.8	&	$	8.2	\pm	2.1	$	&	PL	&	$	3.1	\pm	0.6								$	&		\\
FGC	J1733.3	-3318	&	354.89	&	-0.1	&	57.4	&	$	6.1	\pm	0.9	$	&	PL	&	$	2.9	\pm	0.2								$	&	Liller 1~\cite{GBC:catalog}	\\
FGC	J1729.9	-3423	&	353.61	&	-0.105	&	57.2	&	$	4.2	\pm	1.2	$	&	PL	&	$	2.3	\pm	0.3								$	&		\\
FGC	J1725.9	-3429	&	353.05	&	0.534	&	55.7	&	$	3.1	\pm	0.8	$	&	PL	&	$	2.4	\pm	0.2								$	&		\\
FGC	J1800.9	-2353	&	6.09	&	-0.394	&	54.6	&	$	1.3	\pm	0.3	$	&	PL	&	$	1.2	\pm	0.2								$	&		\\\hline
\multicolumn{8}{p{18cm}}{\textbf{Notes.} Point source candidate localizatios and total band TS-values for $19+2$ degrees of freedom $-$ note that a $\sim 4\sigma$ significance detection corresponds to $\mbox{TS}\geq 55$. Best-fit fluxes obtained in the $1-100$ GeV energy range are denoted by $F_{1-100\,{\rm GeV}}$. The last column displays 1FIG~\cite{TroyandSimona:2015} associations as well as spatial overlaps with tentative multi-wavelenght counterparts. We use the globular cluster catalog~\cite{GBC:catalog}, ATNF pular catalog~\cite{ATNF:catalog} and the Green's SNR catalog~\cite{Green:SNRcatalog}.}\\
\label{tab:Sources_3fgl}
\end{longtable}
\normalsize

\section{Conclusions}
We have shown that the GCE can be explained by standard astrophysical processes. Firstly the gas maps of the GC needed to be improved by using the hydrodynamical approach of Pohl. et al. \cite{Pohl2008} rather than the traditional interpolation approach. Once this was done it became apparent that the excess had an x-shaped morphology which was well correlated with the previously-known\cite{Nataf2010,NessLang2016} X-shaped stellar over-density in the Galactic bulge. We also find a strong correlation with 20 cm continuum emission associated with bremsstrahlung emission in the galactic ridge.


\bibliography{GCEuncertainty}
